\documentclass[twocolumn, trackchanges]{aastex6}

\shorttitle{INFRARED PROPERTIES OF \textit{SWIFT}/BAT AGN}
\shortauthors{Ichikawa et al.}

\usepackage{amsmath}
\usepackage{color}
\usepackage{url}
\usepackage{ulem}
\usepackage{multirow}
\usepackage{longtable}

\usepackage{graphicx}
\usepackage{epstopdf}

\begin{document}


\title{The Complete Infrared View of Active Galactic Nuclei from the 70-month \textit{Swift}/BAT Catalog}


\author{Kohei~Ichikawa\altaffilmark{1,2,3}, 
Claudio Ricci\altaffilmark{4,5},
Yoshihiro~Ueda\altaffilmark{6},
Kenta~Matsuoka\altaffilmark{6},
Yoshiki~Toba\altaffilmark{7},
Taiki~Kawamuro\altaffilmark{6},
Benny Trakhtenbrot\altaffilmark{8},
and Michael J. Koss\altaffilmark{8}
}
\altaffiltext{1}{National Astronomical Observatory of Japan, 2-21-1 Osawa, Mitaka, Tokyo 181-8588, Japan}
\altaffiltext{2}{Department of Physics and Astronomy, University of Texas at San Antonio, One UTSA Circle, San Antonio, TX 78249, USA}
\altaffiltext{3}{Department of Astronomy, Columbia University, 550 West 120th Street, New York, NY 10027, USA}
\altaffiltext{4}{Institute of Astrophysics, Pontificia Universidad Catolica de Chile, Avenida Vicua Mackenna 4860, 7820436, Chile}
\altaffiltext{5}{Kavli Institute for Astronomy and Astrophysics, Peking University, Beijing 100871, China}
\altaffiltext{6}{Department of Astronomy, Kyoto University, Kitashirakawa-Oiwake-cho, Sakyo-ku, Kyoto 606-8502, Japan}
\altaffiltext{7}{Academia Sinica Institute of Astronomy and Astrophysics, P.O. Box 23-141, Taipei 10617, Taiwan}
\altaffiltext{8}{Institute for Astronomy, Department of Physics, ETH Zurich, Wolfgang-Pauli-Strasse 27, CH-8093 Zurich, Switzerland}
\email{k.ichikawa@astro.columbia.edu}




\begin{abstract}

We systematically investigate the near- (NIR)  to far-infrared (FIR) photometric properties of 
a nearly complete sample of local active galactic nuclei (AGN) detected in the \textit{Swift}/Burst
 Alert Telescope (BAT) all-sky ultra hard X-ray (14--195 keV) survey. 
Out of 606 non-blazar AGN in the \textit{Swift}/BAT 70-month catalog at high galactic
 latitude of $|b|>10^{\circ}$, we obtain IR photometric data of 604 objects by 
cross-matching the AGN positions with catalogs from the \textit{WISE}, \textit{AKARI},
 \textit{IRAS}, and \textit{Herschel} infrared observatories.
We find a good correlation between the ultra-hard X-ray and mid-IR (MIR) luminosities
 over five orders of magnitude ($41 < \log (L_{14-195}/{\rm erg}~{\rm s}^{-1})< 46$).
Informed by previous measures of the intrinsic spectral energy distribution of AGN,
we find FIR pure-AGN candidates whose FIR emission is thought to be AGN-dominated
 with low starformation activity.
We demonstrate that the dust covering factor decreases with the bolometric
AGN luminosity, confirming the luminosity-dependent unified scheme.
We also show that the completeness of the \textit{WISE} color-color cut in
 selecting \textit{Swift}/BAT AGN increases strongly with 14--195~keV luminosity.
\end{abstract}





\section{INTRODUCTION}
Understanding the cosmic evolution of supermassive black holes (SMBHs) in galactic
 centers and their connections with the evolution of their host galaxies is one of the 
 main goals in modern astronomy. 
 Active galactic nuclei (AGN) are the fundamental laboratories in those studies because
  they are in the stage where the surrounding gas is accreting onto the SMBHs by releasing 
 their gravitational energy into radiation.
It is known that the central engines of AGN are surrounded by a dusty ``torus'' \citep{kro86}. 
Since optical and ultraviolet emission is easily absorbed by the torus,
a complete survey of AGN including obscured populations is crucial to
 elucidate the growth history of SMBHs.

The ultra-hard X-ray ($E > 10$~keV) band is extremely useful
 for detecting the whole population of AGN because they have
 1) stronger penetrating power than optical/UV and even hard 
 ($E < 10$~keV) X-ray radiation and
 2) very little contamination from the starburst emission.
Ultra-hard X-ray detectors such as \textit{Swift}/Burst Alert 
Telescope \citep[BAT,][]{bar05},
IBIS/ISGRI on board \textit{INTEGRAL} \citep{win03}, FPMA/FPMB
on board \textit{NuSTAR}~\citep{har13} are therefore well suited
 for those studies.
Among them, \textit{Swift}/BAT provides the most sensitive ultra-hard
 X-ray survey of the whole sky in the 14--195 keV range.

Since most of the \textit{Swift}/BAT sources are local objects, they
 have been observed by a large number of 
multi-wavelength facilities, which allow us to study their properties.
Follow-up studies below 10~keV have
 shown  that the fraction of obscured ($N_{\rm H} \ge 10^{22}$~cm$^{-2}$) AGN
 highly depends on the intrinsic X-ray luminosities \citep[e.g.,][]{bec09,bur11,
  ric14, kaw16a}, and also proved to be an effective tool to identify previously 
  missed class of AGN with small opening angle tori \citep[e.g.,][]{ued07, 
  win09, egu09,egu11, ric11}, and Compton-thick AGN \citep{gan15, ric15, tan16}.
 Studies carried out by optical spectroscopy enable us to investigate 
 the properties of extended ($>100$~pc) narrow line regions
 \citep[NLR; e.g., ][]{hai13, hai14a}
  through analysis of the [OIII]$\lambda5007$ emission line
   \citep{win09, ued15} and also offer the opportunity to estimate the black
    hole masses through the broad line regions or velocity dispersion measurements.
\textit{Swift}/BAT AGN Spectroscopic Survey (BASS) is in progress to complete 
the first large ($>$500) sample of BAT detected AGN with optical spectroscopy, 
which enables us to constrain the nature of the NLR \citep{kos16, ber15, oh16}.

Cross-matching the \textit{Swift}/BAT AGN with all-sky mid-infrared
 (MIR\footnote{Here we define near-IR (NIR) as
  $\lambda<5$~$\mu$m and MIR as $5$~$\mu$m~$ < \lambda \le 25$~$\mu$m 
  since all of the all-sky IR surveys used here cover IR bands in 
  $5$~$\mu$m~$ < \lambda \le 25$~$\mu$m, whereas only the \textit{WISE} survey
   covers IR bands at $\lambda < 5$~$\mu$m.}) catalogs can provide information
   on the dust surrounding the central engine.
While sometimes the MIR suffers contamination from the star formation, 
for luminous AGN the MIR is dominated by the torus dust re-emission 
with $T\sim200$--$300$~K.
This fact is used for new diagnostics identifying various AGN population \citep{mat12b},
 and it has shown that clumpy torus models \cite[e.g.,][]{nen02, nen08a, nen08b, 
 hon06, hon10, kaw10,kaw11, sch08,sta12,sie15}
 are favored to explain that MIR emission of AGN is almost isotropic
  \citep{mul11,ich12, asm15, gar16b} rather than the smooth torus 
  models \citep{pie92,pie93,efs95}.
 
 Near-IR (NIR) observations ($\lambda<5$~$\mu$m) are useful for identifying
 luminous obscured AGN because the NIR colors trace well the hot dust
   emission which cannot be reproduced by starburst galaxies \citep{lac04,ste05,hic07,ima10,mat12a,don12,ste12,ass13,ich14}.
 However, the color-color plots often miss the known X-ray selected 
 obscured/Compton-thick AGN due to the strong contamination from the
  host galaxies in the NIR bands \citep[e.g.,][]{gan14,gan15}, especially 
 at the low-luminosity end \citep{kaw16b}. 
 Thus we are motivated to evaluate the NIR two-color selection efficiency
  as a function of AGN luminosity, using a complete sample including 
  Compton-thick and low-luminosity AGN.

 On the other hand, far-IR (FIR; $\lambda \ge 60$~$\mu$m) data shed light
 on the starburst emission in the host galaxies of AGN.
Using IR Astronomical Satellite (\textit{IRAS}) FIR bands, \cite{rod87} found 
that the FIR 60~$\mu$m to 100~$\mu$m colors of nearby AGN and starburst 
galaxies are indistinguishable,
suggesting that most of the FIR emission of nearby AGN must originate from 
star formation processes \citep[see also; ][]{net07,mul11}.
Using the clumpy torus model, \cite{ich15} demonstrated that torus model 
emission is one order of magnitude smaller than the observed
\textit{Herschel} 70~$\mu$m data points, suggesting starburst emission is necessary
 in order to reproduce them.
Utilizing \textit{Herschel}/PACS 70/160~$\mu$m bands, \cite{mel14} and
 \cite{mus14} found that the FIR emission of most AGN is dominated
by the nuclear starburst within the $\sim2$~kpc scale, while there are 
exceptions in which the emission is dominated by the AGN torus \citep[e.g.,][]{mat15a,gar16}.
\cite{hat10} also found that the \textit{Spitzer}/MIPS and \textit{Herschel}/SPIRE
 two-color plot ($f_{250}/f_{70}$ and $f_{70}/f_{24}$) can separate AGN and 
 starburst galaxies because the 24~$\mu$m flux is dominated by the torus emission.
  However, the SPIRE colors alone do not differ from those of non-AGN galaxies.
Thus, combining the MIR and FIR as well as the hard X-ray band enables 
us to investigate the properties of torus, host galaxies, and 
accretion processes in AGN,
all of which are the key components to understand SMBH/host galaxy connection.

We report here the NIR to FIR (3--500~$\mu$m) properties of
 ultra-hard X-ray  selected AGN from the \textit{Swift}/BAT 70-month catalog
 \citep{bau13},
by cross-matching the AGN positions with the \textit{WISE}, \textit{AKARI},
 \textit{IRAS} all-sky surveys as well as the \textit{Herschel} archived data.
 The main advantage of the BAT 70 month survey compared to previous
  \textit{Swift}/BAT surveys includes better sensitivity resulting
from a complete reprocessing of the data with an improved data reduction
 pipeline and more exposure time. 
Throughout the paper, we adopt $H_{0}=70.0$~km~s$^{-1}$~Mpc$^{-1}$, 
$\Omega_{\rm M}=0.3$, and $\Omega_{\Lambda}=0.7$.

\begin{figure*}
\begin{center}
\includegraphics[width=4.6cm]{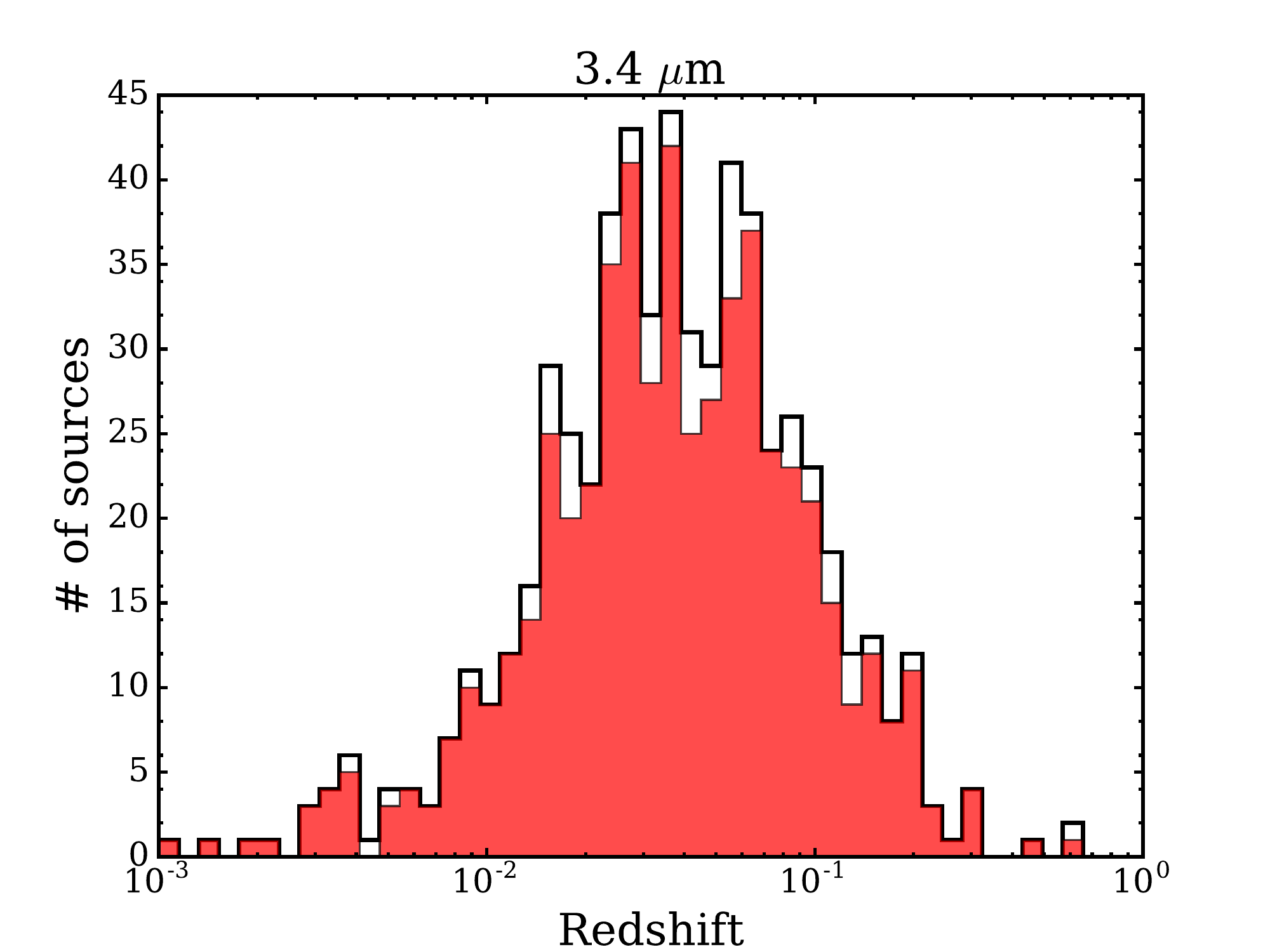}~
\includegraphics[width=4.6cm]{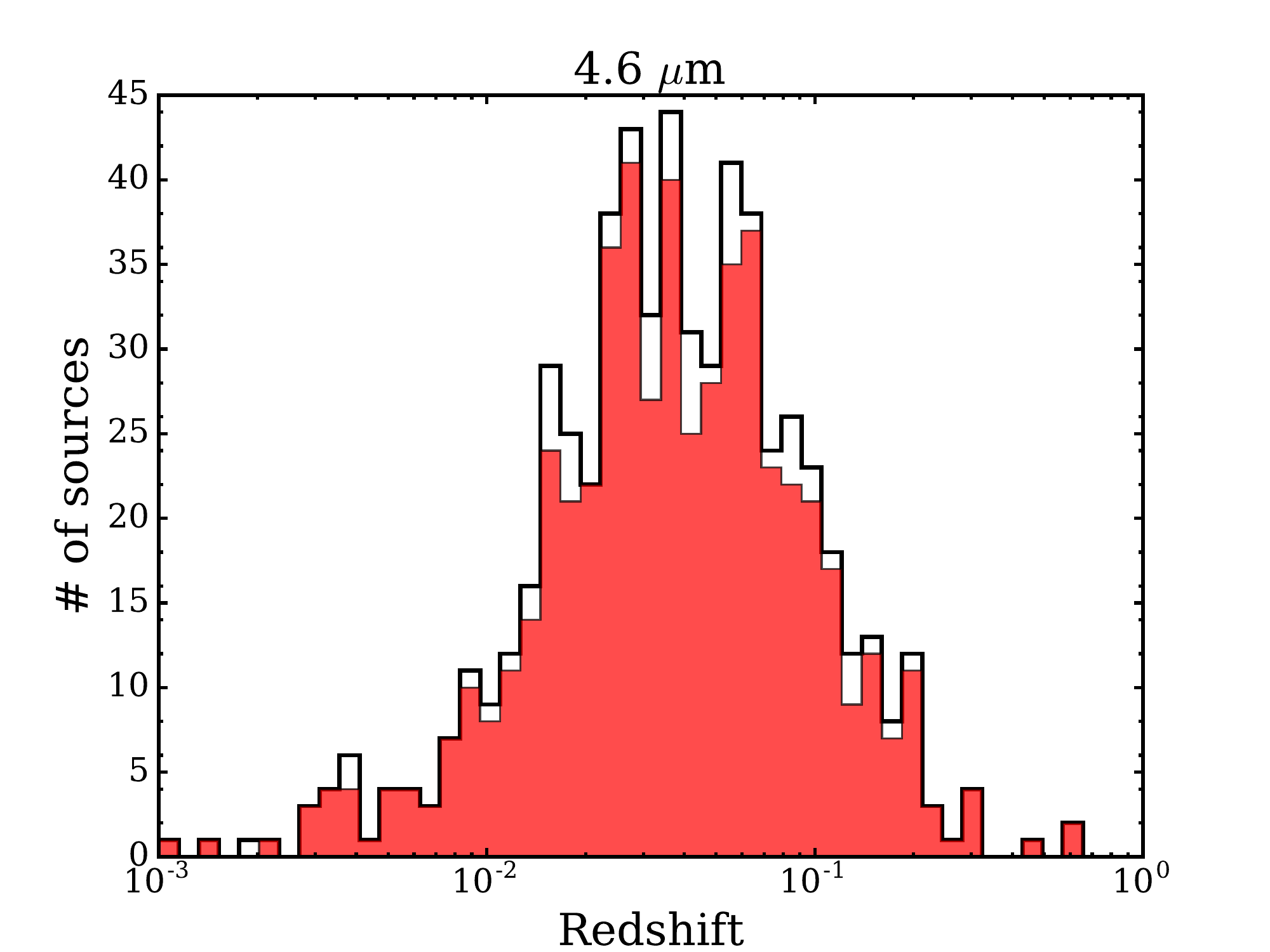}~
\includegraphics[width=4.6cm]{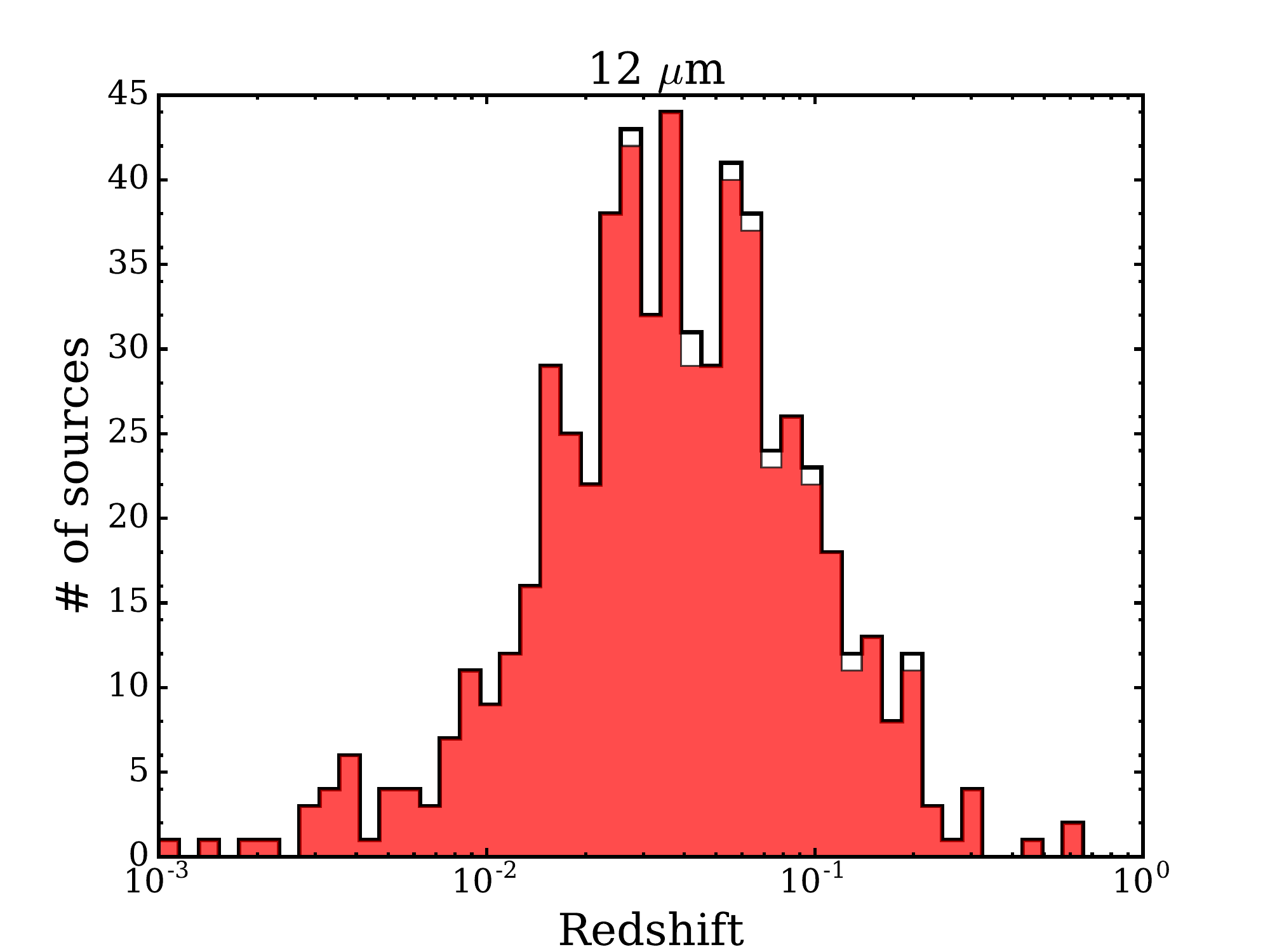}~
\includegraphics[width=4.6cm]{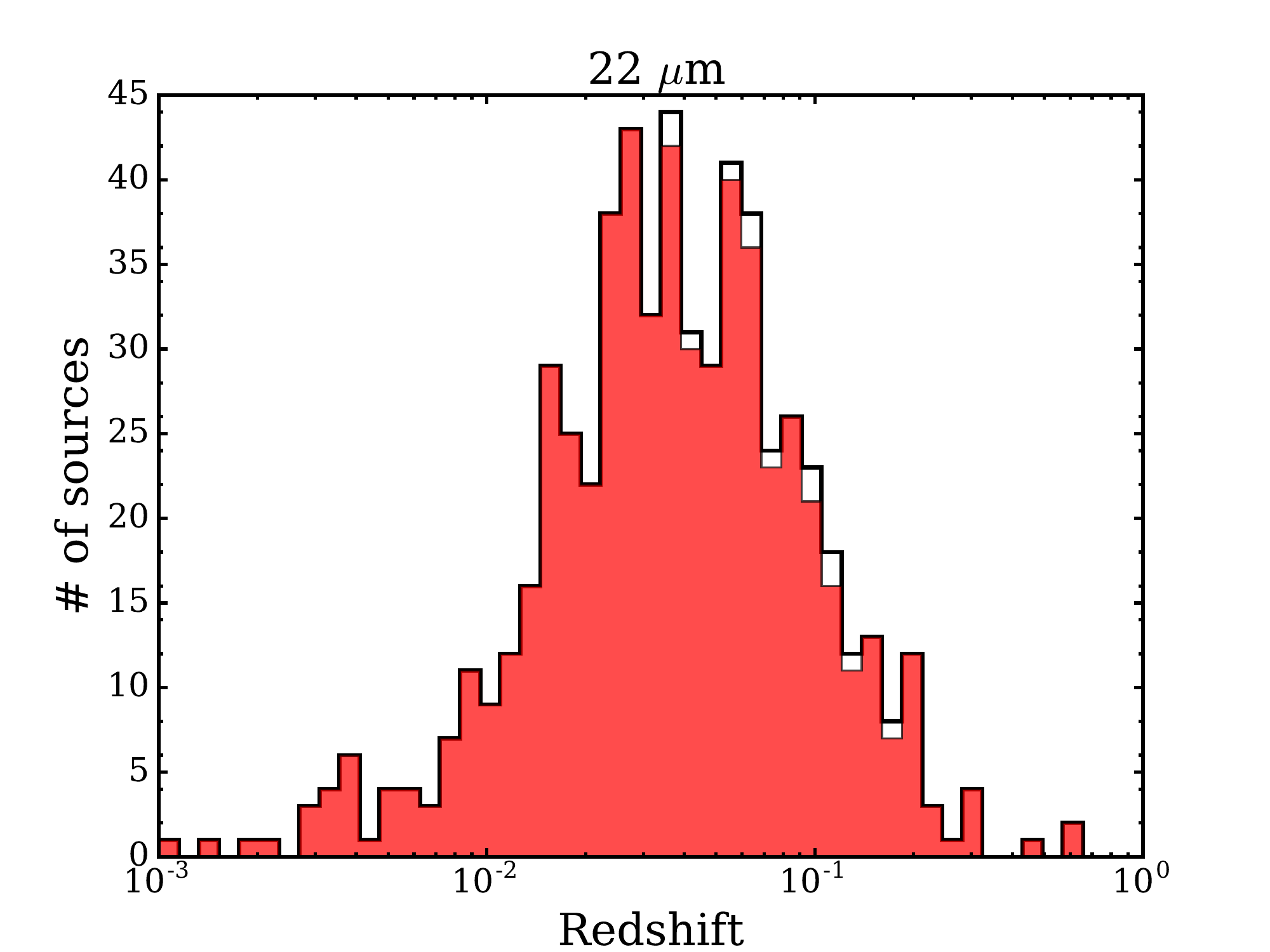}\\
\includegraphics[width=4.6cm]{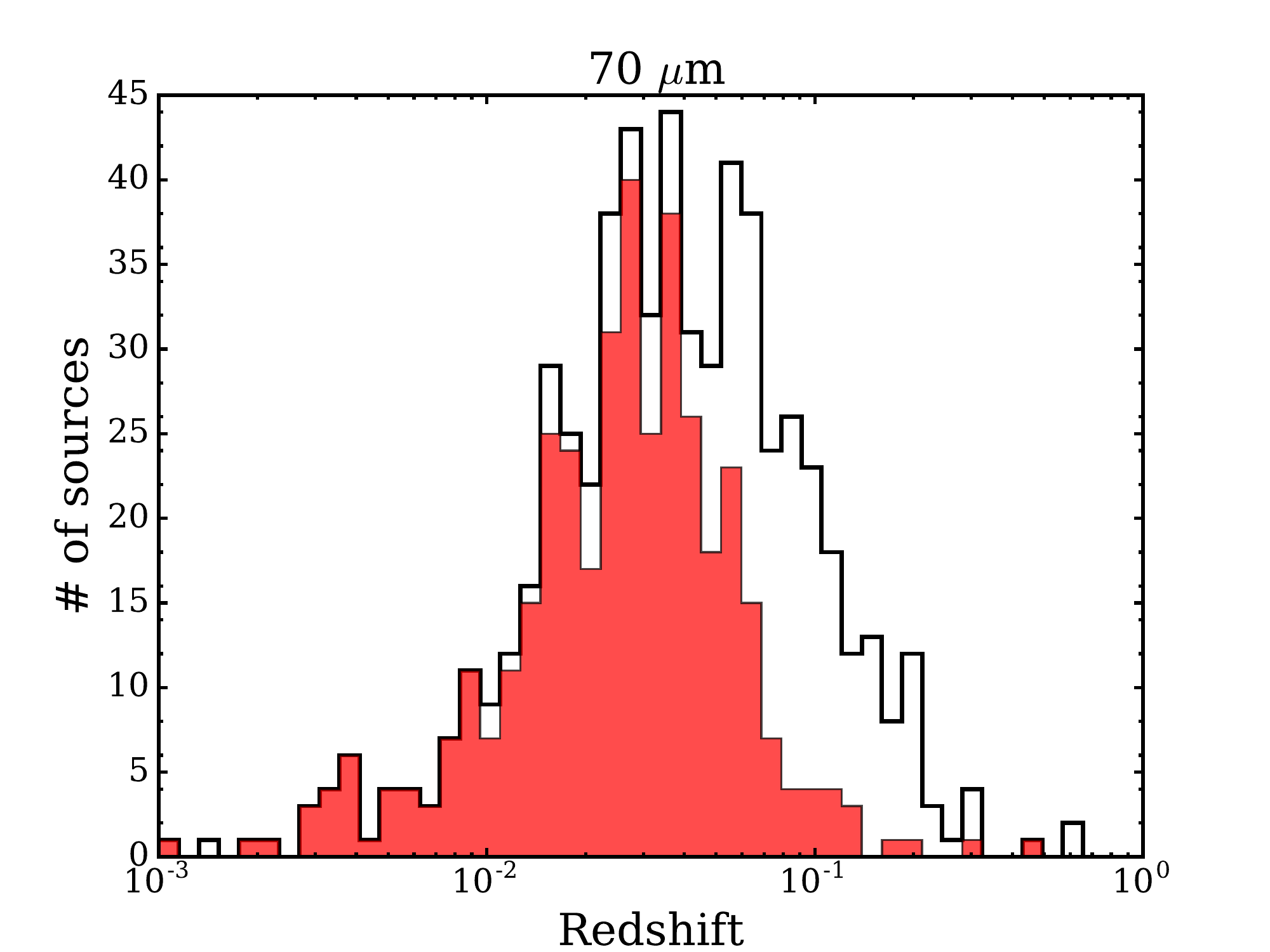}~
\includegraphics[width=4.6cm]{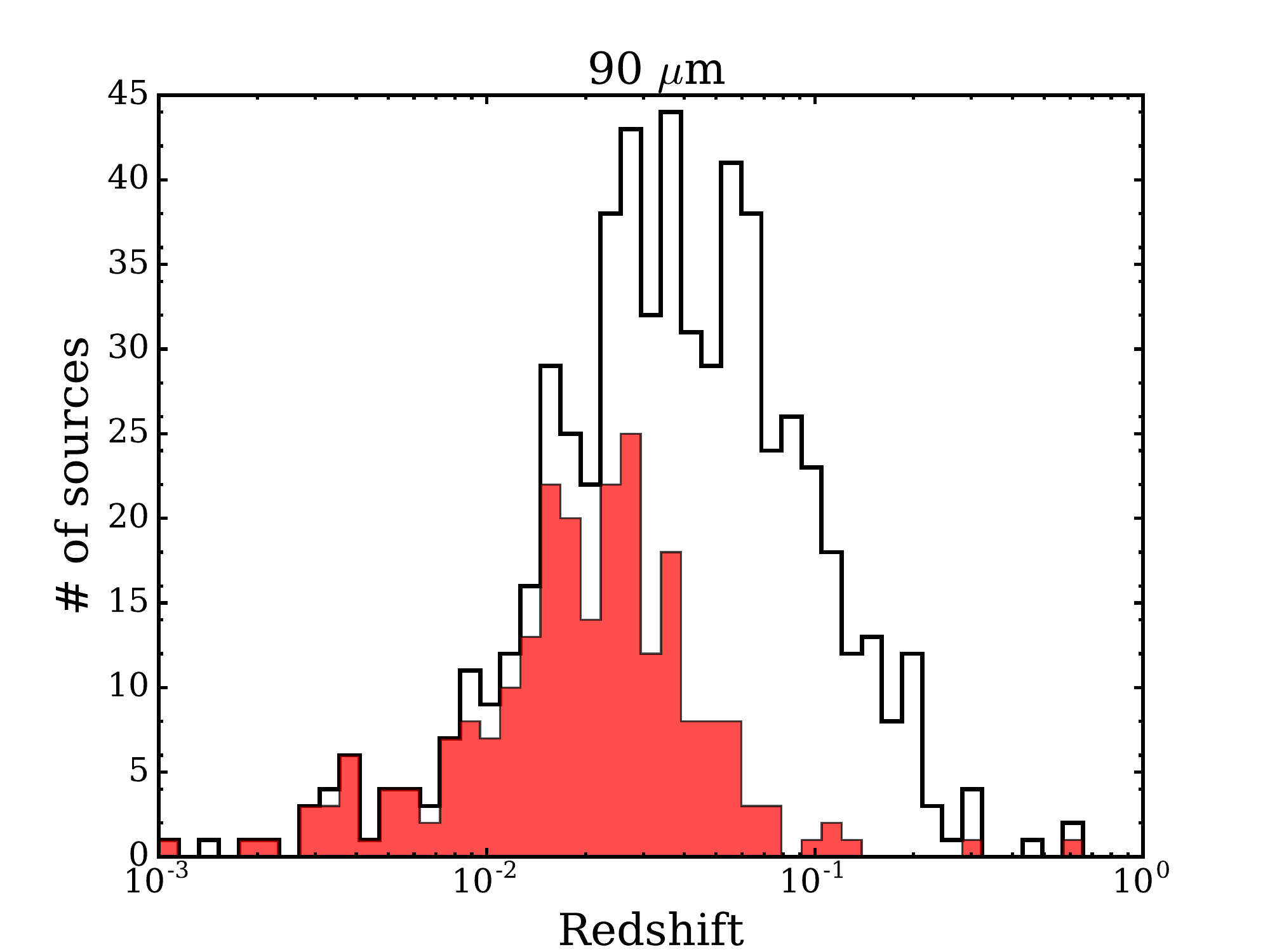}~
\includegraphics[width=4.6cm]{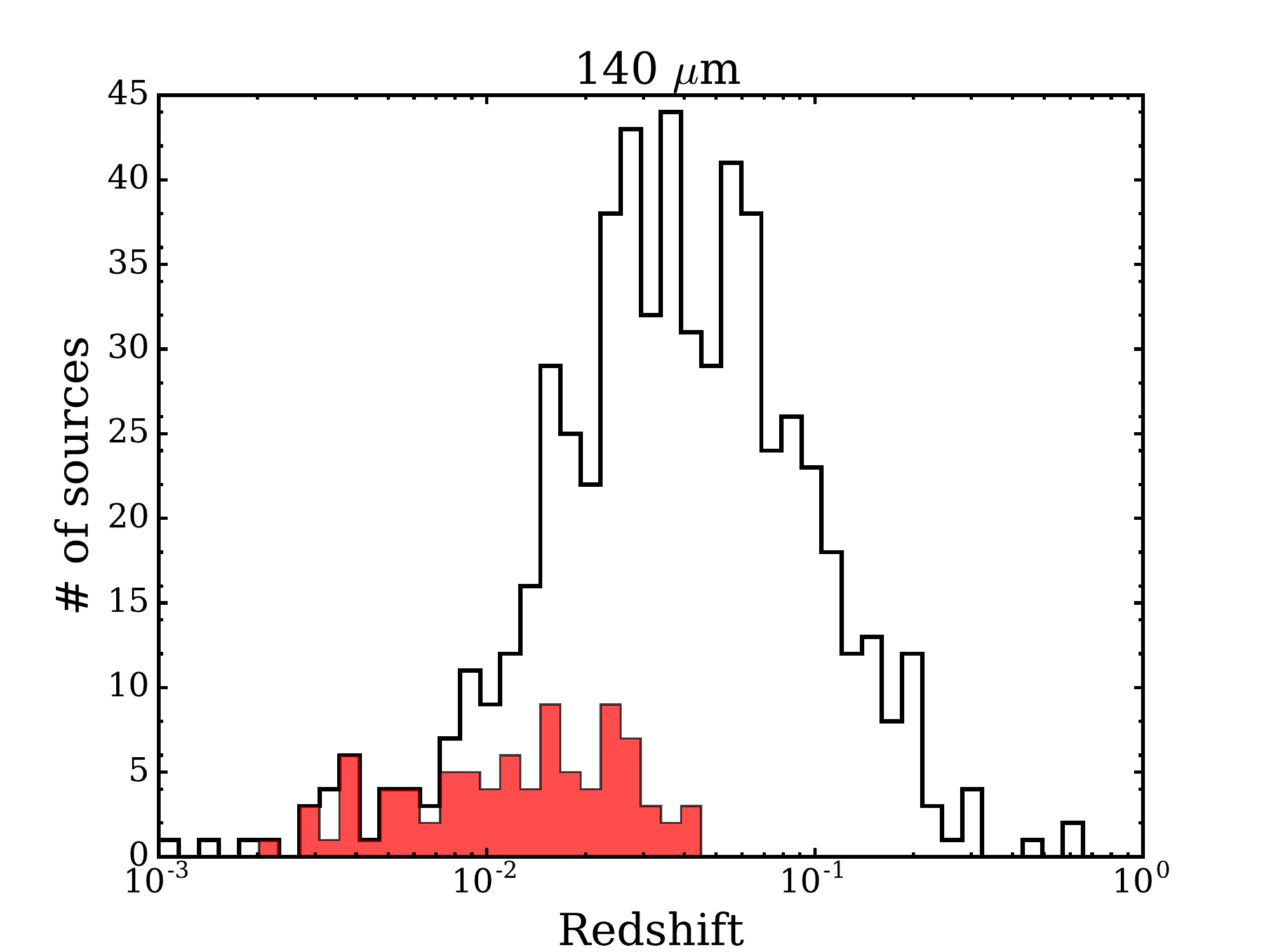}~ 
\includegraphics[width=4.6cm]{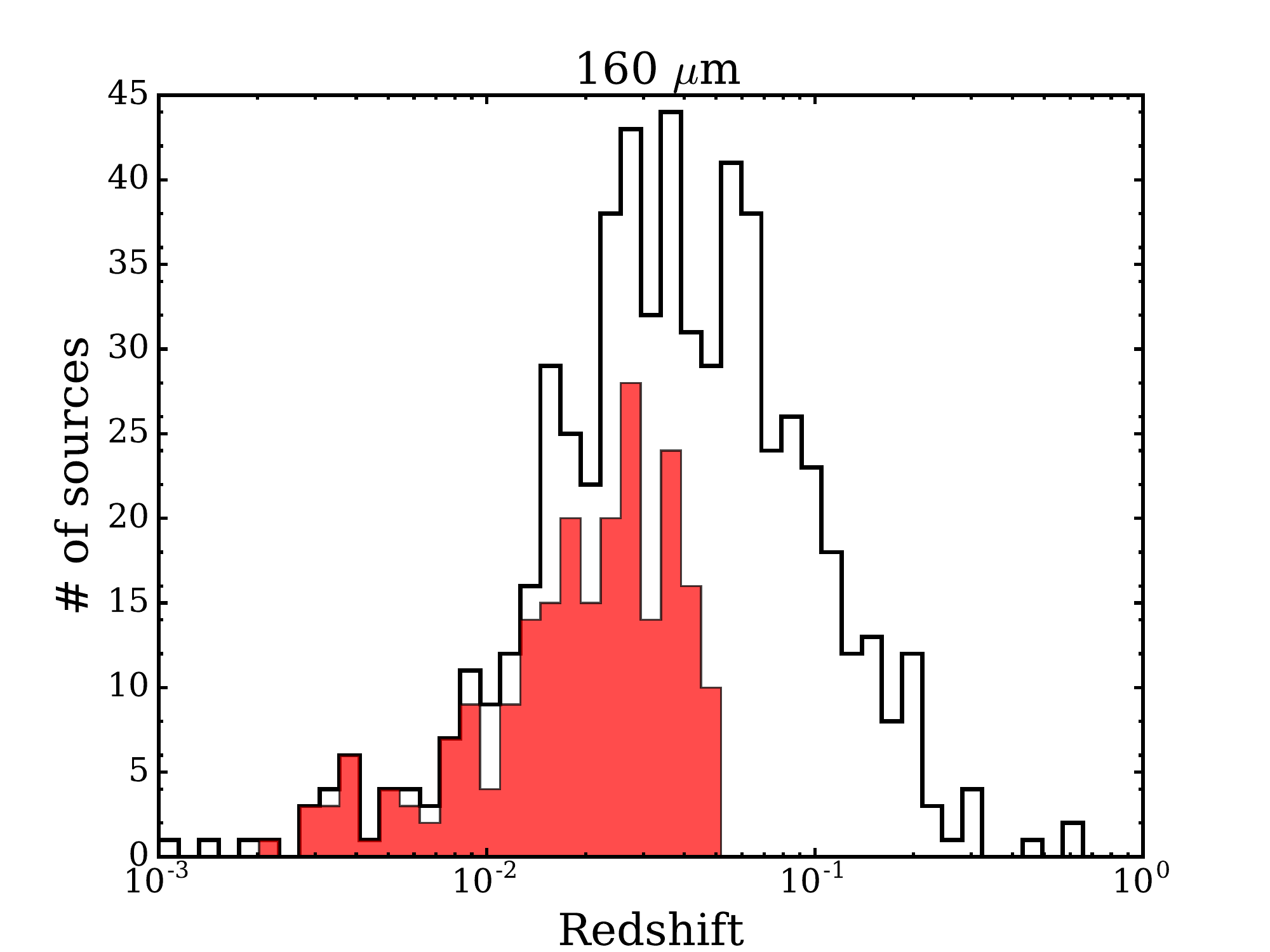}\\
\includegraphics[width=4.6cm]{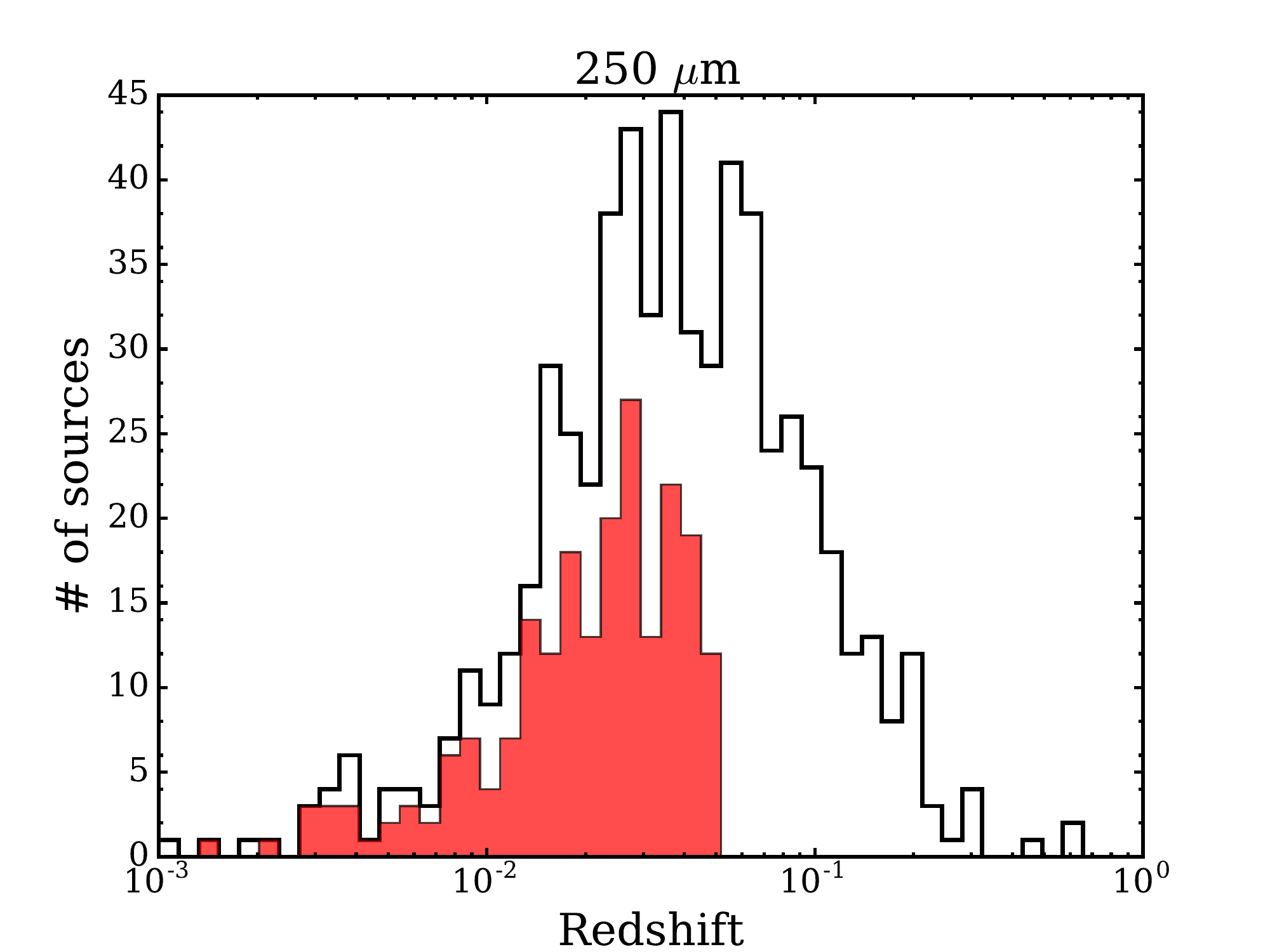}~
\includegraphics[width=4.6cm]{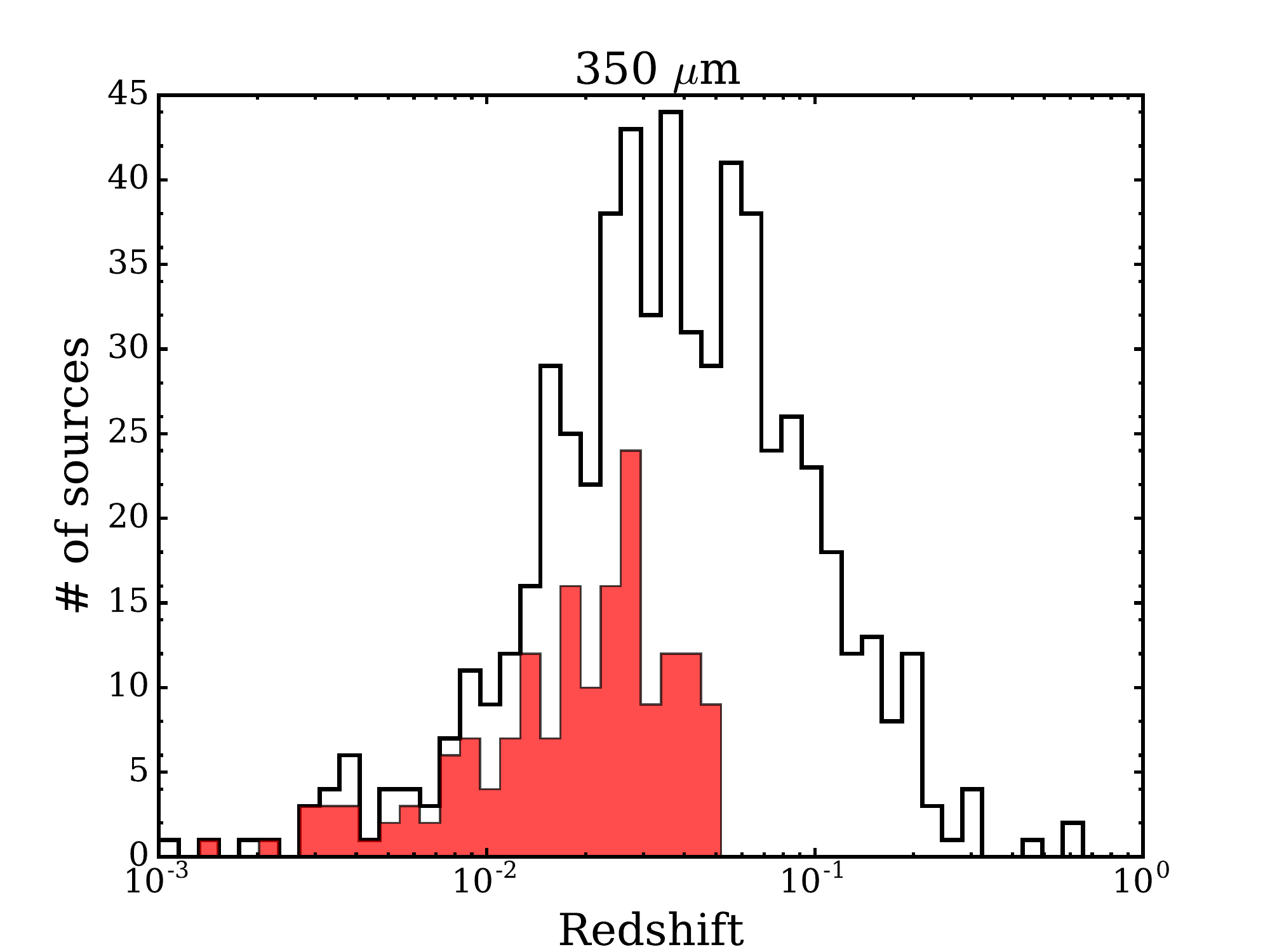}~
\includegraphics[width=4.6cm]{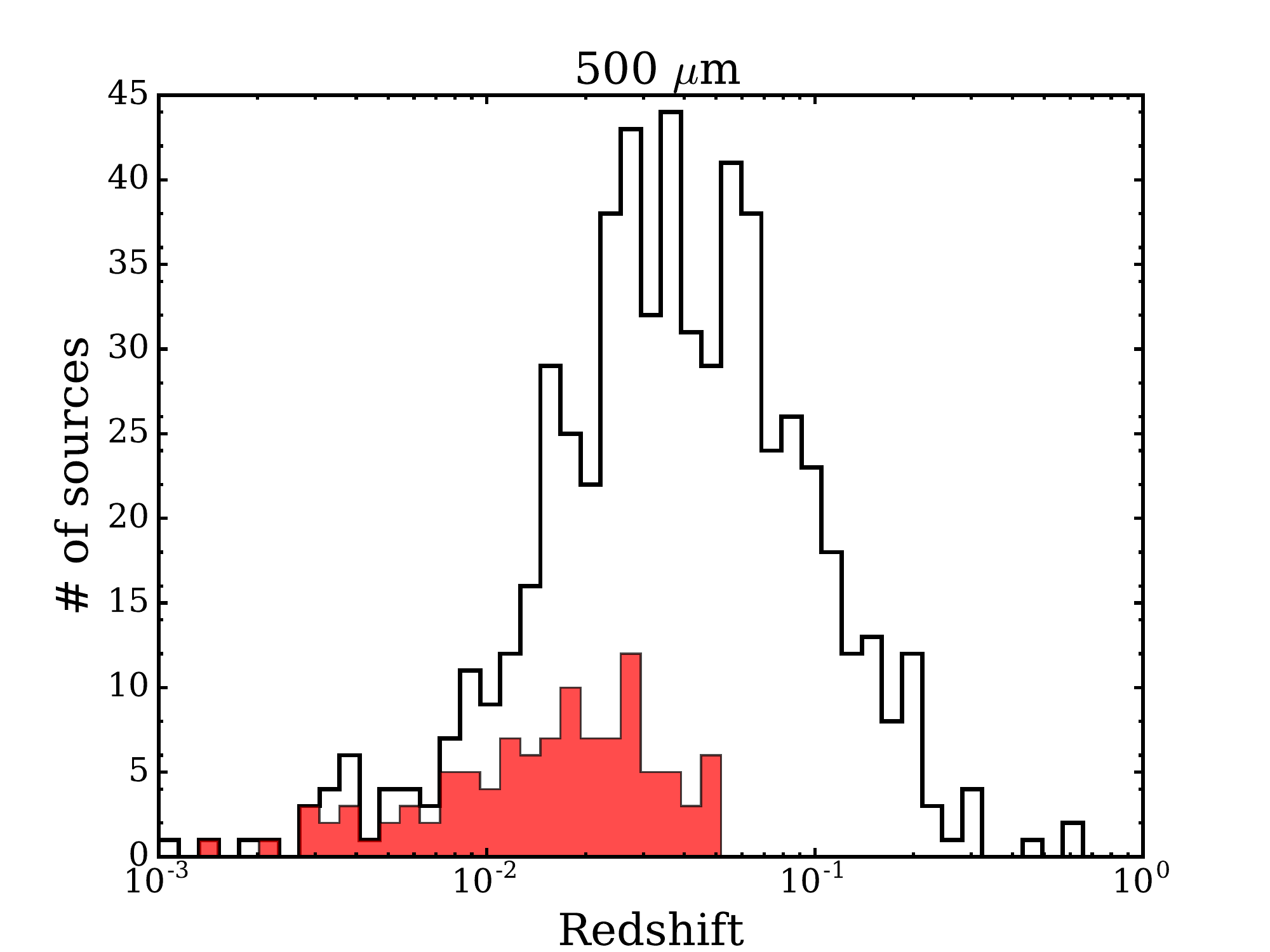}~
\includegraphics[width=4.6cm]{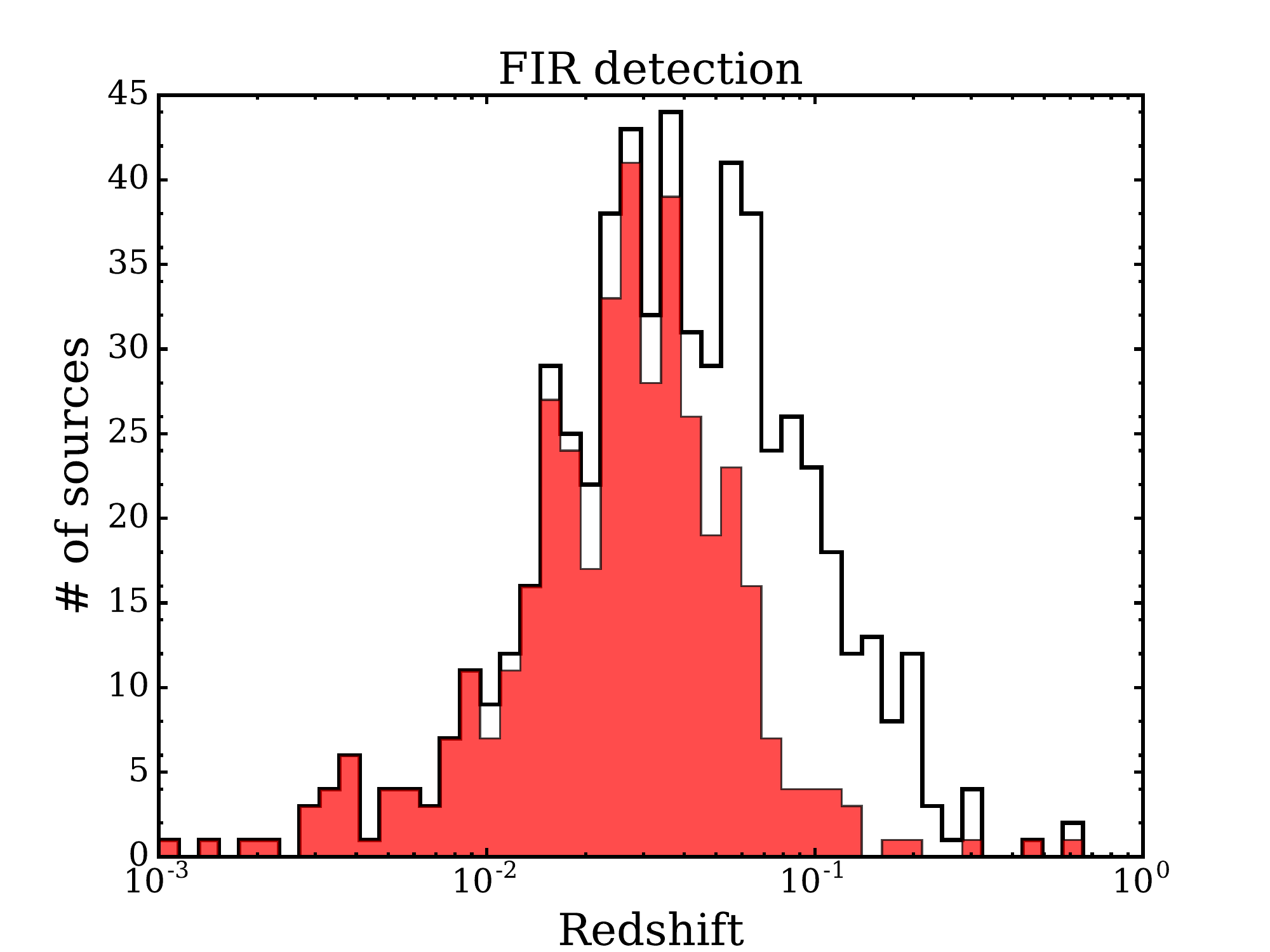}
\caption{
Redshift distribution of AGN in the \textit{Swift}/
BAT 70-month catalog (black solid line: 606 objects) and of those with IR
counterparts (red color area) at each wavelength. ``FIR detection'' represents
 the counterparts detected in any of the FIR bands.
The number of detected sources at each IR band is compiled in Table~2.}\label{fig:zdist}
\end{center}
\end{figure*}
 
\section{Sample}

\subsection{{\it Swift}/BAT Hard X-ray Catalog}
Our initial sample contains the 834 AGN reported in the 70-month
 \textit{Swift}/BAT catalog \citep{bau13,ric16}, of which
105 are blazars. Blazars were identified based on the Rome
BZCAT \citep{mas15} and on recent literature \citep{ric16}.
Of the remaining 729 sources, 697 sources have secure redshift information
as presented in \cite{ric16}.
Next, we removed galaxy pairs or interacting galaxies not resolved
 in the BAT survey because the BAT catalog in \cite{bau13} only 
 provides the counterpart name of the galaxy pair, not the galaxy itself,
which makes us to obtain the IR counterpart very difficult for those sources. 
Out of 697 sources, 684 fulfilled this criterion.
Further, the 606 sources located at higher galactic latitude with $|b|>10^{\circ}$
were selected to reduce the contamination in the crowded region through 
IR catalog matching.
 In the following we refer only to these 606 non-blazar AGN as the parent sample. 
 The sample is local, with an average redshift of $\left<z\right>=0.055$ as shown
  in Figure~\ref{fig:zdist} (black solid lines)\footnote{
M~81 is not shown in the Figures due to its low redshift ($z<10^{-3}$).}. 
 \cite{ric16} collected the X-ray spectra below 10~keV,
 including the $\sim60$~unknown objects in the \textit{Swift}/BAT 70-month catalog,
 then derived the best estimated line of sight column density ($N_{\rm H}$) and 
 absorption corrected BAT 14--195~keV luminosity ($L_{14-195}$).
Even the energy band of the \textit{Swift}/BAT survey, the observed flux is 
affected by obscuring material if the column density of the target exceeds
 $N_{\rm H}> 10^{24}$~cm$^{-2}$ \citep[e.g., see Figure~1 of ][]{ric15}.
Thus, we use absorption corrected 14--195~keV luminosity ($L_{14-195}$) 
in this study and all the values of $L_{14-195}$ and $N_{\rm H}$ will be
tabulated in \cite{ric16}.

\subsection{IR Catalogs}
The available NIR to FIR data were obtained as follows.

\subsubsection{ALLWISE Catalog}\label{sec:allwise_catalog}

The \textit{WISE} mission mapped the all-sky in 3.4 (W1), 4.6 (W2),
 12 (W3), and 22~$\mu$m (W4) bands.
In this study, we obtained the data from the latest \textsc{Allwise} catalog \citep{cut13}
that achieved better sensitivity than the \textit{WISE} all-sky data release \citep{wri10}
thanks to an improved data processing pipeline.
The catalog tabulates the pipeline-measured magnitudes based on the profile fitting 
on $\sim6$~arcsec scale.
In this study, we use this instrumental profile-fit photometry magnitude.
The \textsc{Allwise} achieved 5$\sigma$ sensitivity at 3.4, 4.6, 12, and 22~$\mu$m 
is 0.054, 0.071, 1, and 6~mJy, respectively.
The positional accuracy based on cross-matching with the 2MASS catalog
 is $\sim2$~arcsec at $3\sigma$ level.
We only use the sources with the flux quality \verb|ph_qual=A|, with a signal-to-noise
 ratio larger than 10.
We also check sources of contamination and/or biased flux, due to the proximity
 to an image artifact (e.g., diffraction spikes, scattered-light halos, and/or optical 
 ghosts) using the flag name \verb|ccflag|.
A source that is unaffected by known artifacts is flagged as \verb|ccflag=0|.
We thus only use sources with \verb|ccflag=0| for each band.

\subsubsection{\textit{AKARI} Point Source Catalogs}

To further obtain the IR properties of the \textit{Swift}/BAT AGN, 
we use the \textit{AKARI} All-Sky Survey Point Source Catalogs
(AKARI-PSC). \textit{AKARI} carries two instruments, the infrared camera
\citep[IRC;][]{ona07} operating in the 2--26 $\mu$m band (centered at 9 $\mu$m
and 18 $\mu$m) and the Far-Infrared Surveyor \citep[FIS;][]{kaw07} operating in
the 50--200 $\mu$m band (centered at 65, 90, 140, and 160 $\mu$m). 
The \textit{AKARI} catalogs cover the brightest sources ($> 1$~Jy at $12$~$\mu$m band) 
whose fluxes \textsc{Allwise} could not trace properly due to for saturation.
The AKARI-PSC achieved the flux
sensitivities of 0.05, 0.09, 2.4, 0.55, 1.4, and 6.3 Jy with position accuracies of
6 arcsec at the 9, 18, 65, 90, 140, and 160 $\mu$m bands,
respectively.  In our study, we only utilize sources with the quality
flag of \verb|fqual=3|, whose flux measurements are reliable\footnote{
See the release note of the AKARI/FIS catalog for the details of \texttt{fqual}.
It is recommended not to use the flux data when \texttt{fqual <= 2} for a reliable scientific analysis.\\
\url{http://irsa.ipac.caltech.edu/data/AKARI/documentation/AKARI-FIS\_BSC\_V1\_RN.pdf}}.

\subsubsection{\textit{IRAS} Catalogs}

The \textit{IRAS} mission performed an unbiased all sky survey in the 12, 25,
60, and 100 $\mu$m bands. The typical position accuracy at 12
and 25 $\mu$m is 7 arcsec and 35 arcsec in the scan and cross scan
direction, respectively \citep{bei88}. In this paper we use two largest catalogs,
the \textit{IRAS} Point Source Catalog (\textit{IRAS}-PSC) and the \textit{IRAS} Faint Source
Catalog (\textit{IRAS}-FSC). \textit{IRAS} achieved 10$\sigma$ point source
sensitivities better than 0.7 Jy over the whole sky. The
\textit{IRAS}-FSC  contains even fainter sources with fluxes of $>$0.2
Jy in the 12 and 25~$\mu$m bands. 
We use only \textit{IRAS} sources with \texttt{fqual=3} (the highest quality)
\footnote{
see \cite{bei88} for the definition of \texttt{fqual} in the \textit{IRAS} catalogs.
False detections may be included when \texttt{fqual <= 2}. }.

\subsubsection{\textit{Herschel} BAT AGN Catalog}

The \textit{Swift}/BAT AGN were also observed 
with \textit{Herschel}/Photodetector Array Camera and 
Spectrometer \citep[PACS;][]{pog10} and
Spectral and Photometric Imaging Receiver \citep[SPIRE;][]{gri10}.
\cite{mel14} compiled a catalog of 313 nearby ($z<0.05$) sources
 observed with \textit{Herschel}/PACS.
The PACS covers the two bands at the center wavelength of
 70~$\mu$m (60--85~$\mu$m) and 160~$\mu$m (130--210~$\mu$m) simultaneously.
The PSF is 1.4 and 2.85~arcsec at 70~$\mu$m and 160~$\mu$m, respectively.
Considering the median redshift ($z \sim 0.025$) of the catalog, PACS 70~$\mu$m
PSF covers $\sim$2.8~kpc, which contains most of the host galaxy component.
\cite{shi16} reported that nearby ($z<0.05$) 293 sources were observed
 with \textit{Herschel}/SPIRE as part of a cycle-1 open time program. 
 In addition, other 20 sources were included from other separate
  programs to complete the sample. 
 The PSF is 18, 24, and 36~arcsec for 250, 350, and 500~$\mu$m, respectively.

\subsection{Cross Matching of BAT AGN with the IR Catalogs}
We first compile the IR counterparts by cross matching the BAT AGN 
positions with IR catalogs. In this study, the IR luminosity $L_{X~\mu{\rm m}}$
 represents the observed frame luminosity $\lambda L_{\lambda} (X\mu{\rm m})$ (erg~s$^{-1}$),
 where $3.4 \le X \le 500$.

\subsubsection{NIR bands}\label{sect:NIRselection}
We determine the NIR (3.4 and 4.6~$\mu$m) counterparts of the
 \textit{Swift}/BAT AGN through the positional matching with the \textsc{Allwise}.
We applied a cross-matching radius of 2~arcsec, informed by the cross-matches
 with the 2MASS catalog as described in Section~\ref{sec:allwise_catalog}.
Using \textsc{Allwise}, we found 591 NIR counterparts out of 
606 sources within the 2~arcsec radius.
Considering the superb sensitivity of \textsc{Allwise} than that of the BAT
 survey (see Appendix A), essentially all of them should be detected. 
 Therefore, we checked again the \textsc{Allwise} counterparts of the 
 remaining 15 non-detected sources by expanding the matching-radius.
As a result, 13 sources have been found within 5~arcsec radius,
and we confirmed that the detections are real based on the visual 
inspection of DSS optical and \textsc{Allwise} images.
One of the remaining two sources not detected, 
the counterpart of NGC~3516 was classified as one of the \textsc{Allwise} reject table sources
\footnote{The sources not selected from the \textsc{Allwise} catalog 
because they are low signal-to-noise ratio or spurious detections
 of image artifacts}.
Another source (3C~59) was not detected even by expanding the
 searching radius up to 15~arcsec. 
 After checking the visual inspection between DSS optical
 and XMM/PN X-ray image, we found that the coordinate of 3C 59
  in the BAT catalog traces the jet lobe component, not the central
   object. We used the coordinate of the
  central object obtained from Simbad (RA, Dec)=(31.7592, 29.512775)
   for this target and we found the \textit{WISE} counterpart successfully.
 In total, 605 counterparts are identified in the \textsc{Allwise} catalog.

Out of the 605 sources, 602 and 603 sources fulfill \texttt{ph\_qual=A} 
at 3.4~$\mu$m and 4.6~$\mu$m.
After selecting the sources which fulfill \texttt{ccflag = 0},
the number of IR counterparts at 3.4~$\mu$m and 4.6~$\mu$m 
turns out to be 549 ($\sim90.6$\%) and 548 ($\sim90.4$\%)
sources, respectively.
The number of IR counterparts in the NIR band (either 3.4 or 4.6~$\mu$m)
 is 560 ($\sim92.4$\%) sources.

\begin{figure*}
\begin{center}
\includegraphics[width=7.4cm]{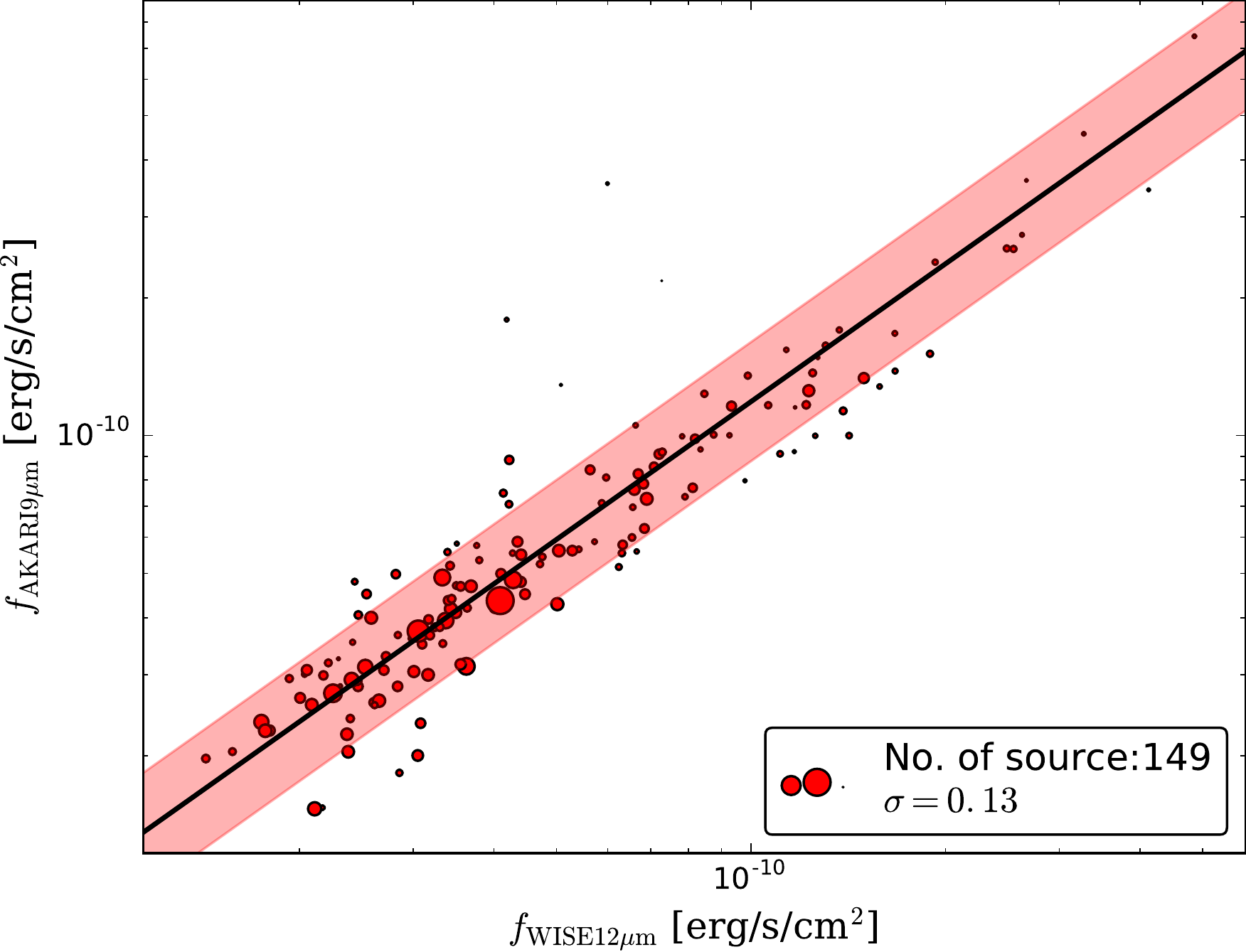}~
\includegraphics[width=7.4cm]{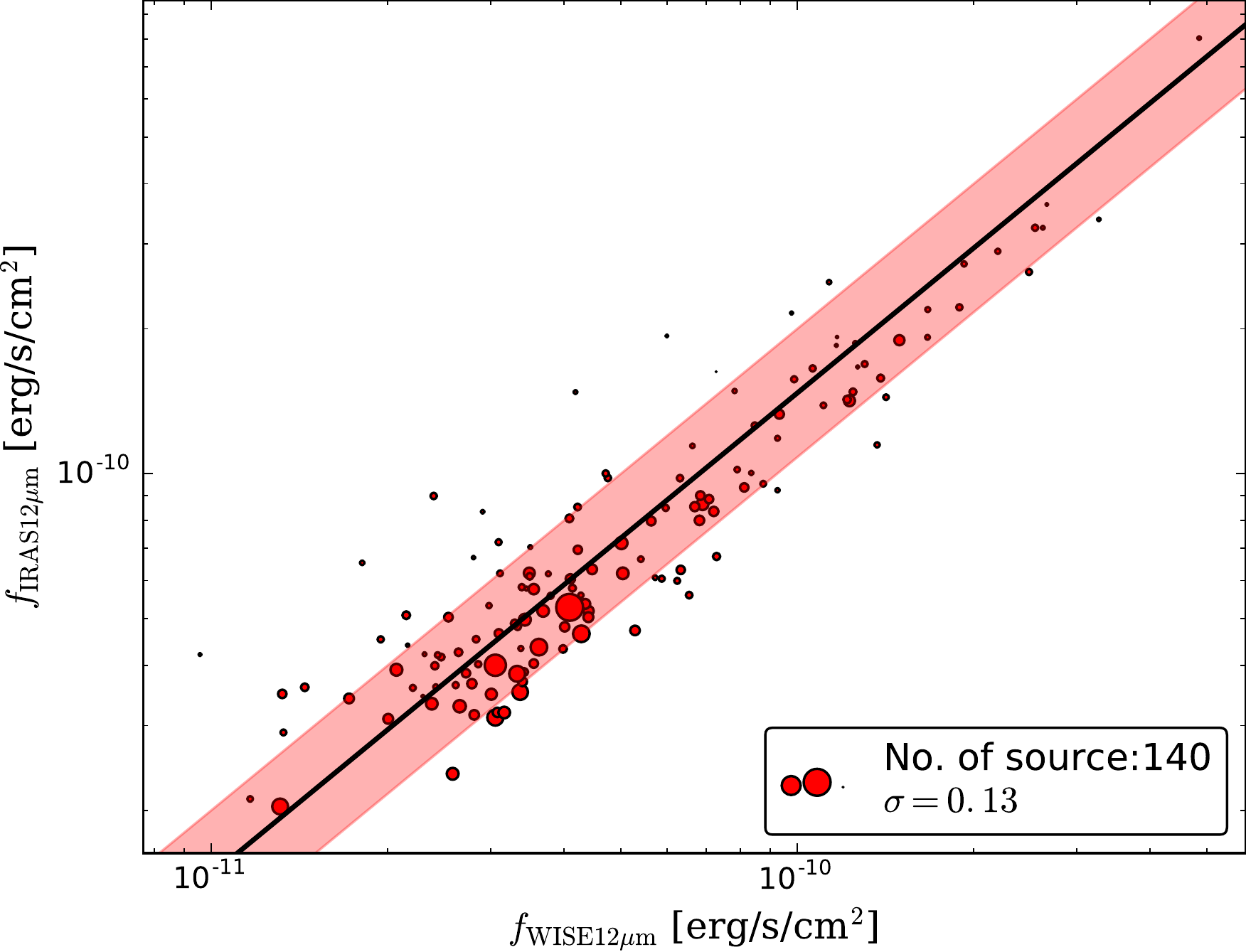}~\\
\includegraphics[width=7.4cm]{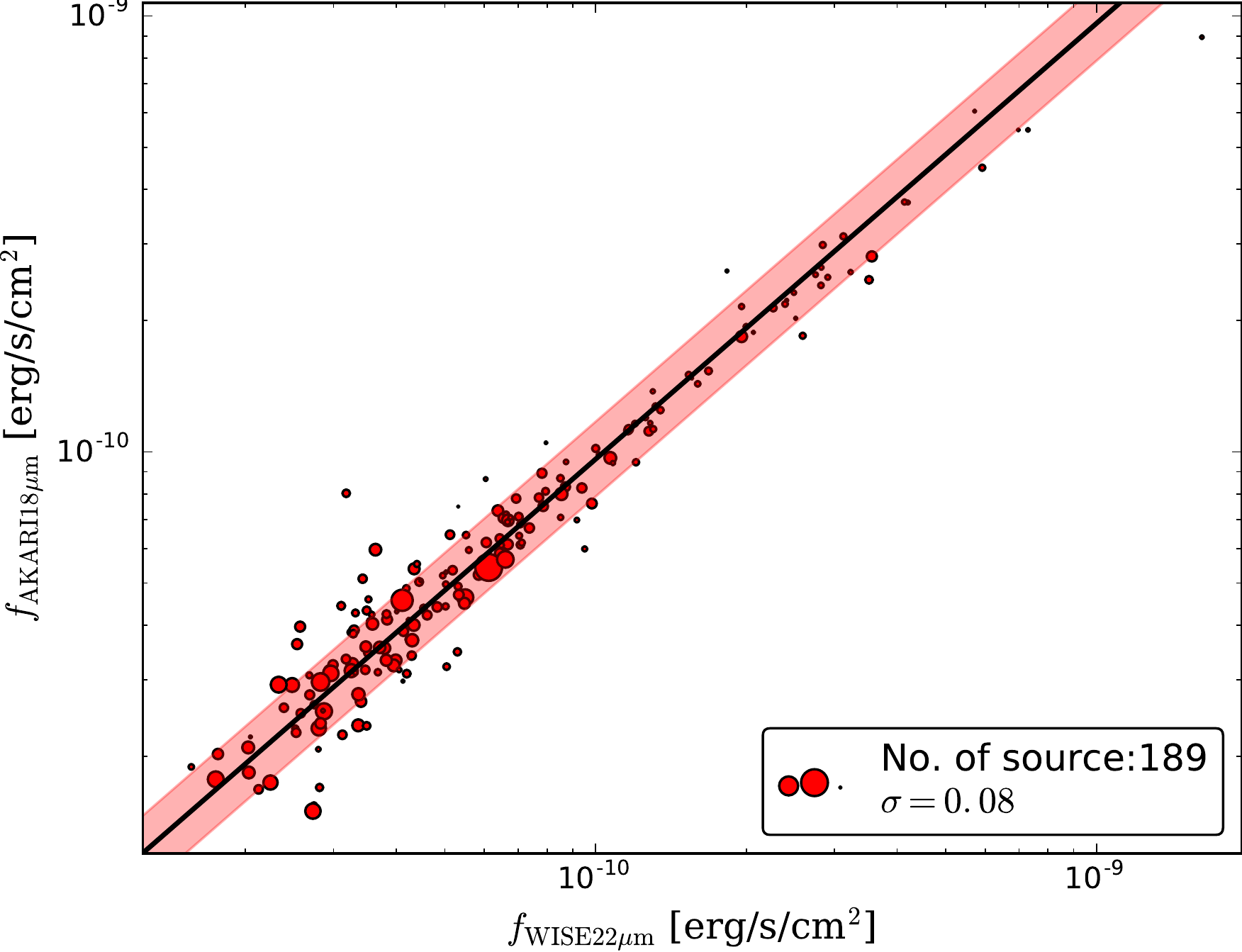}~
\includegraphics[width=7.4cm]{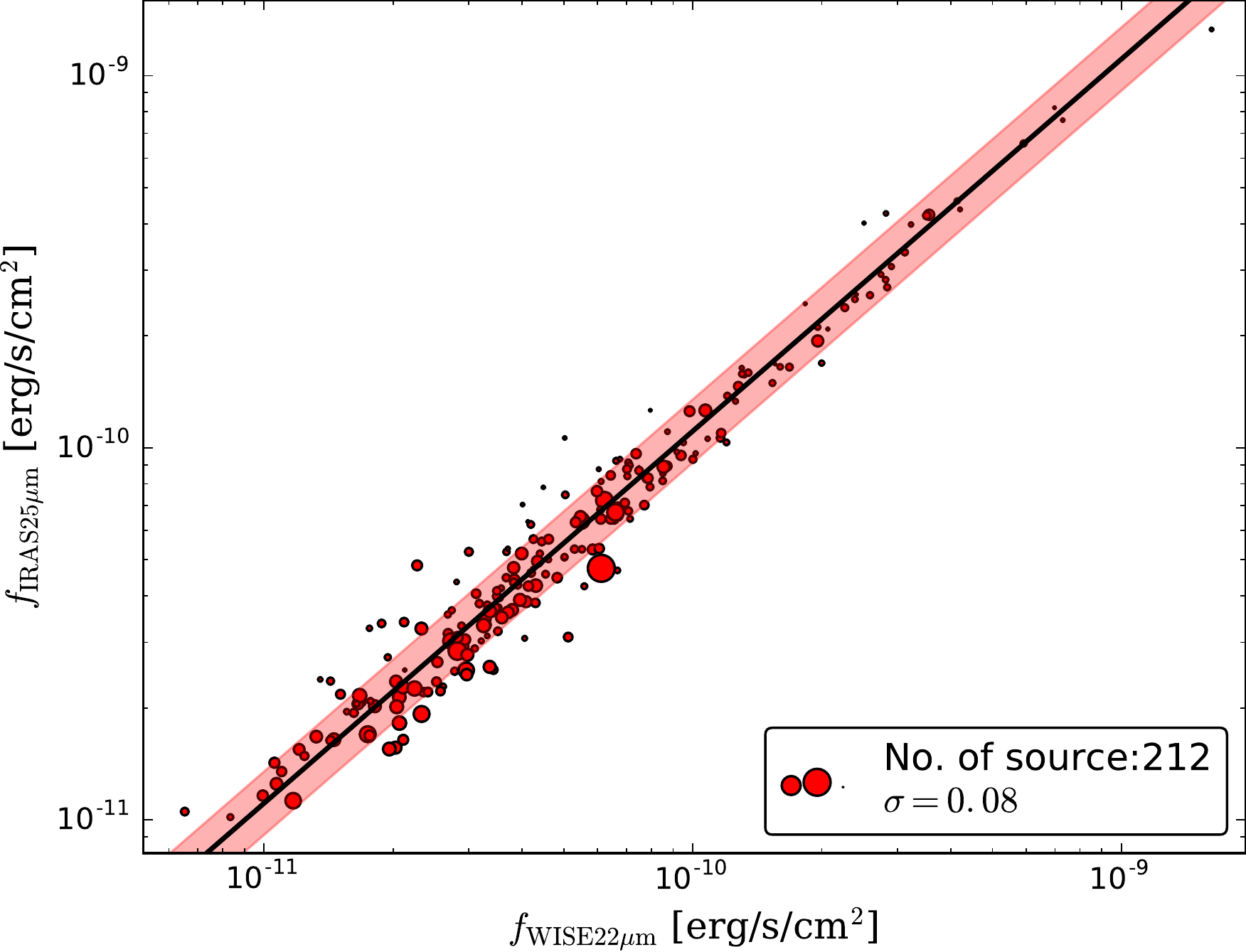}\\
\includegraphics[width=7.4cm]{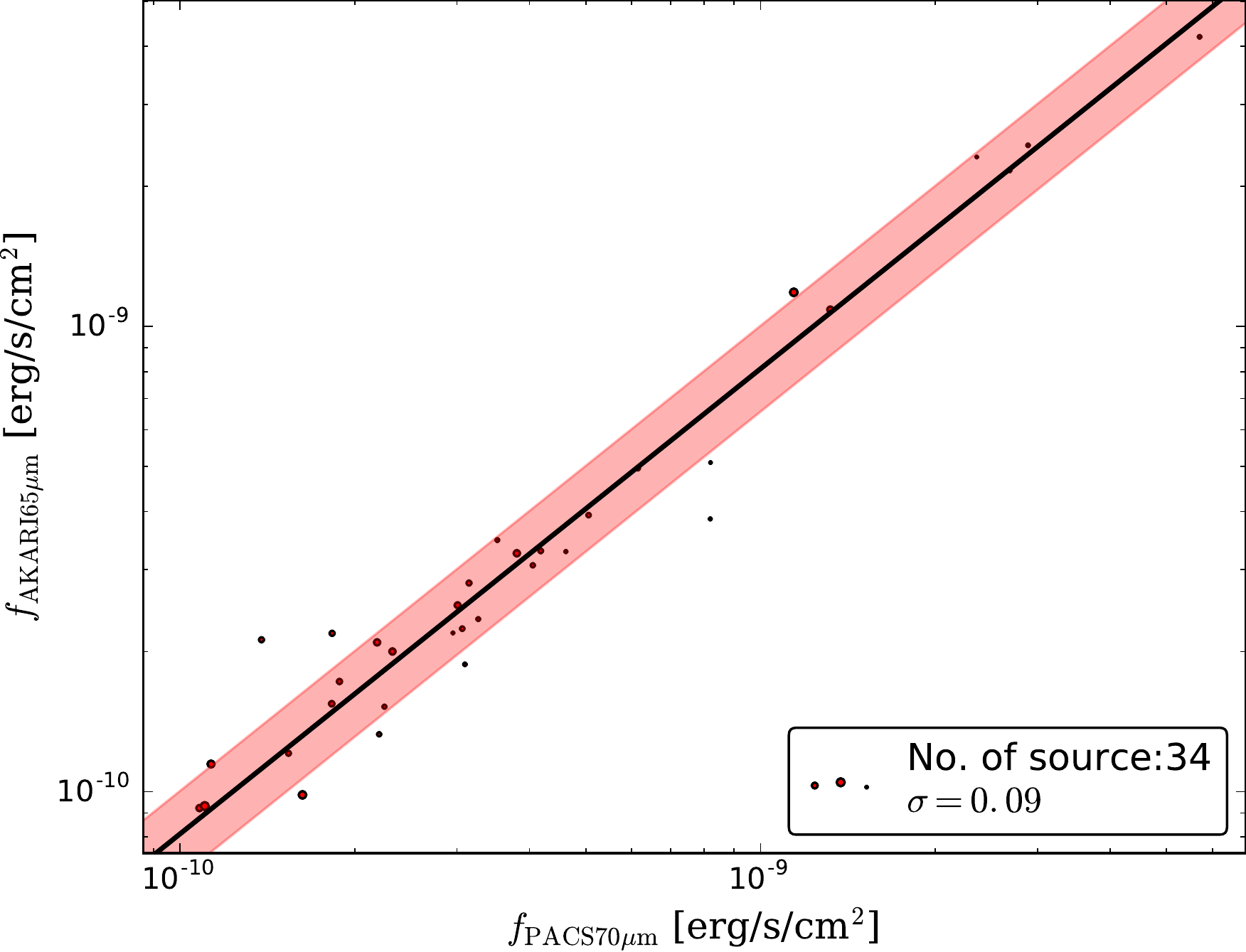}~
\includegraphics[width=7.4cm]{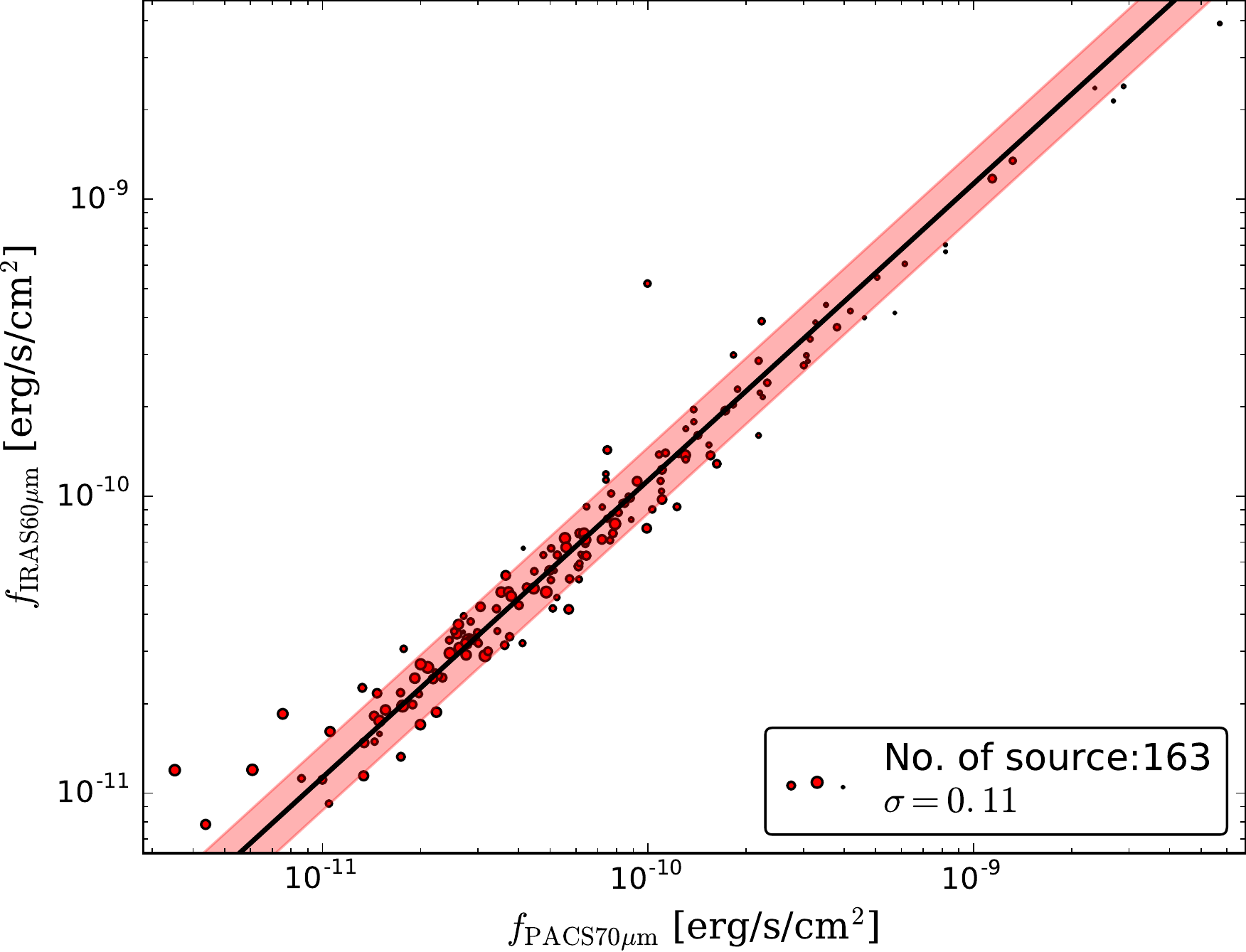}~\\
\includegraphics[width=7.4cm]{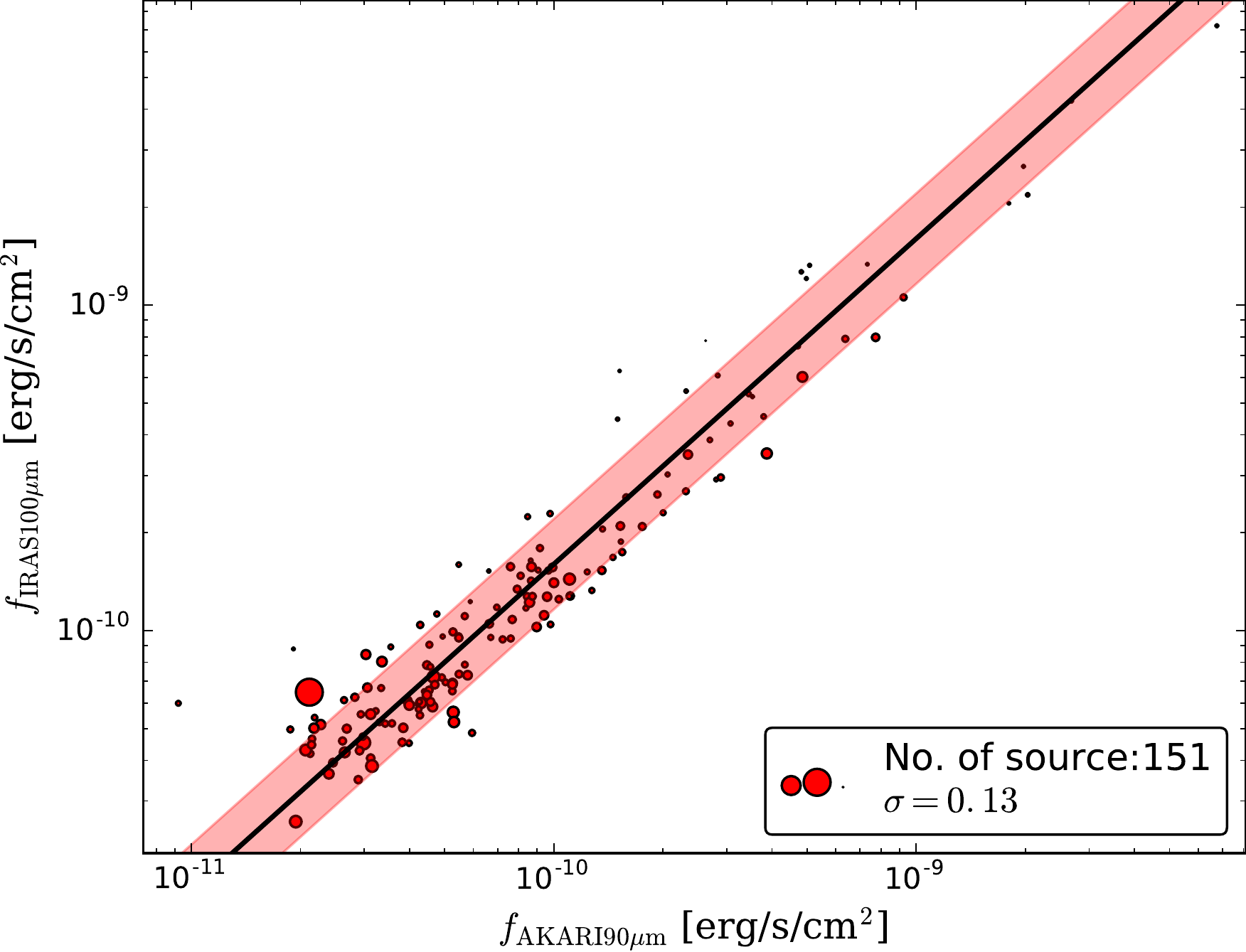}~
\includegraphics[width=7.4cm]{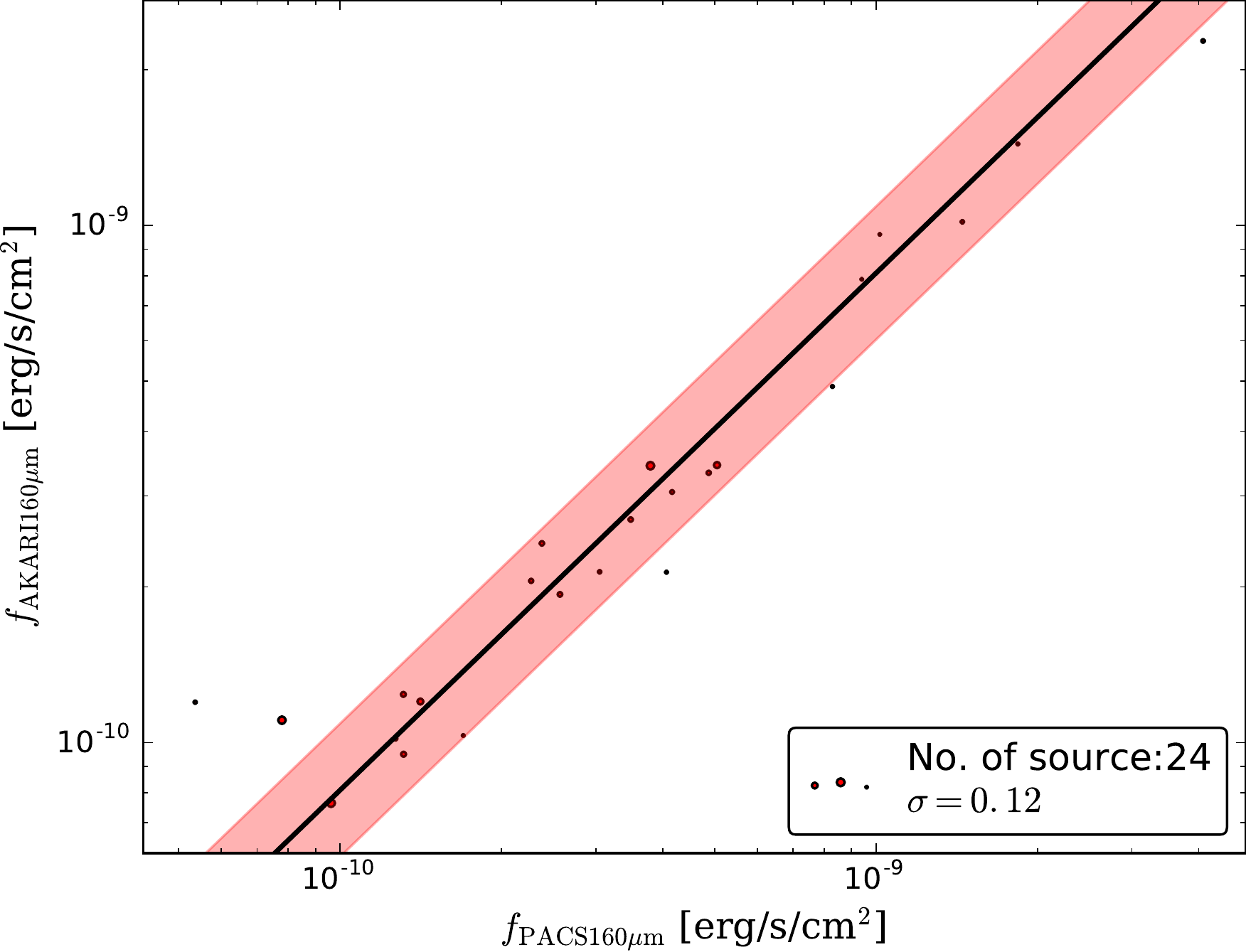}
\caption{
Flux-flux relations of AGN between two IR bands. 
The red color filled circle represents the source detected in both bands. 
The size of the circle is proportional to the redshift of the source.
 The solid line represents the best-fit line and red colored shade are represents
1$\sigma$ dispersion of each linear scaling relation.
The number of sources for the fitting and 1$\sigma$ error is also written in 
the right bottom at each panel.
(Left) From top to bottom, 
\textit{AKARI} 9~$\mu$m vs. \textit{WISE}~12~$\mu$m, 
\textit{AKARI} 18~$\mu$m vs. \textit{WISE}~22~$\mu$m, 
\textit{AKARI} 65~$\mu$m vs. \textit{Herschel}/PACS 70~$\mu$m,
\textit{IRAS} 100~$\mu$m vs. \textit{AKARI} 90~$\mu$m,
(Right) from top to bottom,
\textit{IRAS} 12~$\mu$m vs. \textit{WISE}~12~$\mu$m, 
\textit{IRAS} 25~$\mu$m vs. \textit{WISE}~22~$\mu$m, 
\textit{IRAS} 60~$\mu$m vs. \textit{Herschel}/PACS 70~$\mu$m,
\textit{AKARI} 160~$\mu$m vs. \textit{Herschel}/PACS 160~$\mu$m.}\label{fig:LIRvsLIR}
\end{center}
\end{figure*}

\subsubsection{MIR bands}
We determine the MIR (9--25~$\mu$m) counterparts of the \textit{Swift}/BAT AGN
by cross matching the \textsc{Allwise}, \textit{AKARI}, and \textit{IRAS}
catalogs in this order. 
Our primary goal is to obtain photometric data in the IR band
 as completely as possible for the \textit{Swift/}BAT selected AGN. 
We give the highest priority to the \textsc{Allwise} catalog because of 
its 50~times better sensitivity than \textit{AKARI}, which allows us to
search for fainter sources in the MIR all-sky view.
Then we cross matched the sources undetected by \textsc{Allwise} with \textit{AKARI}.
\textit{AKARI} covers the brighter sources which are saturated due to the high sensitivity
of \textsc{Allwise}, and have the advantage of a 2--4 times
 higher sensitivity than the \textit{IRAS} survey. While all the \textit{IRAS}
  sources should be detected with \textit{AKARI}, the flux quality flags of \textit{AKARI} for 
  very nearby ($z < 0.005$) objects turn out to be bad due to their 
  extended morphology when fitted with a single Gaussian. 
  In such cases, we rather refer to the \textit{IRAS} data with good flux quality, 
  which have $\sim11$ times worse angular resolution than \textit{AKARI}, 
  since we aim to measure the total MIR flux from both nucleus and host galaxy
   in a uniform way for the whole AGN sample. 
   
   The positional matching of the optical counterparts of the \textit{Swift}/BAT AGN
   with IR survey catalogs was already discussed in Section~\ref{sect:NIRselection} 
   for \textsc{Allwise} and in \cite{ich12} for \textit{AKARI} and \textit{IRAS}, 
   and we follow here the same approach.
  For the MIR bands, the number of detections is compiled at the second column
   in Table~\ref{tab:LIRvsLx}.
Here the detection at 12~$\mu$m represents the detection either at
 \textit{AKARI}~9~$\mu$m, \textit{WISE}~12~$\mu$m, or \textit{IRAS}~12~$\mu$m;
  22~$\mu$m represents either at \textit{AKARI}~18~$\mu$m, \textit{WISE}~22~$\mu$m, 
  or \textit{IRAS}~25~$\mu$m;
 the MIR band represents either at 12~$\mu$m or 22~$\mu$m band defined above.
  Finally, we obtained 601 ($\sim99.2$\%) counterparts in at least one MIR band.
Thus, the identification in the MIR bands is almost as complete as in the NIR bands.
The redshift distribution of the IR counterparts at each wavelength is shown in Figure~\ref{fig:zdist}.

   \subsubsection{FIR bands}
  The FIR counterparts of the \textit{Swift}/BAT AGN at $60 \le \lambda \le 160$~$\mu$m
   were gathered by cross-matching the \textit{AKARI}, \textit{IRAS}, and \textit{Herschel}
   in this order. 
   Our goal is to obtain photometric data for the full host galaxy emission in the FIR band.
   We gave \textit{AKARI} counterparts the highest priority because of the
    better sensitivity with respect to \textit{IRAS} surveys.
   Then we matched the potion of the sources undetected by \textit{AKARI} with \textit{IRAS}. 
   Considering the better sensitivity of \textit{AKARI}/FIS, 
   one might expect that \textit{IRAS} would not cover many sources. 
   However, \textit{AKARI} often misses emission from sources with extended morphology
   due to its better angular resolution. 
   In such cases, \textit{IRAS} gives the best quality estimate of flux by measuring the 
   whole FIR flux from the host galaxies. 
   Finally, the remaining distant sources or faint sources which neither \textit{AKARI} nor
   \textit{IRAS} detected were cross matched with the \textit{Herschel}/PACS catalog of \cite{mel14}. 
   We cross matched the sources by referring to the counterpart source names reported by
   \cite{mel14} and the \textit{Swift}/BAT catalog.
  
  For the FIR counterpart at $250 \le \lambda \le 500$~$\mu$m,  only \textit{Herschel}/SPIRE
   catalog can access to those wavelengths. 
  We also cross matched the sources by referring the counterpart source names written
   in \cite{shi16} and the \textit{Swift}/BAT catalog.
   
   For the FIR bands, 388 ($\sim64.2$\%), 241 ($\sim 39.9$\%), 89 ($\sim14.7$\%), 
   229 ($\sim 37.9$\%), 213 ($\sim 35.3$\%), 170 ($\sim28.1$\%), and 107 ($\sim17.7$\%) 
   sources are compiled at
    70 (either at \textit{IRAS}~60~$\mu$m, \textit{AKARI}~65~$\mu$m, or \textit{Herschel}~70~$\mu$m), 
    90 (either at \textit{AKARI}~90~$\mu$m or \textit{IRAS}~100~$\mu$m),
    140 (at \textit{AKARI}~140~$\mu$m), and
    160~$\mu$m (at \textit{Herschel}~160~$\mu$m or \textit{AKARI}~160~$\mu$m),
     250~$\mu$m (at \textit{Herschel}/SPIRE~250~$\mu$m),
     350~$\mu$m (at \textit{Herschel}/SPIRE~350~$\mu$m),
     500~$\mu$m (at \textit{Herschel}/SPIRE~500~$\mu$m),
      respectively. Those numbers are also compiled in the second column of Table~\ref{tab:LIRvsLx}.
   Finally, 402 ($\sim 66.3$\%) IR counterparts are obtained in at least one FIR band.
   Thus, the identification in the FIR bands is not yet complete, but a statistically significant
   sample has been compiled for this analysis.

   \subsection{Luminosity Correlation among IR catalogs}
    
Since the four IR catalogs have slightly different central wavelengths and
 aperture sizes, we investigate the correlation between the
\textit{AKARI/IRAS/WISE/Herschel} luminosities, using only the sources
 detected in two separate observations.
For the MIR bands, we choose \textit{AKARI} 9~$\mu$m and \textit{IRAS}~12~$\mu$m
 for \textit{WISE} 12~$\mu$m, and \textit{AKARI} 18~$\mu$m and \textit{IRAS} 
 25~$\mu$m for \textit{WISE} 22~$\mu$m, respectively, because of the proximity
  of the central wavelengths. 
For the FIR bands, \textit{AKARI} 65~$\mu$m and \textit{IRAS} 60~$\mu$m for
 \textit{Herschel}/PACS 70~$\mu$m, \textit{IRAS} 100~$\mu$m for \textit{AKARI} 
 90~$\mu$m, \textit{AKARI} 160~$\mu$m for \textit{Herschel}/PACS 160~$\mu$m.
Figure~\ref{fig:LIRvsLIR} displays the flux correlations between the two bands, 
showing that the correlation in flux between different IR catalogs are tight and significant.
The standard deviation of the flux-ratio distribution between these two bands
 are written in the caption of Figure~\ref{fig:LIRvsLIR}.

Figure~\ref{fig:LIRvsLIR} also shows that the flux relations are independent 
from the redshift.
Although within the scatter, the flux relations between \textit{WISE} 12,
 22~$\mu$m and \textit{IRAS} 12, 25~$\mu$m show systematic $z$ dependence
  that flux ratio of $f_{\rm IRAS}/f_{\rm WISE}$ is anti-correlated to $z$.
This could be due to greatly larger aperture of \textit{IRAS} than that of \textit{WISE}, 
therefore the MIR emission from the host galaxy slightly contaminates to the \textit{IRAS}
 fluxes of low-$z$ sources.

Based on the flux correlation, we derive the empirical formula to convert
the flux of each band into \textit{WISE}~12~$\mu$m, 22~$\mu$m, 
Herschel/PACS 70~$\mu$m, 160~$\mu$m, and \textit{AKARI} 90~$\mu$m as follows:

{\small
\begin{align*}
\log \left(  \frac{f_{{\rm WISE}~12~\mu{\rm m}}}{{\rm erg}~{\rm s}^{-1}~{\rm cm}^{-2}} \right) &= \log \left( \frac{f_{{\rm AKARI}~9~\mu{\rm m}}}{{\rm erg}~{\rm s}^{-1}~{\rm cm}^{-2}} \right)	-0.074\\
\log \left(  \frac{f_{{\rm WISE}~12~\mu{\rm m}}}{{\rm erg}~{\rm s}^{-1}~{\rm cm}^{-2}} \right) &= \log \left( \frac{f_{{\rm IRAS}~12~\mu{\rm m}}}{{\rm erg}~{\rm s}^{-1}~{\rm cm}^{-2}} \right)	-0.167\\
\log \left(  \frac{f_{{\rm WISE}~22~\mu{\rm m}}}{{\rm erg}~{\rm s}^{-1}~{\rm cm}^{-2}} \right) &= \log \left( \frac{f_{{\rm AKARI}~18~\mu{\rm m}}}{{\rm erg}~{\rm s}^{-1}~{\rm cm}^{-2}} \right)	+0.017\\
\log \left(  \frac{f_{{\rm WISE}~22~\mu{\rm m}}}{{\rm erg}~{\rm s}^{-1}~{\rm cm}^{-2}} \right) &= \log \left( \frac{f_{{\rm IRAS}~25~\mu{\rm m}}}{{\rm erg}~{\rm s}^{-1}~{\rm cm}^{-2}} \right)	-0.045\\
\log \left(  \frac{f_{{\rm PACS}~70~\mu{\rm m}}}{{\rm erg}~{\rm s}^{-1}~{\rm cm}^{-2}} \right) &= \log \left( \frac{f_{{\rm AKARI}~65~\mu{\rm m}}}{{\rm erg}~{\rm s}^{-1}~{\rm cm}^{-2}} \right)	+0.091\\
\log \left(  \frac{f_{{\rm PACS}~70~\mu{\rm m}}}{{\rm erg}~{\rm s}^{-1}~{\rm cm}^{-2}} \right) &= \log \left( \frac{f_{{\rm IRAS}~60~\mu{\rm m}}}{{\rm erg}~{\rm s}^{-1}~{\rm cm}^{-2}} \right)	-0.053\\
\log \left(  \frac{f_{{\rm AKARI}~90~\mu{\rm m}}}{{\rm erg}~{\rm s}^{-1}~{\rm cm}^{-2}} \right) &= \log \left( \frac{f_{{\rm IRAS}~100~\mu{\rm m}}}{{\rm erg}~{\rm s}^{-1}~{\rm cm}^{-2}} \right)	-0.204\\
\log \left(  \frac{f_{{\rm PACS}~160~\mu{\rm m}}}{{\rm erg}~{\rm s}^{-1}~{\rm cm}^{-2}} \right) &= \log \left( \frac{f_{{\rm AKARI}~160~\mu{\rm m}}}{{\rm erg}~{\rm s}^{-1}~{\rm cm}^{-2}} \right)	+0.092\\

\end{align*}
}

\begin{figure*}
\begin{center}
\includegraphics[width=9.0cm]{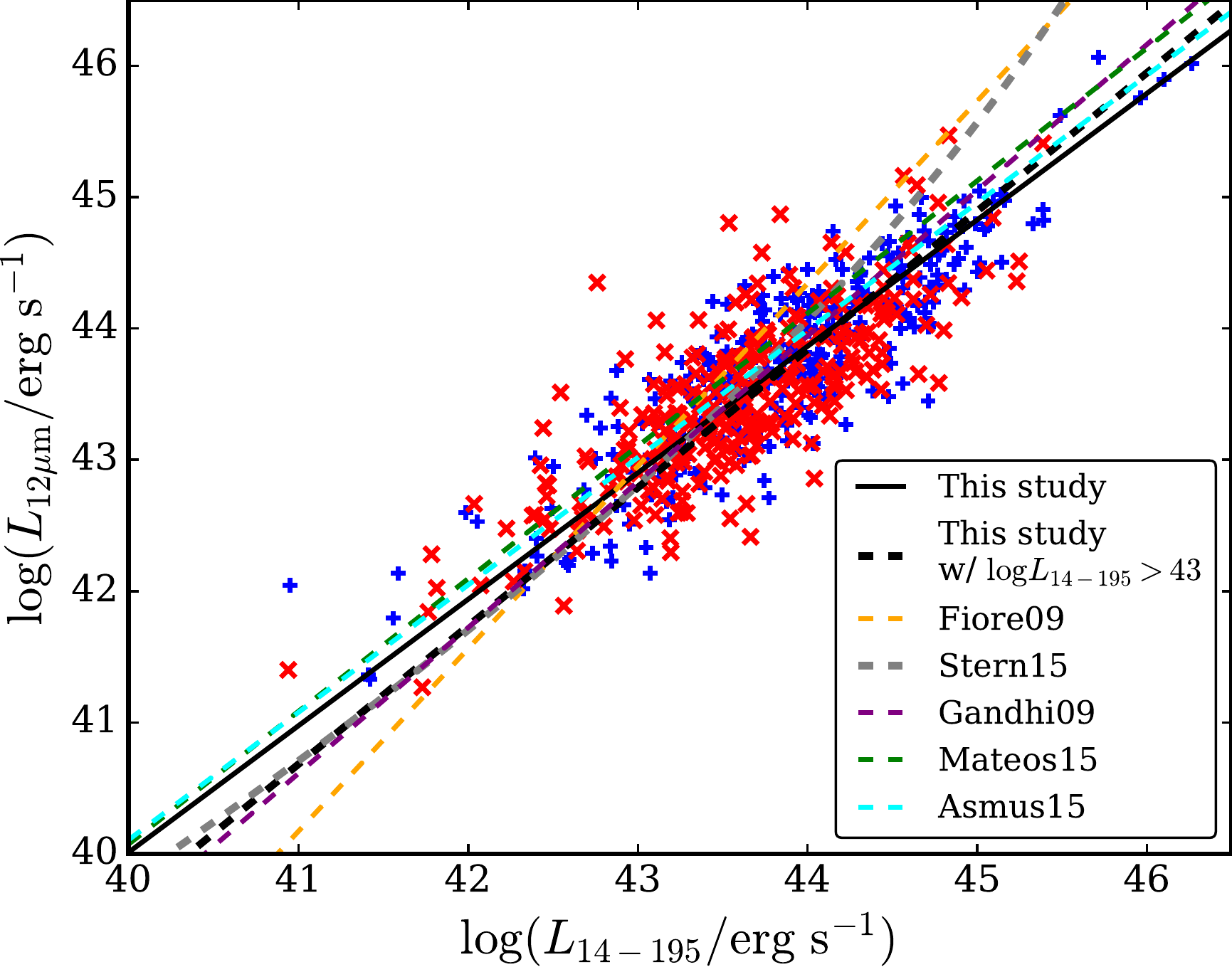}~
\includegraphics[width=9.0cm]{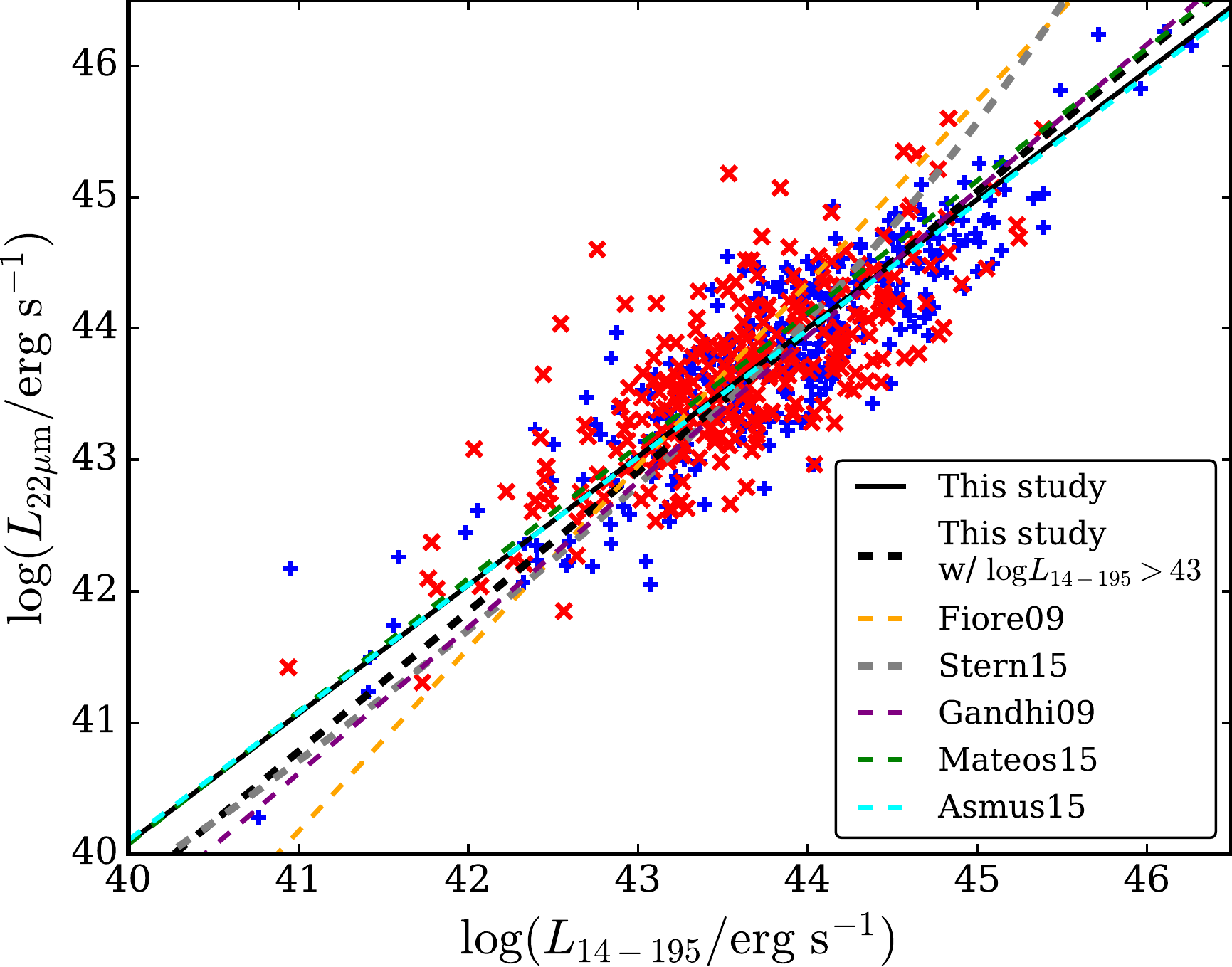}~
\caption{
Luminosity correlations between the luminosities at 12 (left) and 22~$\mu$m (right)  
($L_{12~\mu{\rm m}}, L_{22~\mu{\rm m}}$) and 14--195~keV ($L_{14-195}$).
Blue/red cross represents type-1/-2, respectively.
The black solid line represents the slope of our study in Equation (1) for
 left panel and (2) for right panel.
The black dashed line represents the slope of our study using only high 
luminosity sources with $\log L_{14-195}>43$.
The other dashed line represents the study of \cite{fio09} (orange), 
\cite{ste15} (gray), \cite{gan09} (purple), \cite{mat15b} (green), and \cite{asm15} (cyan)
  respectively.
The studies with local sample (mostly $z<0.1$ and main luminosity range of 
$41<L_{\rm X}<46$) are our study, \cite{gan09}, and \cite{asm15}.
The studies with high-$z$ sample (mostly $0.1 <z < 5$ and $42<L_{\rm X}<46$) 
are \cite{fio09}, \cite{mat15b}, and \cite{ste15}.
The studies with type-1 AGN are \cite{fio09}, \cite{ste15}, and with both type-1 
and type-2 AGN are \cite{gan09}, \cite{mat15b}, \cite{asm15}, and our study.
See Table~\ref{tab:LIRvsLx_wlit} for more details.
}\label{fig:LIRvsLx}
\end{center}
\end{figure*}

Assuming that AGN that are not detected in the highest priority band
(\textit{WISE}~12~$\mu$m, 22~$\mu$m, Herschel/PACS 70~$\mu$m, 160~$\mu$m,
and \textit{AKARI} 90~$\mu$m) but in second or third priority bands
 should follow the same correlations as examined here, we
apply the conversion factors reported above to derive
 the 12, 22, 70, 90, and 160~$\mu$m luminosities. Doing so
  we can discuss the luminosity correlation with the 14--195~keV band in
a uniform way regardless of the matched catalogs.
All the IR properties of the parent sample AGN are summarized in Table~\ref{tab:IRcatalog}.

\subsection{AGN type}
To examine the IR properties of different AGN populations,
we divide the sample into two types based on the column density ($N_{\rm H}$) 
obtained from the X-ray spectral fitting by \cite{ric16}. 
The AGN with $N_{\rm H} < 10^{22}$~cm$^{-2}$ are called
 ``X-ray type-1'' (hereafter type-1), and the AGN with $N_{\rm H} \ge 10^{22}$~cm$^{-2}$
 are called ``X-ray type-2'' (hereafter type-2).
The sample is divided into 311 type-1 AGN and 293 type-2 AGN. 
The AGN type of each source will be tabulated in \cite{ric16}.

\section{Results and Discussion}
\subsection{Correlation between the MIR and Ultra-Hard X-Ray Luminosities}\label{sect:LxvsLIR}

\begin{figure*}
\begin{center}
\includegraphics[width=8.5cm]{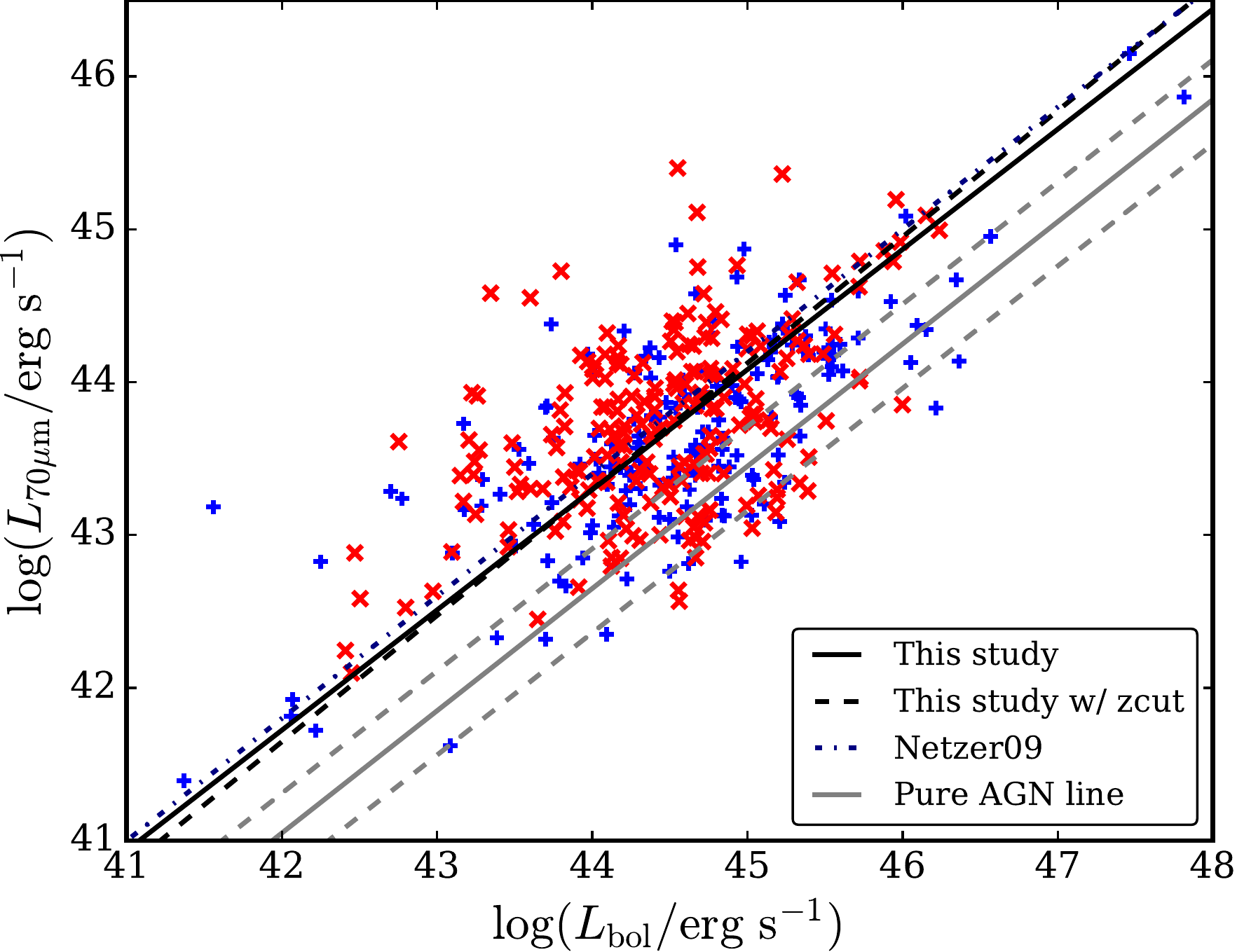}~
\includegraphics[width=8.5cm]{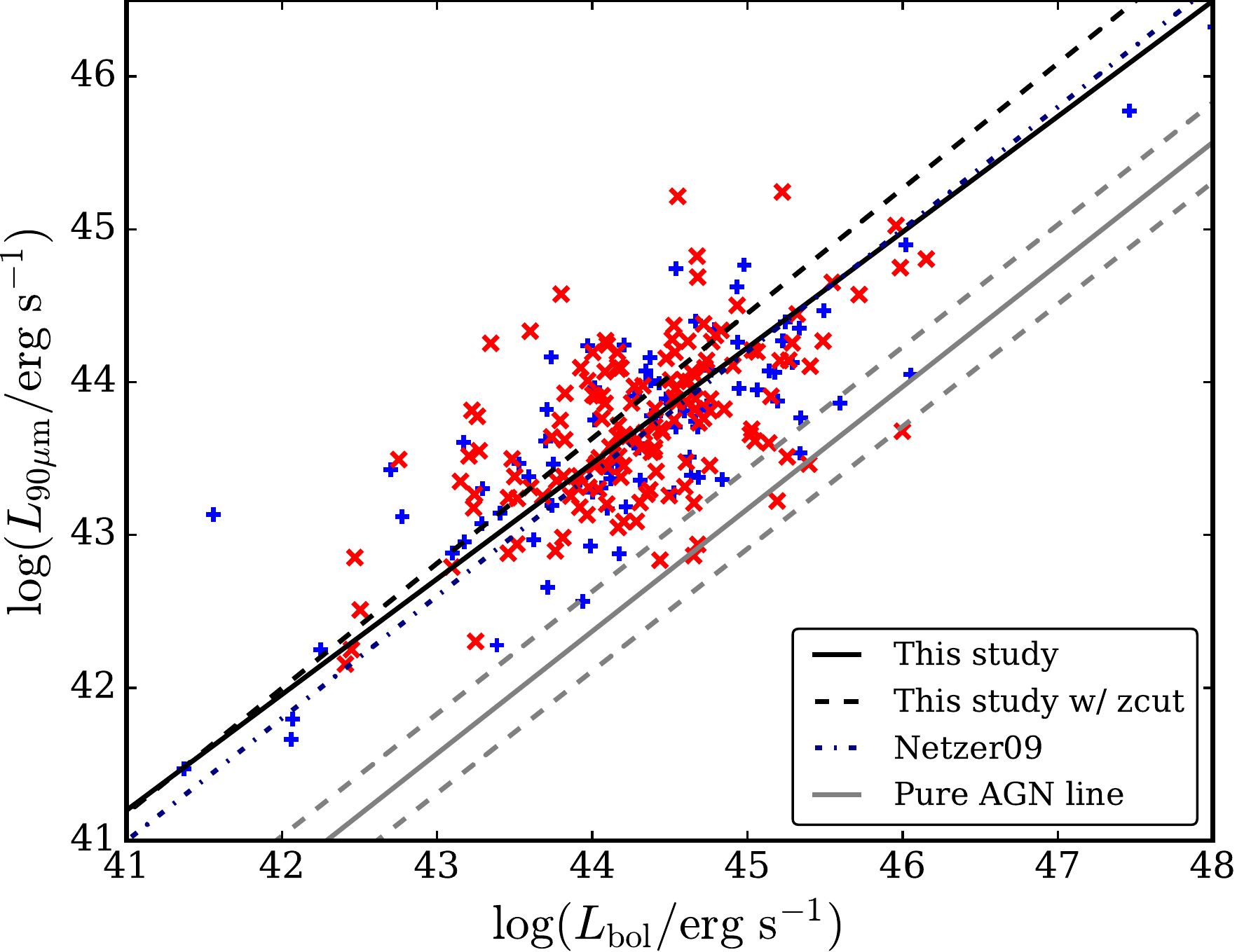}\\
\caption{
Luminosity correlations between the luminosities at 70 and
 90~$\mu$m ($ L_{70~\mu{\rm m}}, L_{90~\mu{\rm m}}$) and 
 bolometric luminosity ($L_{\rm bol}$) estimated from Equation~(5).
Blue/red color represents type-1/-2, respectively.
The black solid line represents the slope of Equation (6)
 and (7), respectively.
The black dashed line represents the slope of Equation (8) and (9),
 respectively.
The dot-dashed line (navy) represents the slope obtained by \cite{net09}.
The gray solid line represents the pure-AGN sequence reported in Equation (11) and (12),
 respectively.
}\label{fig:LFIRvsLbol}
\end{center}
\end{figure*}

Figure~\ref{fig:LIRvsLx} shows the luminosity correlations between the 
MIR (12 and 22~$\mu$m) luminosities ($L_{12\mu{\rm m}}, L_{22\mu{\rm m}}$) 
and $L_{14-195}$ in the luminosity range of  $10^{40} < L_{14-195}<10^{47}$~erg~s$^{-1}$
\footnote{M~81 and NGC~4395 are not shown in Figure~\ref{fig:LIRvsLx} and 
~\ref{fig:LFIRvsLbol} due to their low luminosities
of ($\log L_{12~\mu{\rm m}}, \log L_{22 \mu{\rm m}}, \log L_{90~\mu{\rm m}}, 
\log L_{14-195}) = (39.20, 39.22, 39.76,38.50)$ for M~81 and
$(\log L_{12~\mu{\rm m}}, \log L_{22 \mu{\rm m}}, \log L_{90~\mu{\rm m}}, \log L_{14-195})
 =  (39.88, 40.27, 41.48, 40.77)$ for NGC~4395.}.
Blue and red crosses represent type-1 and type-2 AGN, respectively.
The error-bars are not shown in Figure~\ref{fig:LIRvsLx} and Figure~\ref{fig:LFIRvsLbol} 
since the uncertainties of both infrared luminosities and 14-195 keV luminosity are 
vanishingly small ($<10$\%) in the log-log plot.

Since our motivation is to determine the slope of the luminosity relation between
 $L_{\rm MIR}$ and $L_{14-195}$ as two independent variables,
we apply ordinary least-squares Bisector fits, which minimizes perpendicular 
distance from the slope line to data points \citep{iso90}.
The ordinal least-squares Bisector fits (with the form of
 $[ \log (L_{\rm MIR}/10^{43}~{\rm erg}~{\rm s}^{-1}) =
  (a \pm \Delta a) + (b \pm \Delta b) \log (L_{14-195}/10^{43}~{\rm erg}~{\rm s}^{-1}$)],
where $\Delta a$ and $\Delta b$ is the standard deviation of $a$ and $b$, respectively)
gives the correlations of

{\small
\begin{align}\label{Eq:LbatvsLIR_all}
\log \frac{L_{12 \mu{\rm m}}}{10^{43}~{\rm erg/s}} &= (-0.10 \pm 0.02) + (0.96 \pm 0.02) \log \frac{L_{\rm 14-195}}{10^{43}~{\rm erg/s}}\\
\log \frac{L_{22 \mu{\rm m}}}{10^{43}~{\rm erg/s}} &= (0.02 \pm 0.02) + (0.98 \pm 0.02) \log \frac{L_{\rm 14-195}}{10^{43}~{\rm erg/s}}

\end{align}
}

The significance of the correlations between the two bands luminosities (and fluxes) can be obtained
by performing Spearman's tests. The results are summarized in Table~\ref{tab:LIRvsLx}.
 We find that both luminosity--luminosity and flux--flux correlations
between the NIR, MIR bands and the 14--195~keV bands are highly significant.

In the Seyfert galaxy class with $L_{14-195}<10^{44}$~erg~s$^{-1}$,
the correlation between MIR and X-ray was first reported by using ground 
telescopes with low spatial resolutions \citep{elv78, kra01}, and then
by several authors thanks to the new windows opened by the \textit{ISO} 
satellite \citep{lut04, ram07} and by \textit{Spitzer} \citep{saz12}.
Studies based on the ground-based high spatial resolution MIR photometry
 were first compiled by \cite{hor06}, then expanded independently by \cite{lev09} 
 and \cite{gan09}, and finally by \cite{asm15}. 
 The correlation parameters of \cite{gan09} and \cite{asm15}
  are the most widely used because they include Compton-thick AGN.
\cite{gan09} show steeper results than ours with $b=1.11 \pm 0.04$, but
 \cite{asm15} report the results consistent with our studies within the uncertainties 
 with $b=0.97 \pm 0.03$.
Both slopes are over-plotted in Figure~\ref{fig:LIRvsLx} and also compiled 
in Table~\ref{tab:LIRvsLx_wlit}.
Since both studies used 2--10~keV luminosity as X-ray luminosity, we apply 
the conversion factor of $L_{14-195}/L_{2-10} = 2.1$ under the assumption
 of $\Gamma = 1.9$ for the over-plot in Figure~\ref{fig:LIRvsLx}. 
 Hereafter, we always apply this conversion factor for estimating $L_{14-195}$ from $L_{2-10}$.

The host galaxy contamination in the MIR emission especially in the low-luminosity 
end could affect the slope values of $b=0.96-0.98$ in our study.
If we use only the sources with $L_{14-195}>10^{43}$~erg~s$^{-1}$, 
the luminosity relations become

{\small
\begin{align}
\log \frac{L_{12 \mu{\rm m}}}{10^{43}~{\rm erg/s}} &= (-0.21 \pm 0.03) + (1.05 \pm 0.03) \log \frac{L_{\rm 14-195}}{10^{43}~{\rm erg/s}}\\
\log \frac{L_{22 \mu{\rm m}}}{10^{43}~{\rm erg/s}} &= (-0.09 \pm 0.03) + (1.07 \pm 0.03) \log \frac{L_{\rm 14-195}}{10^{43}~{\rm erg/s}}
\label{Eq:LMIRvsLbat_high},
\end{align}
}
which is slightly steeper, but within 2$\sigma$ uncertainty of \cite{asm15}.
The slope obtained by \cite{gan09} depends on the choice of algorithm and the
 value becomes $b=1.00 \pm 0.08$ when using the same method of \cite{asm15}. 
 Therefore, our results are generally fully consistent with the high spatial resolution 
 results in high luminosity end with $L_{14-195}>10^{43}$~erg~s$^{-1}$.
While our results with poorer spatial resolution suffer from the contamination 
from the host galaxies in the lower luminosity end,
our study has the advantage of the completeness ($\sim98$\%) in the 
MIR bands of the ultra-hard-X-ray flux limited \textit{Swift}/BAT 70 month catalog, 
which is the least bias against absorption up to $N_{\rm H} \simeq 10^{24}$~cm$^{-2}$.

The comparison to the literature from the higher luminosity (and also high-$z$) studies
 with $L_{14-195} \ge 10^{44}$~erg~s$^{-1}$ can also provide important information.
We compile the luminosity correlations of those studies in Table~\ref{tab:IRXcatalog}. 
\cite{fio09} derived the observed rest-frame 6~$\mu$m and 2--10~keV luminosities of
 $\sim80$ X-ray-selected type-1 AGN in the COSMOS and CDF-S fields obtained from
  \textit{Chandra} and \textit{Spitzer} satellites. 
  The slope is quite steep, with $b=1.39$ for $\log L_{6~\mu {\rm m}} \ge 43$.
Although the detailed fitting algorithm were not mentioned in their studies, 
there is a trend of increasing MIR--X-ray ratio at high luminosity end with 
$\log (L_{6~\mu {\rm m}} / {\rm erg}~{\rm s}^{-1} ) > 44$.
Further evidence of this trend is obtained by \cite{ste15} (plotted with gray dotted
 line in Figure~\ref{fig:LIRvsLx}) using SDSS DR5, tracing at high-$z$ QSOs mainly 
 with $2<z<4$. 
They used quadratic function for reproducing the X-ray--MIR luminosity relations.

If the trends above are true, the steeper slope suggests that X-ray emission is 
inefficient in the high-luminosity end.
This is reported by several observations that SED shape of AGN changes with 
luminosity and Eddington ratio \citep[][]{vas07}.
The existence of X-ray weak sources at high bolometric luminosities has been
 recently confirmed by \cite{ric16a}, who found that Hot Dust Obscured Galaxies
  \citep[hot DOGs; ][]{wu12} seem to have X-ray luminosities one or two order
   of magnitudes below the value expected by the local X-ray -- MIR correlation.
Since our sample is ultra-hard X-ray selected, our studies might miss those X-ray
 saturated sources in the high-luminosity end.
 Those sources could be located faint end in 14--195~keV flux, but not in MIR fluxes.
 Since BAT 14--195~keV flux limit is over one order of magnitude shallower than those of MIR fluxes,
a deeper survey is necessary to assess the luminosity relation between MIR and 14--195~keV
luminosity at high luminosity end (see also Figure~\ref{fig:fIRvsfx_supp}).
Another suggestion from this trend is that AGN might have large obscuring fraction
 in the high luminosity regime and/or in the high-$z$ universe \citep{buc15}.
This might be true considering the X-ray studies that the fraction of Compton-thick 
AGN increase with redshift from $z=0$ to $z=2$ \citep{bri12}.
We do not attempt to solve this question at high luminosities here 
but it is valuable to mention another possibility of the slope differences 
among those studies of high-luminosity end.
One possibility which make the slope steeper originates from the
 6~$\mu$m bands instead of 12~$\mu$m.
\cite{asm15} pointed out that hot dust component could dominate
 around 6~$\mu$m \citep{mor12} rather than the typical torus warm
  component  which peaks around 20--30~$\mu$m \citep[e.g.,][]{mul11}. 
  The contamination from the host galaxies at 6~$\mu$m
 is also possible concern. \cite{mat15b} revised the 
 $L_{6 \mu {\rm m}}$--$L_{2-10}$ luminosity relations
by using a complete and flux limited sample of $>200$ AGN
 from the Bright Ultra hard \textit{XMM-Newton} Survey and \textit{WISE}.
They obtained absorption corrected X-ray luminosities and also 
derived the 6~$\mu$m AGN luminosity by the spectral decomposition
 of the torus and host galaxies.
They applied the Bayesian approach to linear regression with errors
 in both X and Y-axis by using the IDL command \verb|linmix_err| 
 with the X-ray luminosity as independent variable, which is the same
  method used by \cite{asm15}.
They report a slope of $b=0.99 \pm 0.03$ (over plotted as green 
dotted line in Figure~\ref{fig:LIRvsLx}) up to luminosities of
 $L_{2-10} \sim 10^{46}$~erg~s$^{-1}$. 
 This agrees well with the results of \cite{asm15} and ours ($b=0.96\pm0.02$).
Note that studies by \cite{mat15b} also might miss very luminous sources
due to the limited survey volume in X-ray, which is
possibly making the slope shallower.

Equations~(1) and (2) also show that intercept $a$ of $L_{22\ \mu{\rm m}}$
 is higher than that of $L_{12 \mu{\rm m}}$.
This tendency can be explained by two possibilities,
1) the torus emission peaks in $\nu F_{\nu}$ unit at 20--40~$\mu$m 
rather than $\sim10$~$\mu$m, which is suggested by both of the observation
\citep{wee05,buc06,mul11,asm11,asm14, ich15, ful16} and clumpy torus 
models \citep{nen08a,hon10, sch08}.
2) the star formation component contaminates more at longer 
wavelengths \citep[e.g.,][]{net07,mul11}.
The former contributes more strongly at high luminosity end
 ($L_{14-195} > 10^{44}$~erg~s$^{-1}$) because the relative star formation 
 contamination could be smaller considering the slope of $L_{\rm FIR}$--$L_{14-195}$
  is shallower ($b<0.94$). 
On the other hand, the latter contributes strongly to lower luminosity sources
 ($L_{14-195} < 10^{44}$~erg~s$^{-1}$). 
 This will be discussed again in Section~\ref{Sect:WISEcolor}.


\begin{figure*}
\begin{center}
\includegraphics[width=8.5cm]{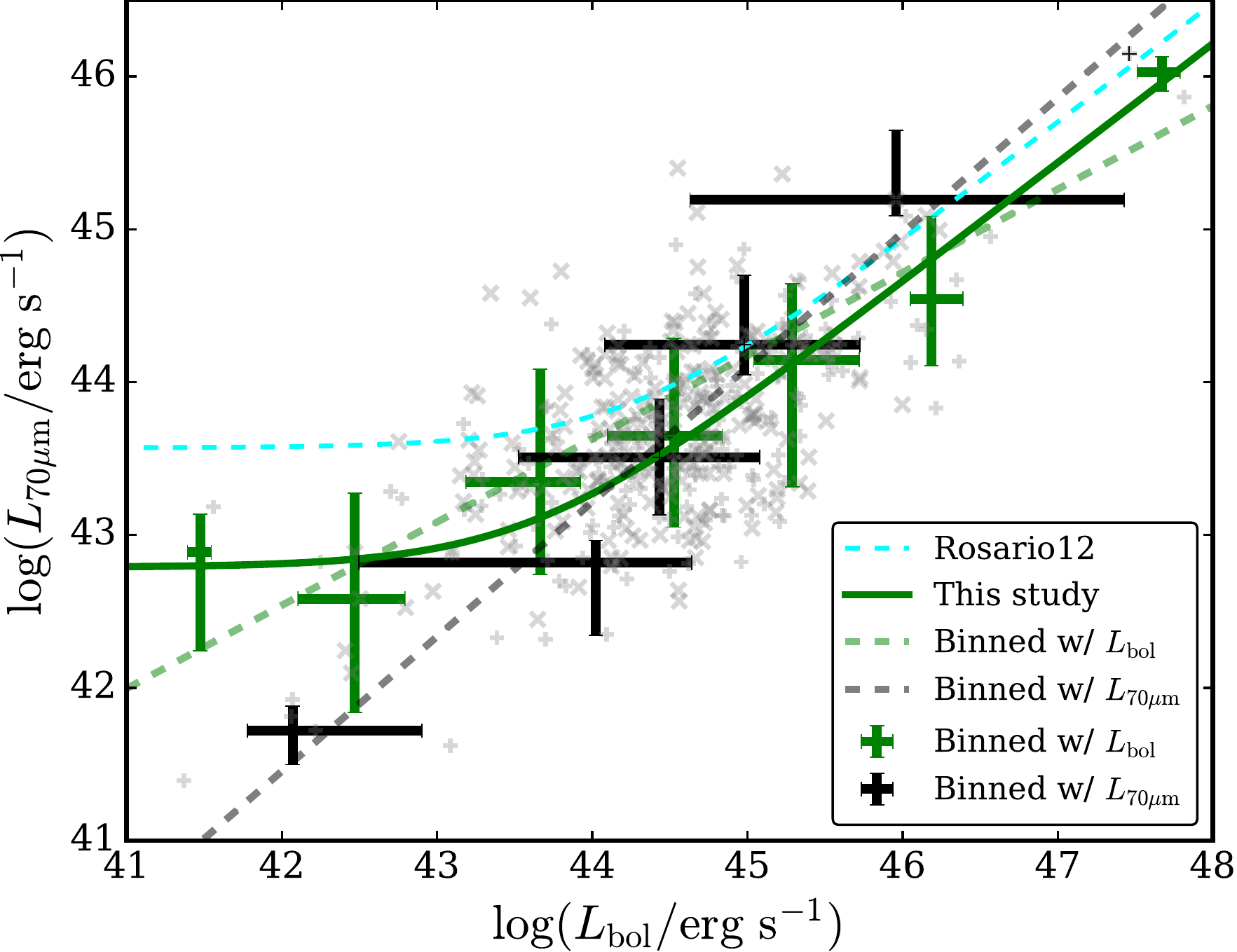}~
\includegraphics[width=8.5cm]{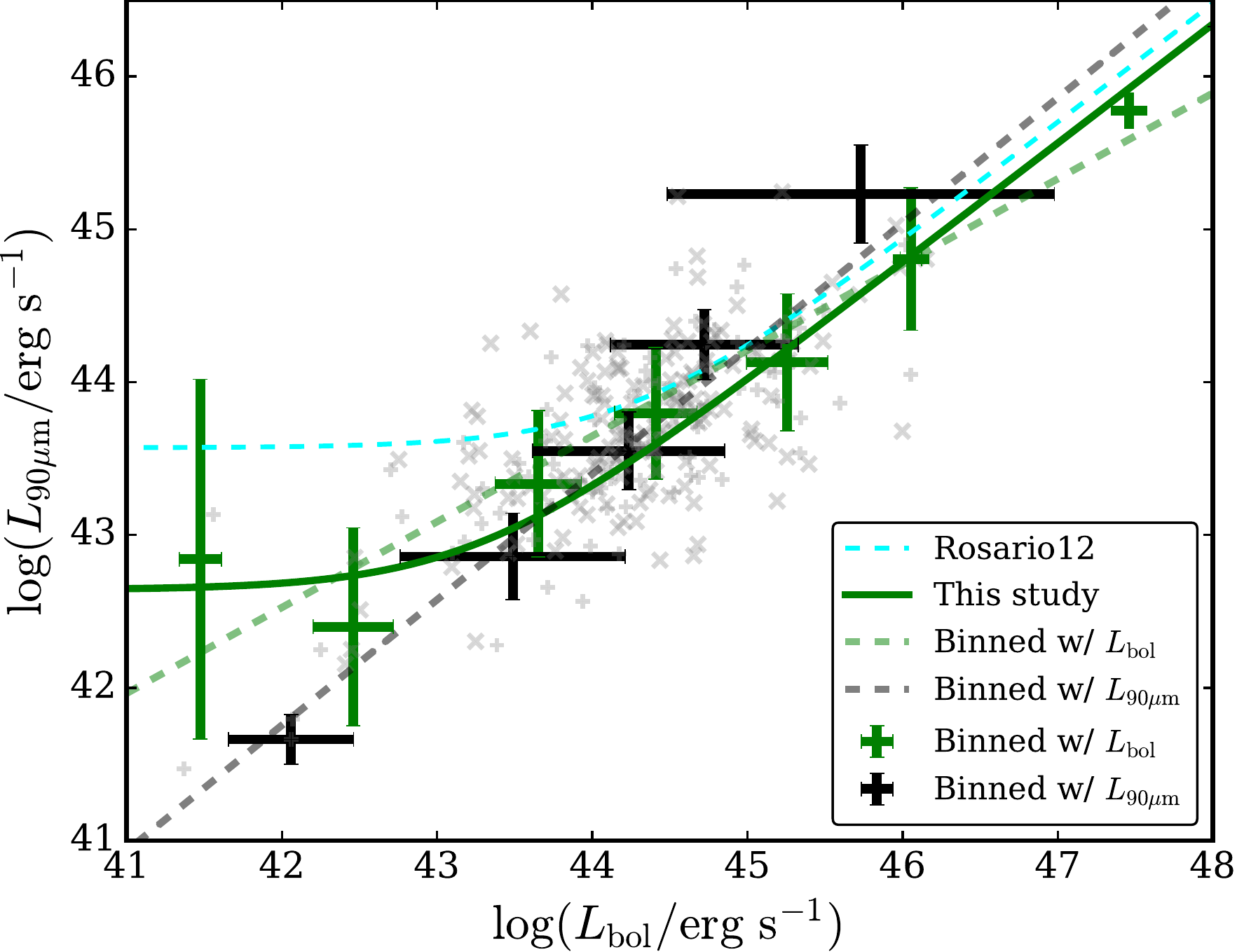}\\
\caption{
Mean luminosity correlations between the luminosities at 70 and 
90~$\mu$m ($L_{70~\mu{\rm m}}, L_{90~\mu{\rm m}}$) and 
bolometric luminosity ($L_{\rm bol}$) estimated from Equation (5).
Green color bin represents the mean measurements of 70 and
 90~$\mu$m luminosity as a function of bolometric luminosity.
Black solid bin represents the mean measurements of bolometric 
luminosity as a function of 70 and 90~$\mu$m luminosity, respectively.
The gray points represent AGN detected in both bands and are the
 same data points shown in Figure~\ref{fig:LFIRvsLbol}.
The green/black dashed line represents the slope obtained by the leas-bisector fit for
the green/black bin sample, respectively.
The cyan dashed line represents the fitted line of local X-ray selected AGN 
obtained from \cite{ros12}.
 The solid green line represents a fit to the relationship using the same function
 from as \cite{ros12}, but applied to our binned data shown by the green points.
}\label{fig:LFIRvsLbol_bin}
\end{center}
\end{figure*}

\subsection{Correlation between the FIR and AGN bolometric luminosities}\label{Sect:LFIRvsLbol}

The correlation between FIR and AGN bolometric luminosity ($L_{\rm bol}$) could shed the light on the link
between the star formation activity of the AGN host galaxies and the accretion rate of AGN.
Since the accretion disk emission cannot be directly obtained for all the sources of our sample,
 the bolometric correction should be applied to $L_{14-195}$ to estimate the bolometric luminosity.
\cite{mar04} account for variations in AGN SEDs by using the well-known anti-correlation
between the optical-to-X-ray spectral index ($\alpha_{\rm OX}$). 
Then, they renormalize the template SED to a particular $\alpha_{\rm OX}$ to obtain 
the bolometric correction with AGN luminosity.
Therefore, they assume a varying relation between optical/UV and X-ray luminosity, 
not a constant value \citep[e.g.,][]{elv94}.
 A similar approach is followed by \cite{hop07} who, however used a template SED
  generated from the averages of real SEDs in different wavebands.
There is a systematic difference that the one of \cite{hop07} is roughly 
a factor of $\sim1.5$ larger than that of \cite{mar04}.
This is because \cite{hop07} defines $L_{\rm bol}$ as
 the integral of the observed template SED including the
reprocessed emission in the MIR from the accretion disk, 
whereas \cite{mar04} only integrates the emission of optical-UV and X-ray radiated
 by the accretion disk itself and hot corona, respectively.
Since the accretion rate is better related to the total luminosity directly 
produced by the accretion process, $L_{\rm bol}$ defined
by \cite{mar04} is better suited for our study. Hence,
we apply the bolometric correction of \cite{mar04} with

 \begin{align}\label{Eq:LbolvsL14_195}
 \log L_{\rm bol} = 0.0378 (\log L_{\rm 14-195})^2 - 2.03 \log L_{\rm 14-195} + 61.6.
 \end{align}
 
 Figure~\ref{fig:LFIRvsLbol} shows that FIR luminosities
  ($L_{70~\mu{\rm m}},L_{90~\mu{\rm m}}$) are plotted against $L_{\rm bol}$.
The least-squares Bisector fits to the FIR versus bolometric AGN luminosity 
with a power-law give the correlations of

{\small
 \begin{align}
\log \frac{L_{70\mu{\rm m}}}{10^{43}~{\rm erg/s}} &= (-0.49 \pm 0.05) + (0.79 \pm 0.03) \log \frac{L_{\rm bol}}{10^{43}~{\rm erg/s}}\\
\log \frac{L_{90\mu{\rm m}}}{10^{43}~{\rm erg/s}} &= (-0.29 \pm 0.05) + (0.76 \pm 0.03) \log \frac{L_{\rm bol}}{10^{43}~{\rm erg/s}}
\label{Eq:LbolvsLFIR_eq}.
\end{align} 
}

Since our relations are obtained based on the FIR detected sample, which is not complete
 (65\% for 70~$\mu$m band and 45\% for 90~$\mu$m band) as shown in Figure~\ref{fig:zdist},
we check the dependence of the completeness by restricting
the redshift down to $z<0.076$ for 70~$\mu$m and $z<0.022$ for 90~$\mu$m band to achieve 80\% completeness of
the IR counterparts, respectively.
The relation at each band is given as

{\small
\begin{align}
\log \frac{L_{70\mu{\rm m}}}{10^{43}~{\rm erg/s}} &= (-0.53 \pm 0.06) + (0.83 \pm 0.04) \log \frac{L_{\rm bol}}{10^{43}~{\rm erg/s}}\\
\log \frac{L_{90\mu{\rm m}}}{10^{43}~{\rm erg/s}} &= (-0.19 \pm 0.08) + (0.82 \pm 0.05) \log \frac{L_{\rm bol}}{10^{43}~{\rm erg/s}}
.
\end{align}
}
The slope here is slightly steeper than those reported in Equation (6) and (7),
but the slopes are consistent within the 1$\sigma$ uncertainties (see also Figure~\ref{fig:LFIRvsLbol})
Therefore, we conclude that dependence of the completeness is weak.

In addition, we also estimate the effective luminosity range based on
 the limited volume of the \textit{Swift}/BAT AGN sample.
This is because AGN with the highest luminosity are rare and
so might be found in the limited \textit{Swift}/BAT survey volume.
Likewise, faint AGN will be missing from the sample because of the
\textit{Swift}/BAT ultra-hard X-ray flux limits.
First, based on the flux limit of \textit{Swift}/BAT survey of $f_{14-195} = 1.34
 \times 10^{-11}$~erg~s$^{-1}$~cm$^{-2}$ \citep{bau13},
and the conversion factor of $f_{2-10} = 2.1 \times f_{14-195}$,
we estimate the survey volume as a function of the flux limited luminosity
 $V(L_{2-10}) = (4/3) \pi D_{L_{2-10}}^3$~Mpc$^3$.
Next, we calculate the expected number of AGN detection as a function of
 $L_{2-10}$ using the 2-10~keV luminosity function from \cite{ued14} with a
  $z$-dependence of $\propto (1+z)^4$. 
Then, we define the effective luminosity range in which 
the expected number of detected AGN per dex in $L_{2-10}$ is greater than 10,
which is sufficient to measure the luminosity relation.
   The result is $40.8 < \log L_{2-10} < 45.5$ 
   which is equivalent to $41.1 < \log L_{14-195} < 45.8$ and $41.8 < \log L_{\rm bol} < 47.7$.
Therefore, a deeper and/or wider survey is needed to measure the relations between
FIR and AGN luminosity both at $\log L_{\rm bol} < 41.8$ and $\log L_{\rm bol}>47.7$.

In Figure~\ref{fig:LFIRvsLbol}, we also show the relation of \cite{net09} (navy dashed line). 
Our local sample reproduces well the relation of \cite{net09} using local optical type-2 AGN ($z \le 0.2$).
 The difference between our study and that of \cite{net09} is that while we used
  $L_{\rm FIR}$ as a proxy for star formation, \cite{net09} used the break at 4000~\AA~(D4000)
   for estimating $L_{\rm FIR}$.
 \cite{mat15a} reported that while D4000-based SFR is not well determined at lower 
 SFR since the calibration was based on starburst galaxies, the systematic difference
  between D4000-based SF luminosity and $L_{\rm FIR}$ is small enough compared to
   the broad distribution between SF luminosity and $L_{\rm AGN}$.
 Even with the $L_{\rm FIR}$, we find a consistent result with \cite{mat15a}, 
 using a sample of SDSS DR7 local AGN at $z<0.2$.

Recent studies have reported that ``mean'' or ``binned'' $L_{\rm FIR}$--$L_{\rm bol}$ 
show a flattened (or even horizontal) pattern in each redshift bin \citep[e.g.,][]{sha10,ros12,sta15} 
for $0<z<2.5$.
However, such flattened pattern is not detected in our sample when we use individual 
luminosity measurements instead of the mean luminosities.
To check this, the binned analysis is also applied to our 70~$\mu$m or 90~$\mu$m 
detected sources and the results are shown in Figure~\ref{fig:LFIRvsLbol_bin}.
The plotted bin is the median value in each luminosity bin with errorbars showing
 the interpecentage range containing 80\% of the sample.
Green points represent the mean measurements of 70 and 90~$\mu$m 
luminosity averaged in bins of bolometric luminosity, while black points represent
 the mean bolometric luminosity averaged in bins of 70 and 90~$\mu$m luminosity, 
 respectively.
 The dashed line represents the estimated relation based on the least-square Bisector fits.
 As shown in the Figure, the slope ($b=0.54 \pm 0.08$ for 70~$\mu$m 
 and $b=0.56 \pm 0.07$ for 90~$\mu$m) of green dashed line (binned with $L_{\rm bol}$)
  is significantly shallower than that of the black dashed line (binned with $L_{\rm FIR}$;
  $b=0.88 \pm 0.05$ for 70~$\mu$m and $b=0.82 \pm 0.06$ for 90~$\mu$m).
To further model the trend of green points, we apply curve fit used in \cite{ros12}.
The function is written as
 \begin{align}
 \log L_{\rm FIR} = \log \left( 10^{b \log L_{\rm bol} + \log L_{\rm b} - b \log L_{\rm c}} + 10^{\log L_{\rm b}}  \right)
 \end{align} 
with three free parameters ($b$, $\log L_{\rm b}$, $\log L_{\rm c}$) 
but we fix the slope $b=0.78$ by following \cite{ros12}.
$L_{\rm b}$ is a constant value mainly determined by the constant $L_{\rm FIR}$
value where $L_{\rm AGN}$ is small.
$L_{\rm c}$ represents the value of $L_{\rm AGN}$ where the function becomes equal to $L_{\rm b}$.
We fit the green points using a non-linear least squares fitting procedure (curve\_fit in Python).
The result is shown with the green solid line in Figure~\ref{fig:LFIRvsLbol_bin}.
The model nicely reproduces the flattened relation, but systematically smaller value
($\log L_{\rm b}=42.79 \pm 0.02$ for 70~$\mu$m and $\log L_{\rm b}=42.64 \pm 0.03$
for 90~$\mu$m) than the line of local AGN ($\log L_{\rm b}= 43.57 \pm 0.08$) in \cite{ros12}. 
This would be because our sample contains deeper data from \textit{Herschel},
 whereas \cite{ros12} use shallower \textit{IRAS}-FSC data
for the local AGN and do not apply stacking analysis of non-detection
 sources for this local sample.
 
Overall, whereas the IR averaging (in bins of bolometric luminosity, shown by the green
points) nicely reproduces the flattened trend as
 reported in the literature \citep{sha10, ros12},
the black solid bin still holds rising trend almost same as the relation 
obtained from the individual objects.
This could be originated from the different time scale of SF and AGN activity.
\cite{hic14} calculated mean $L_{\rm FIR}$--$L_{\rm bol}$ in two ways.
\cite{hic14} constructed a simple model population of SF galaxies in 
which SF and the BH growth are correlated in galaxies
across a range of $0.25<z<1.25$, with a $z$ dependent distribution 
in SFR from the FIR luminosity function derived by \cite{gru13}.
They assigned an observed average SFR to BH accretion rate of 
3000 \citep[e.g.,][]{raf11, mul12, che13}, and also assumed that the
 instantaneous accretion rate relative to the average is distributed from
  the given fiducial luminosity distribution.
They first derived the averaged $L_{\rm bol}$ for galaxies in each
 $L_{\rm FIR}$ bin and compared the results obtained by
  \cite{sym11} and \cite{che13} for a range of $0.25<z<1.25$.
This reproduces well the rising relation as shown in our study.
They next computed the average $L_{\rm FIR}$ as a function of $L_{\rm bol}$.
This then reproduces well the flattened relation. 
This result strongly suggests a picture in which SF
and BH accretion are closely connected over long timescales,
but this correlation is sometimes hidden at low to moderate $L_{\rm bol}$
 due to the short-term AGN variability.
Note that there are clear difference of the sample used in the aforementioned
 studies and ours. 
 They included all FIR detected galaxies whereas we focused only AGN 
 host galaxies with the both detections in FIR and X-rays.
 Further studies using the spectral decomposition of IR SEDs will be 
 discussed in a forthcoming paper (K. Ichikawa et al. in prep).

\begin{figure}
\begin{center}
\includegraphics[width=7cm]{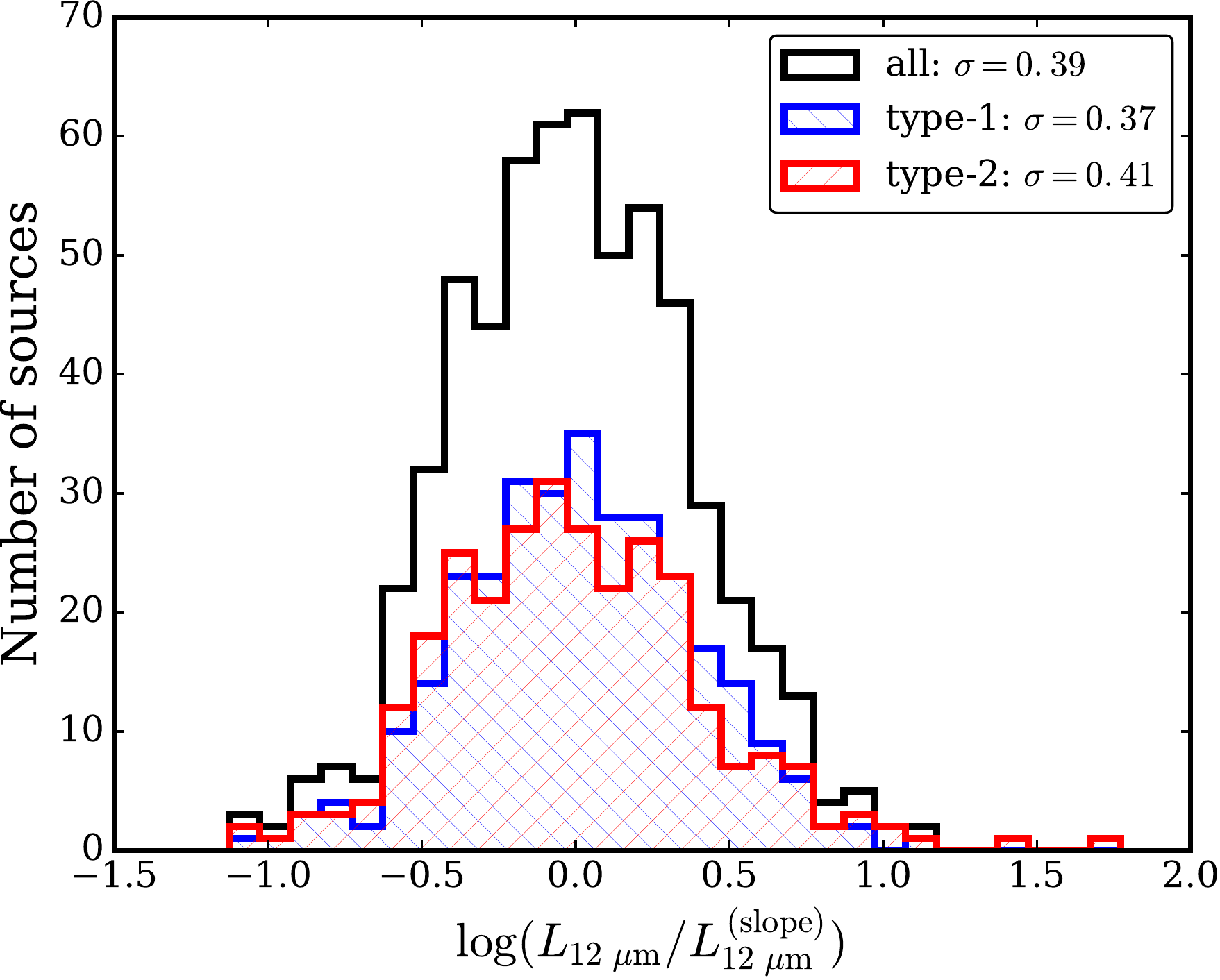}\\
\includegraphics[width=7cm]{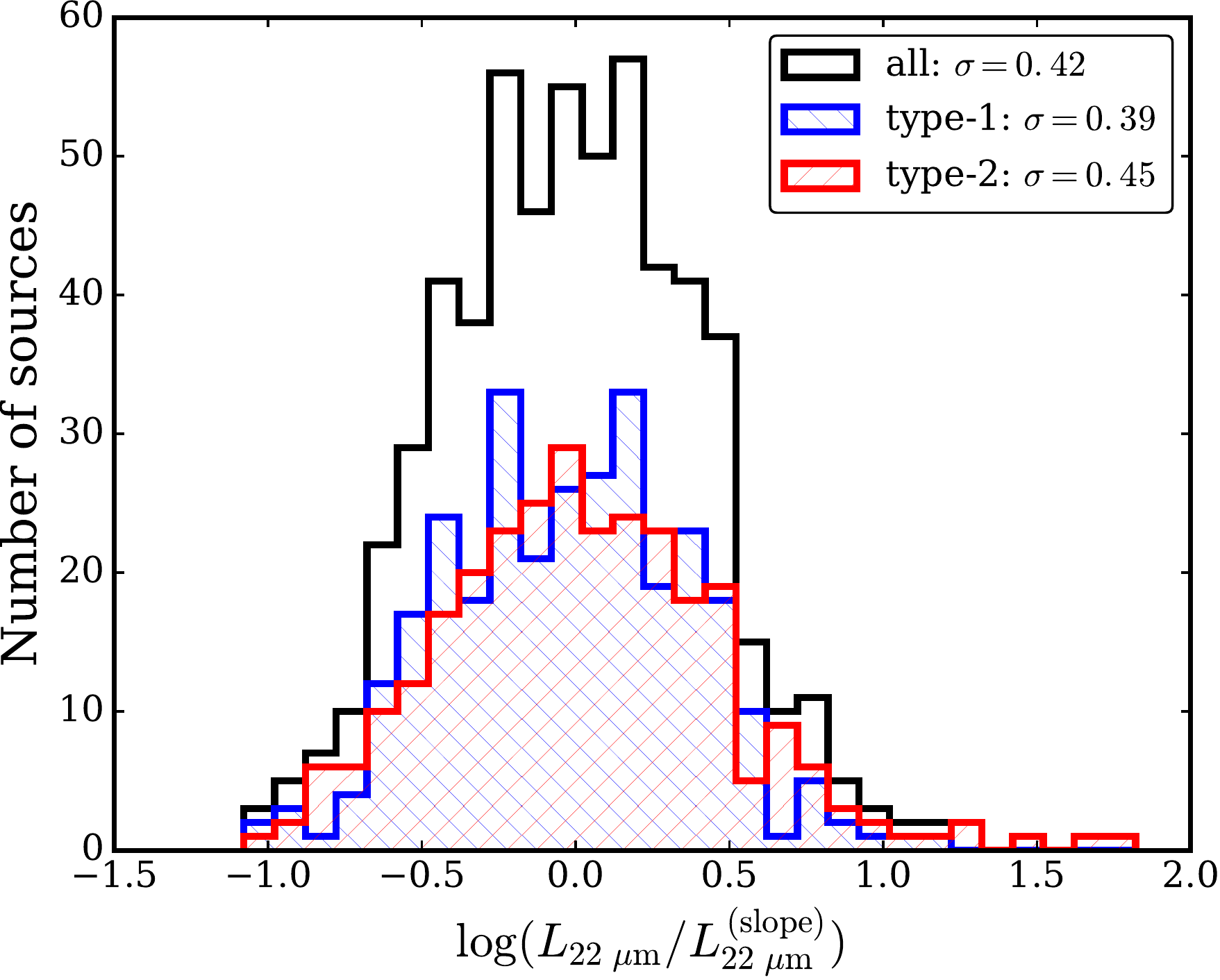}
\caption{
Histograms of $r_{12,22} = \log (L_{12~\mu{\rm m}, 22~\mu{\rm m}}/
L_{12~\mu{\rm m}, 22~\mu{\rm m}}^{\rm (slope)})$ 
(top and bottom panel, respectively).
The solid black/shaded blue/shaded red line represents the 
total, type-1, and type-2 sample, respectively.}\label{fig:LMIRovLx}
\end{center}
\end{figure}

\subsection{FIR Pure-AGN candidates}
If there are luminous AGN hosted by low-SF galaxies, we may find
 those ``FIR pure-AGN'' candidates at the bottom right in Figure~\ref{fig:LFIRvsLbol}.
\cite{mat15a} investigated the FIR pure-AGN sequence between
 $L_{\rm FIR}$--$L_{\rm bol}$ by adopting a typical AGN SED template of \cite{mul11}. 
The FIR pure-AGN sequences are given with
\begin{align}
\log \frac{L_{70~\mu{\rm m}}}{{\rm erg~s}{^{-1}}} = (7.45 \pm 0.26) + 0.80\log \frac{L_{\rm bol}}{{\rm erg~s}{^{-1}}}\\
\log \frac{L_{90~\mu{\rm m}}}{{\rm erg~s}{^{-1}}} = (7.17 \pm 0.26) + 0.80\log \frac{L_{\rm bol}}{{\rm erg~s}{^{-1}}}.
\end{align}
In Figure~\ref{fig:LFIRvsLbol}, the estimated FIR pure-AGN sequence
 is also over plotted with gray lines. 
In this study we define the FIR pure-AGN candidate if the source 
is located under the FIR pure-AGN sequence in Figure~\ref{fig:LFIRvsLbol}.
As a result, 50 and 4 sources fulfill the criterion at 70 and 90~$\mu$m, respectively.
There is a clear number difference between 70 and 90~$\mu$m even 
when we consider the ratio of the FIR pure-AGN to the sample
 ($50/388 \sim 13$\% for 70~$\mu$m while $4/241 \sim 2$\% for 90~$\mu$m criterion). 
 One reason could originate from the sensitivity difference.
 The sensitivity at 70~$\mu$m is better than at 90~$\mu$m
 because of the inclusion of the \textit{Herschel}/PACS detected
  sources at 70~$\mu$m. 
Of the 50 objects selected at 70~$\mu$m, 42 are not detected at 90~$\mu$m.
It is also likely that, at shorter wavelengths, the typical AGN contribution becomes
stronger while the SF contribution becomes weaker \citep[e.g., ][]{mul11}.
To support this, all the four pure-AGN candidates at 90~$\mu$m also fulfill
 the criterion at 70~$\mu$m, while only 50\% (4/8) of 70~$\mu$m selected 
 pure-AGN candidates with 90~$\mu$m detection fulfill the criterion at 90~$\mu$m.

The names of 90~$\mu$m selected five sources are NGC~1194, 
ESO~506-G027, NGC~5252, and CGCG~164-019.
All are type-2 AGN and average luminosity is high with
 $\langle \log (L_{\rm 14-195}/{\rm erg}~{\rm s}^{-1}) \rangle=44.1$.
While those pure-AGN population is quite small with $\sim2$\% 
(4 out of 274), they are good sample to construct the pure-AGN IR
 SEDs including the FIR end, and to examine the extrapolation to 
 FIR luminosities from the intrinsic-AGN SED is correct. 
They also could be in the stage that SF is suppressed because 
AGN feedback is in action \citep[e.g.,][]{woo16}.
The future X-ray satellite e-ROSITA \citep{mer12} will discover 
over 3-million AGN and cross-matching those with the FIR catalogs
 would reveal pure-AGN in large numbers.

\begin{figure*}
\begin{center}
\includegraphics[width=8.5cm]{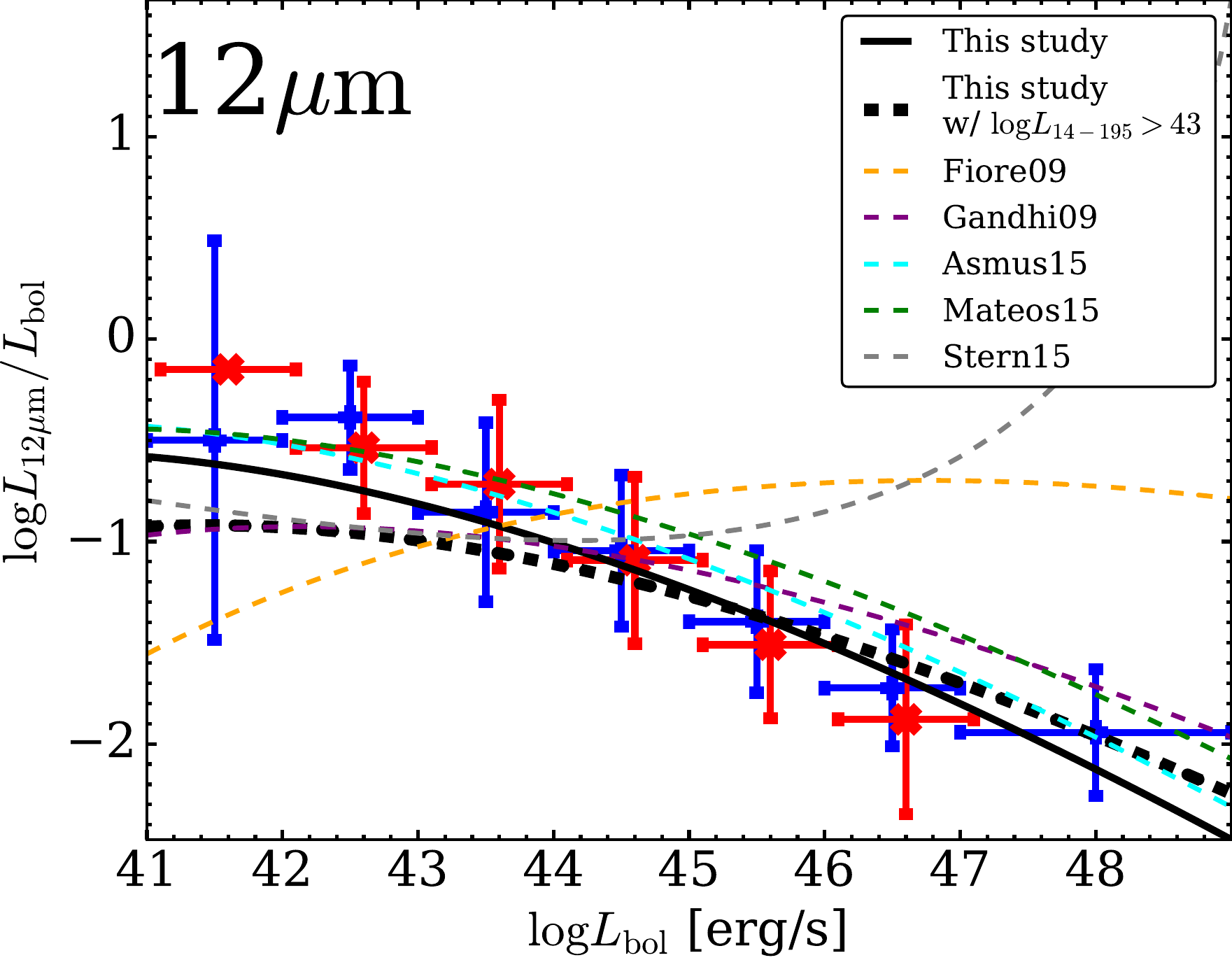}~
\includegraphics[width=8.5cm]{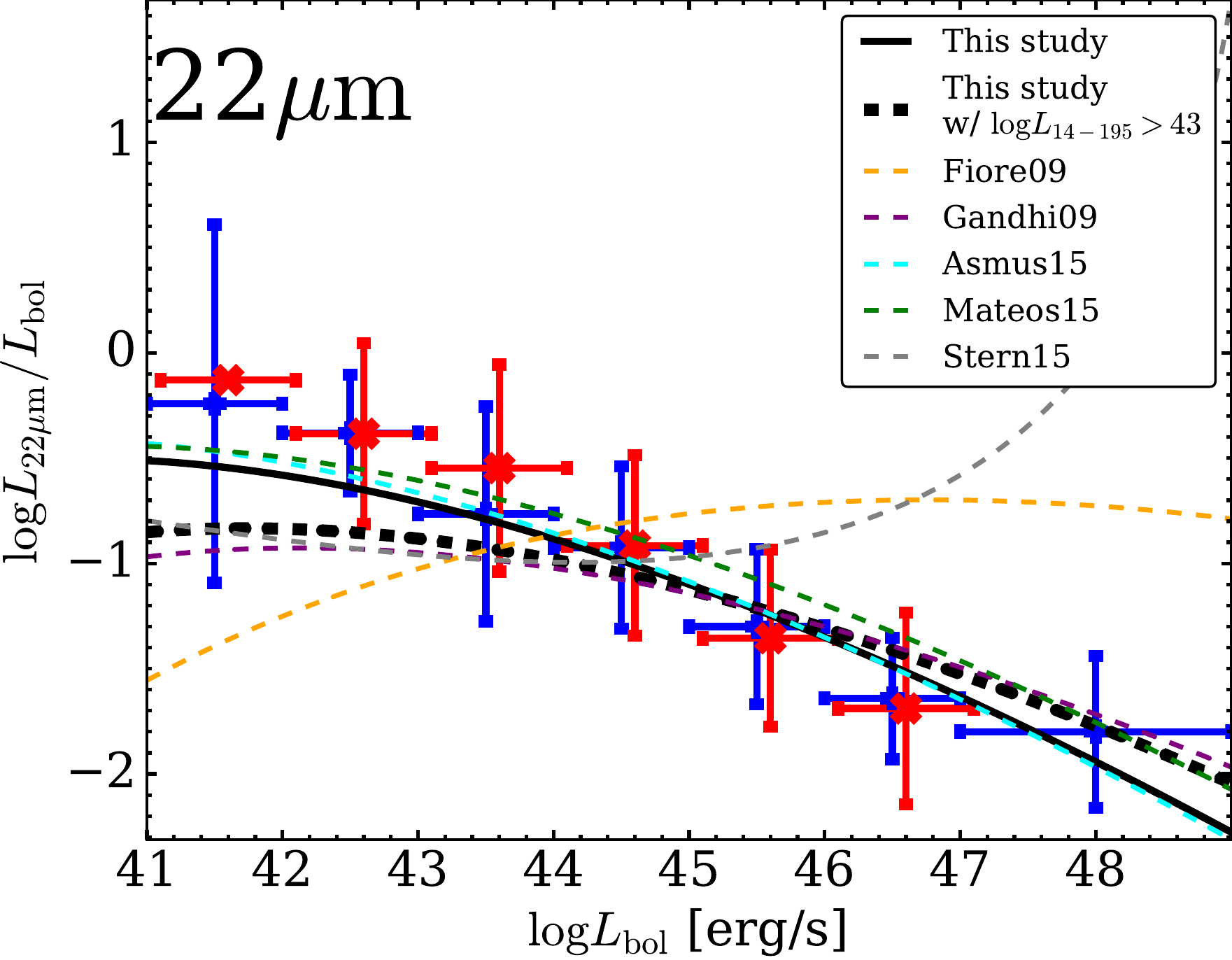}
\caption{MIR luminosity to bolometric luminosity ratio as a function 
of bolometric luminosity.
$L_{\rm bol}$ is estimated from Equation~\ref{Eq:LbolvsL14_195} 
for our sample, while from Equation~\ref{Eq:LbolvsL2_10} for the other
 studies from the literature.
(Left) $\log L_{12 \mu{\rm m}}/L_{\rm bol}$ versus $\log L_{\rm bol}$. 
(Right) $\log L_{22 \mu{\rm m}}/L_{\rm bol}$ versus $\log L_{\rm bol}$.
Blue cross/red X-shape shows type-1/type-2, respectively. 
The red crosses are shifted to the right by 0.1 dex for clarity.
The main AGN type used in \cite{fio09}, \cite{ste15} are type-1, and
that of \cite{gan09}, \cite{mat15b}, \cite{asm15}, and our study are both
 of type-1 and type-2 AGN.
}\label{fig:LIRovLbolvsLbol}
\end{center}
\end{figure*}

\begin{figure*}
\begin{center}
\includegraphics[width=8.5cm]{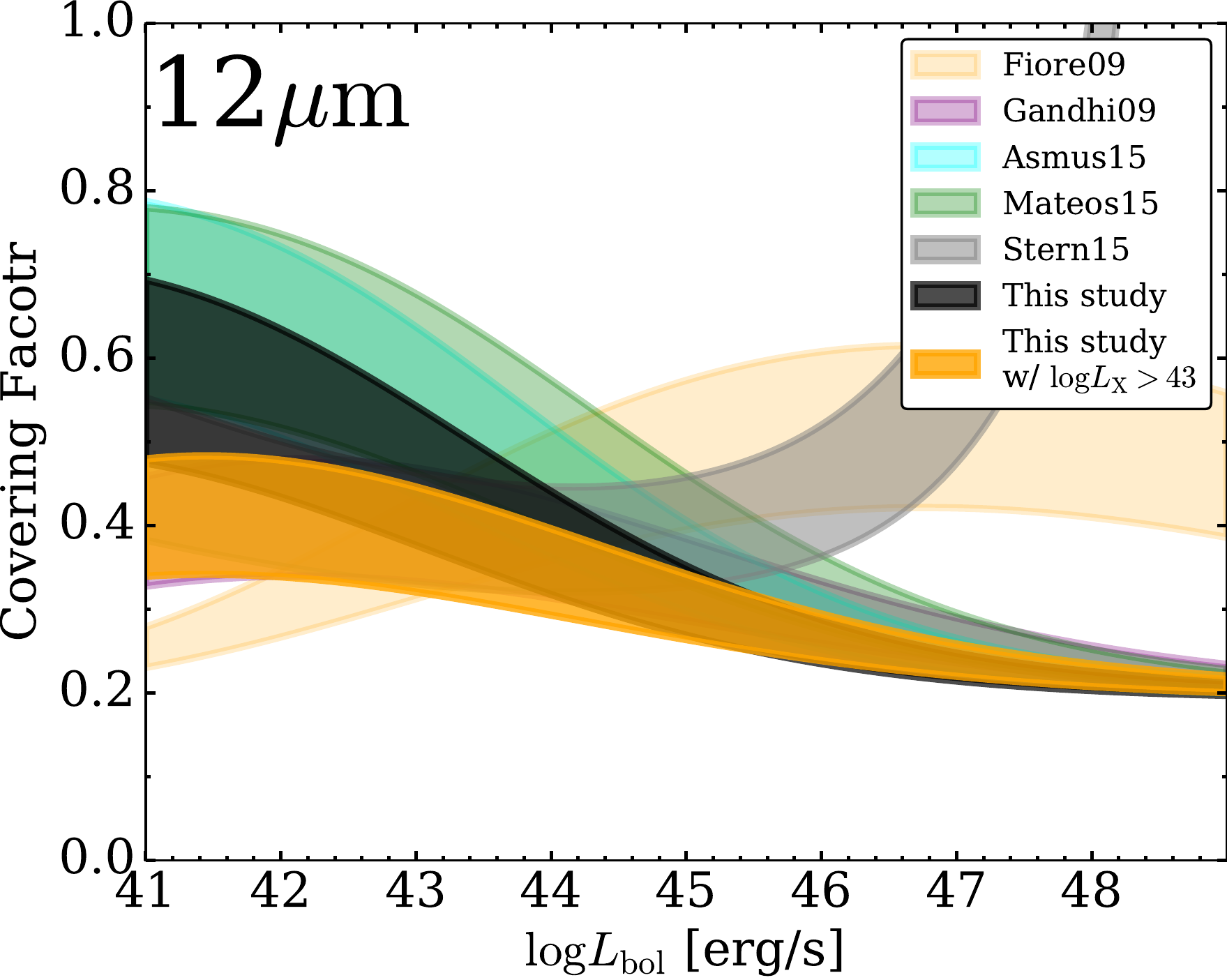}~
\includegraphics[width=8.5cm]{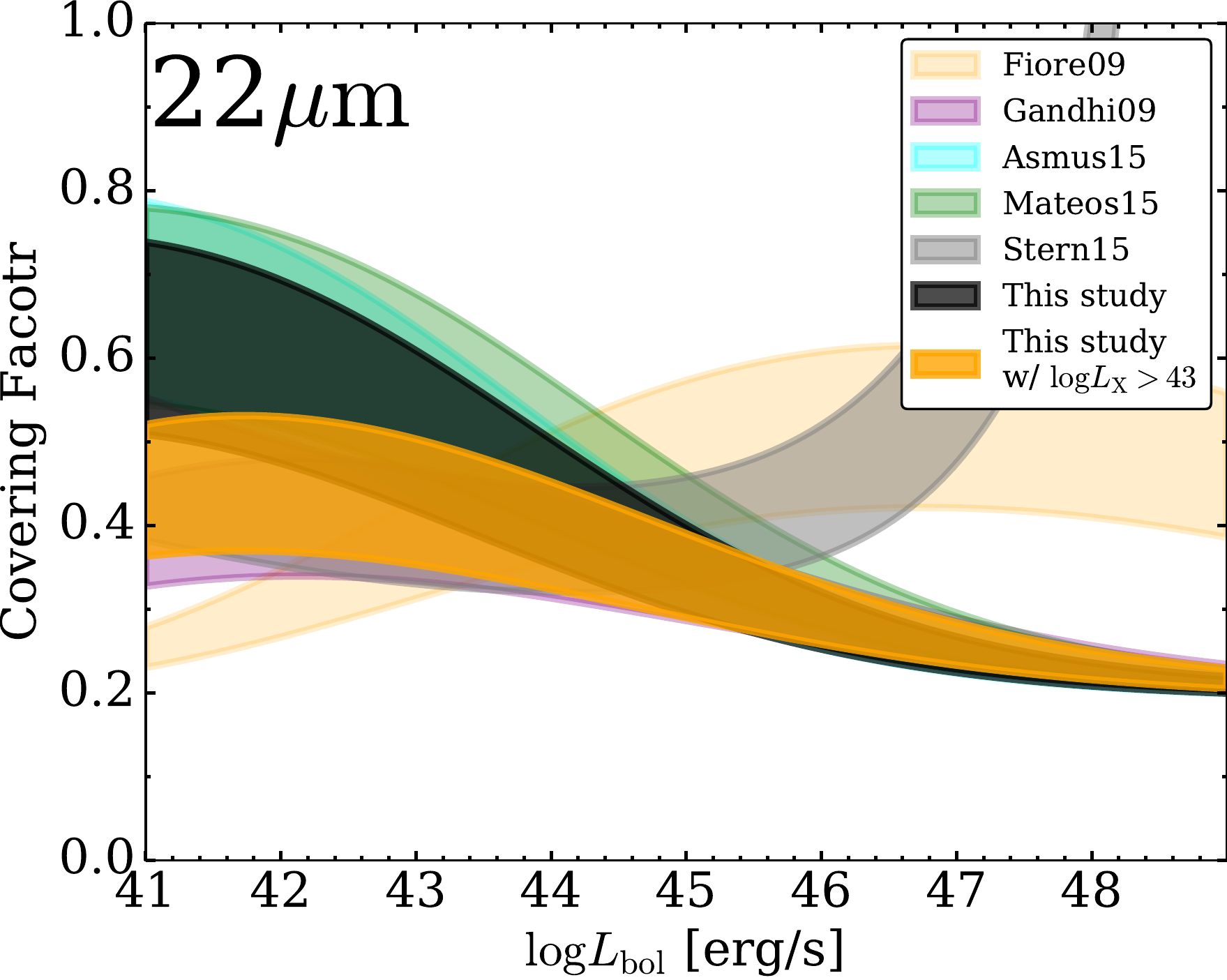}
\caption{ Covering factor as a function of bolometric luminosity obtained
 from 12~$\mu$m (left) and 22~$\mu$m band (right). 
 The correction from $L_{\rm MIR}/L_{\rm bol}$ is made using the correction
  function of \cite{sta16}.
Each filled area represents the range of the possible covering factor at 
each $L_{\rm bol}$. Since the corrected function of type-1 AGN always
 gives the higher value than that of type-2 AGN, we assign an upper limit 
under the assumption that all sources in this study are type-2 AGN, and a lower limit 
 assuming that all sources are type-1 AGN.
The main AGN type used in \cite{fio09}, \cite{ste15} are type-1, and
that of \cite{gan09}, \cite{mat15b}, \cite{asm15}, and our study are both 
of type-1 and type-2 AGN.
}\label{fig:CFvsLbol}
\end{center}
\end{figure*}

\subsection{Distribution of $r_{12,22}$}
Figure~\ref{fig:LMIRovLx} shows the histograms of the ratio
defined as $r_{12,22} = \log (L_{12~\mu{\rm m}, 22~\mu{\rm m}}/
L_{12~\mu{\rm m}, 22~\mu{\rm m}}^{\rm (slope)})$,
where $L_{12~\mu{\rm m}, 22~\mu{\rm m}}^{\rm (slope)}$ represents
 the expected MIR luminosities obtained from $L_{14-195}$ using the 
 slopes between $L_{\rm MIR}$ and $L_{14-195}$.
The standard deviation for each sample is compiled in Table~\ref{tab:LIRovLx}.
The standard deviation for the full sample is $\sigma=0.39$ at 12~$\mu$m and
 $\sigma=0.42$ at 22~$\mu$m.
This value is slightly larger than that of \cite{asm15} with $\sigma=0.32$,
In 12~$\mu$m band, we obtain $\sigma = 0.40 \pm 0.03$ for type-1
and $\sigma = 0.43 \pm 0.04$ for type-2 AGN.
The scatter is consistent between the type-1 and type-2 AGN in our sample,
within the statistical uncertainties, so we find no evidence for a difference in the
scatter of the MIR to 14--195~keV X-ray ratio between AGN types.
This result might support recent observations that 
most of the MIR emission comes from the polar extended region
with $\le 10$~pc scale \citep[e.g., ][]{hon12, hon13, lop16} or from even
larger $\simeq 100$ pc scales \citep{asm16} since an extended geometry of dust
 more easily produces isotropic MIR emission compared to traditional torus models.

\subsection{Luminosity Dependence of Covering Factor}
We investigate the relation between the AGN and its surrounding dusty torus.
Since MIR emission originates from the re-radiation from the dusty torus, 
 one can naturally expect that the ratio of the MIR to AGN luminosity corresponds 
 to the solid angle of the sky covered by the dust 
 (i.e., covering factor; $C_{\rm T}$ and $L_{\rm MIR} \propto C_{\rm T}L_{\rm bol}$). 
 Figure~\ref{fig:LIRovLbolvsLbol} shows the luminosity dependence
  of $L_{\rm MIR}/L_{\rm bol}$.
 The black solid and dashed line in Figure~\ref{fig:LIRovLbolvsLbol} represents
   the estimated line converted from the
  $L_{\rm MIR}$--$L_{14-195}$ luminosity relations 
  of Equations (1), (2) for the full sample, and Equation (3), (4)
  for the high-luminosity sample, respectively.
We apply Equation~\ref{Eq:LbolvsL14_195} for the bolometric correction.
Figure~\ref{fig:LIRovLbolvsLbol} shows that $L_{\rm MIR}/L_{\rm bol}$ is declining when $L_{\rm bol}$ 
increases, being consistent with the trend so called ``luminosity-dependent unified models''.
This model can describe the decrease of covering factor by receding 
the sublimation radius with the AGN luminosity \citep{law82}. 

However, \cite{lus13} found that corrections for the anisotropy for the dust 
emission are necessary for using $L_{\rm MIR}/L_{\rm bol}$ as a proxy of
 the covering factors.
 In addition, using a 3D Monte Carlo radiation code, \cite{sta16} reported
  that the tori of type 1 (viewed from face-on) AGN make
   $L_{\rm MIR}/L_{\rm bol}$ underestimate low covering factors
and overestimate high covering factors. 
Type 2 (viewed from edge-on) AGN always underestimates 
covering factors. 
They also provide the correction functions to account for anisotropy and
 obtain corrected covering factors.
Thus, we derive the corrected covering factor using the combination of 
the ratio $L_{12~\mu{\rm m}} / L_{\rm bol}$, 
$L_{22~\mu{\rm m}} / L_{\rm bol}$ and the correction function by \cite{sta16}.
We use the correction function of

{\small
\begin{equation}
C_{\rm T}=
\begin{cases}
-0.178R^4+0.875R^3-1.487R^2+1.408R+0.192\ {\rm (type1)}\\
2.039R^3-3.976R^2+2.765R+0.205\ {\rm (type2)}
\end{cases}
\end{equation}
}
where $R=L_{\rm MIR}/L_{\rm bol}$ and 
the estimated optical thickness of torus at 9.7~$\mu$m is $\tau_{9.7}=3.0$
(see Table~1 of \cite{sta16} for more details).
Figure~\ref{fig:CFvsLbol} shows corrected $C_{\rm T}$ derived
 from the slopes tabulated in Table~\ref{tab:LIRvsLx_wlit} as a 
 function of $L_{\rm bol}$ (black solid area).
It still holds that $C_{\rm T}$ is a declining function of $L_{\rm bol}$, 
confirming the trend of ``luminosity-dependent unified models''.

It is principle possible that the luminosity-dependent trend may be due
largely to host galaxy contamination, since the emission from 
the host galaxy contributes significantly to MIR emission in the
 low luminosity end as discussed in Section~\ref{sect:LxvsLIR}.
To check this effect, in Figures~\ref{fig:LIRovLbolvsLbol} and 
\ref{fig:CFvsLbol} we also show the $L_{\rm MIR}/L_{\rm bol}$ 
and the corrected $C_{\rm T}$ using the slope with high luminosity
 sample with $\log L_{14-195}>43$, then extrapolating them to
  the lower luminosity end.
The luminosity dependence of the $C_{\rm T}$ mitigates, but 
still holds the relations.
This idea has been gaining observational evidence from radio
 \citep{gri04}, IR \citep{mai07, tre08, mor09, alo11, ich12b, tob13, tob14},
optical \citep{sim05}, and X-ray \citep{ued03,bec09, ued11, ric13, lus13, 
ued14}
 studies of AGN.
On the other hand, in the high-$z$ universe with $z=2-3.5$, \cite{net16} 
reported in their sample infer covering factors consistent with 
no-evolution with AGN luminosity within the uncertainties for bolometric
 correction factor.

\begin{figure}
\begin{center}
\includegraphics[width=8.0cm]{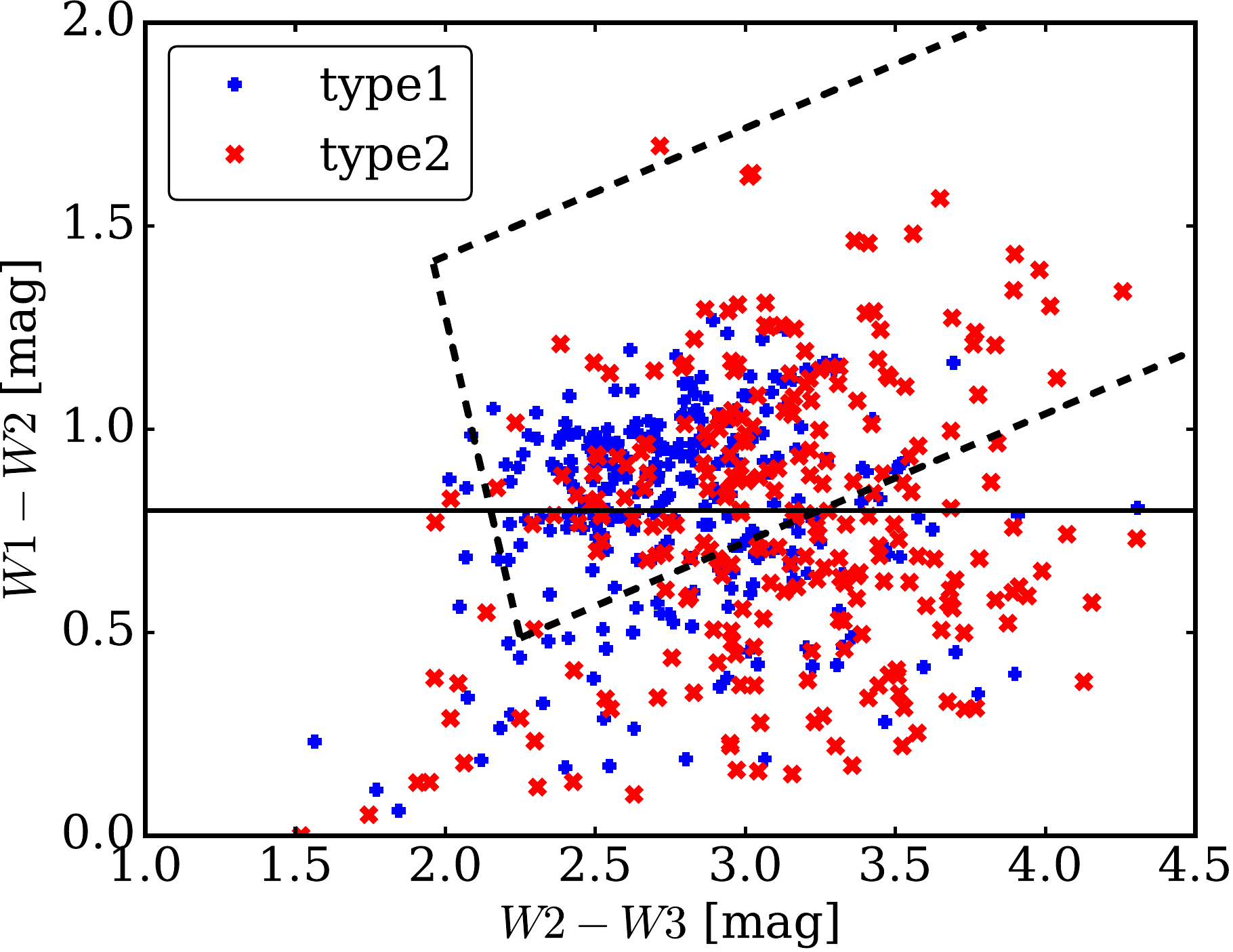}\\
\caption{W1--W2 versus W2--W3 two color diagram in the 
unit of Vega magnitude. The two color diagram for each type
highlighted with blue (type-1) and red (type-2).
}\label{fig:IR2color}
\end{center}
\end{figure}

\begin{figure*}
\begin{center}
\includegraphics[width=8.0cm]{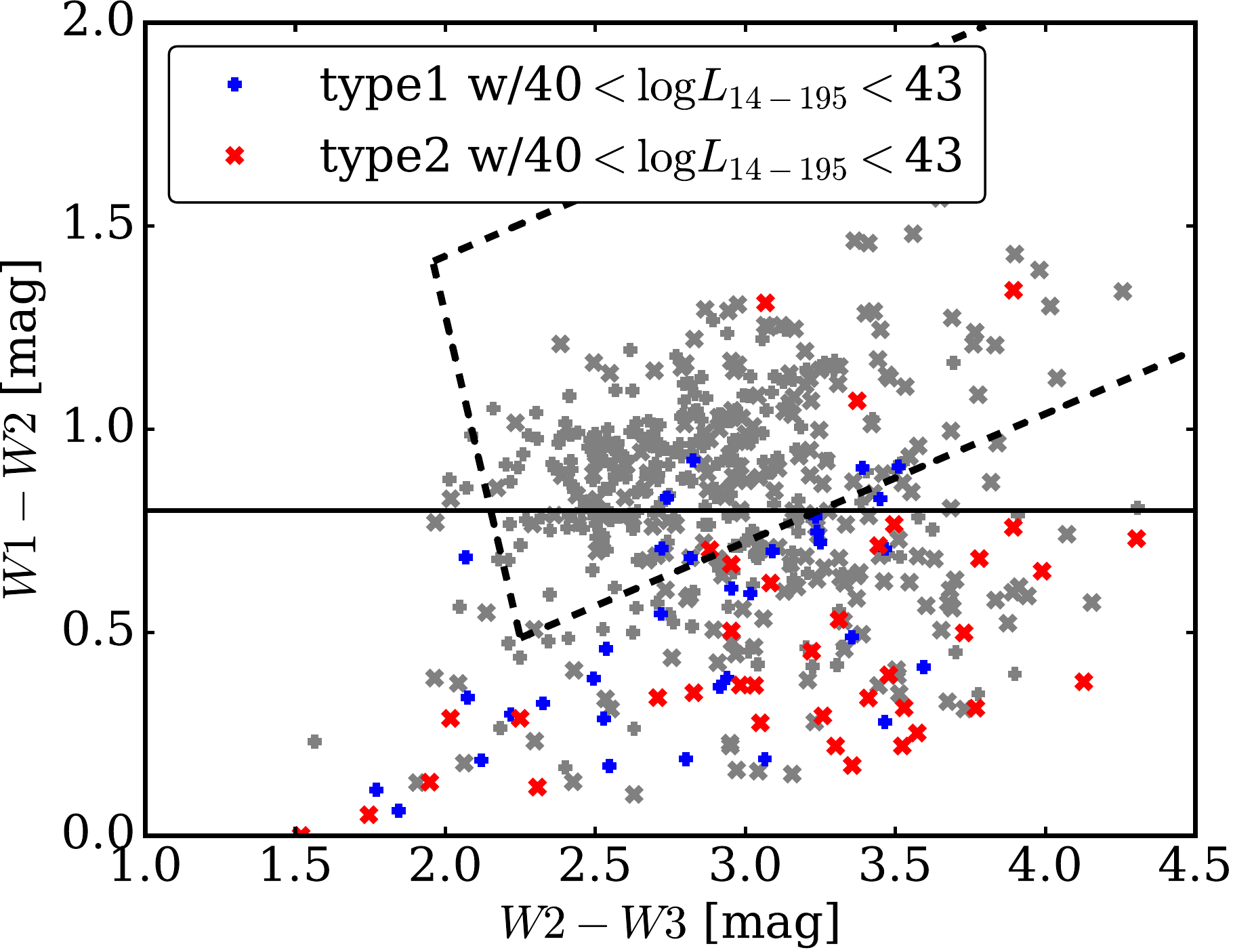}~
\includegraphics[width=8.0cm]{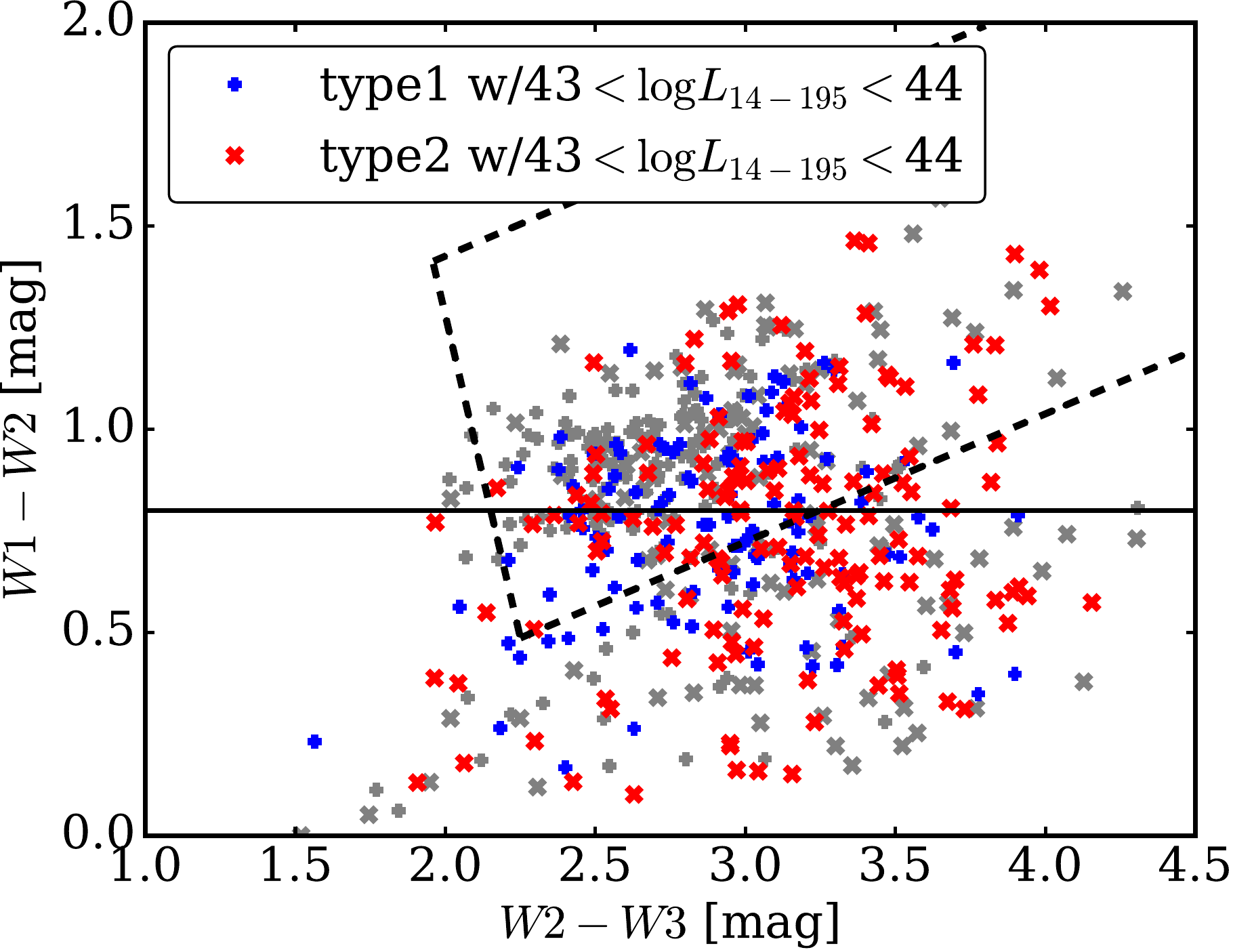}\\
\includegraphics[width=8.0cm]{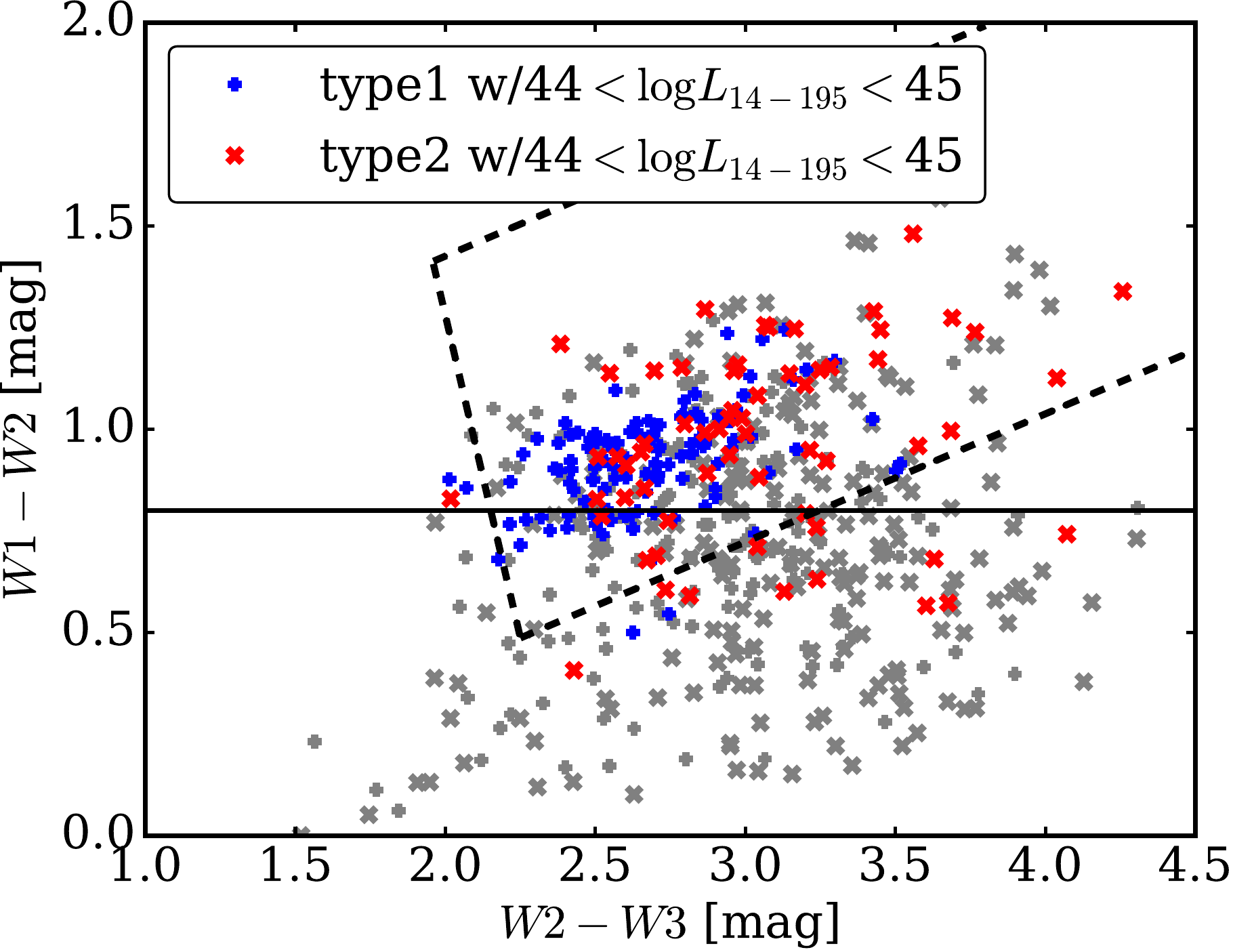}~
\includegraphics[width=8.0cm]{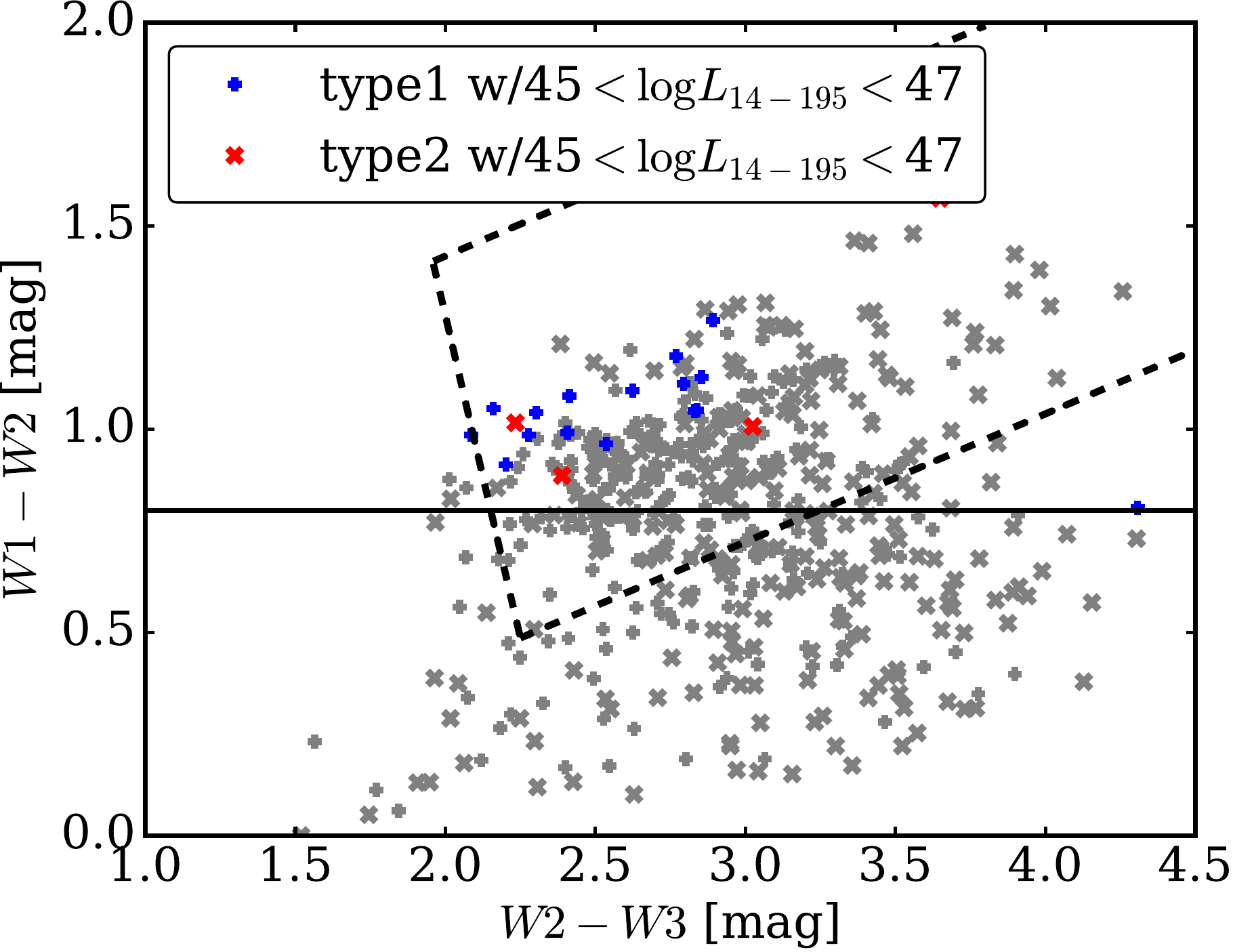}\\
\caption{W1--W2 versus W2--W3 two color diagram in the unit of Vega magnitude with different BAT luminosity populations,
 highlighted with blue (type-1) and red (type-2) over plotted with the total sample with gray.
}\label{fig:IR2color_type}
\end{center}
\end{figure*}

Again, the previous studies from the literature are also over
 plotted in Figure~\ref{fig:LIRovLbolvsLbol} and \ref{fig:CFvsLbol} 
 using the bolometric correction of \cite{mar04}:
\begin{align}\label{Eq:LbolvsL2_10}
\log L_{\rm bol} = 0.0378 (\log L_{2-10})^2 - 2.00 \log L_{2-10} +60.5
\end{align}
for 2--10~keV luminosity ($L_{2-10}$).
As shown in Figure~\ref{fig:LIRovLbolvsLbol} and \ref{fig:CFvsLbol}, studies in the local universe \citep{gan09,asm15} found a decrease of the covering factor with the AGN luminosity.
Even the high luminosity sample of \cite{mat15b} in the high-$z$ universe, same trend can be observed.
On the other hand, the studies carried out using high-$L$ (and high-$z$) sources by \cite{fio09} and \cite{ste15} strongly contradict the luminosity dependent unified models.
Considering their rare population of high luminosity AGN in the local universe 
as discussed in Section~\ref{Sect:LFIRvsLbol}, 
further investigation with deep survey is necessary for solving this controversy at the high luminosity end.

\subsection{WISE color-color distribution of Hard X-ray Selected AGN}\label{Sect:WISEcolor}

\begin{figure}
\begin{center}
\includegraphics[width=7.0cm]{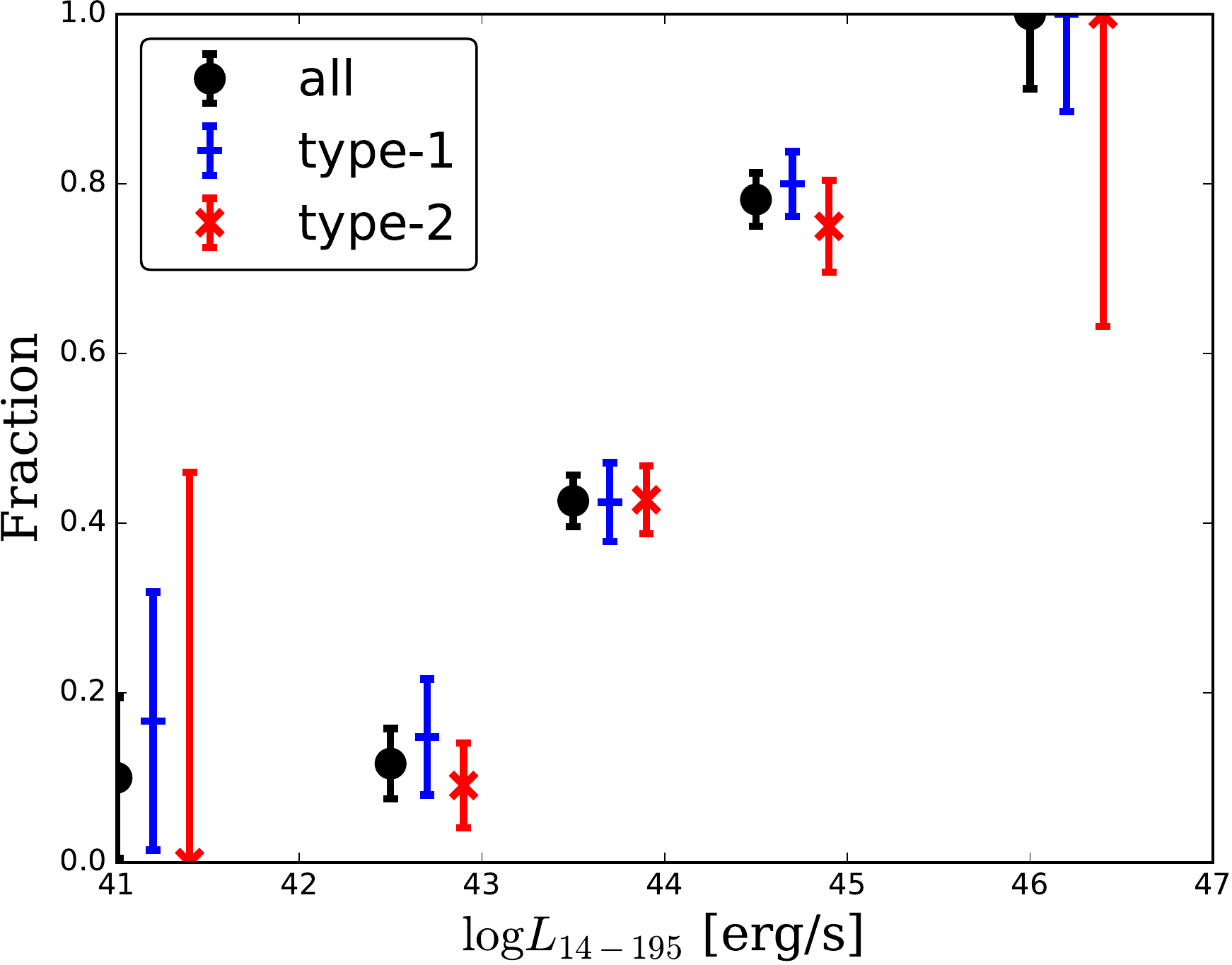}\\
\includegraphics[width=7.0cm]{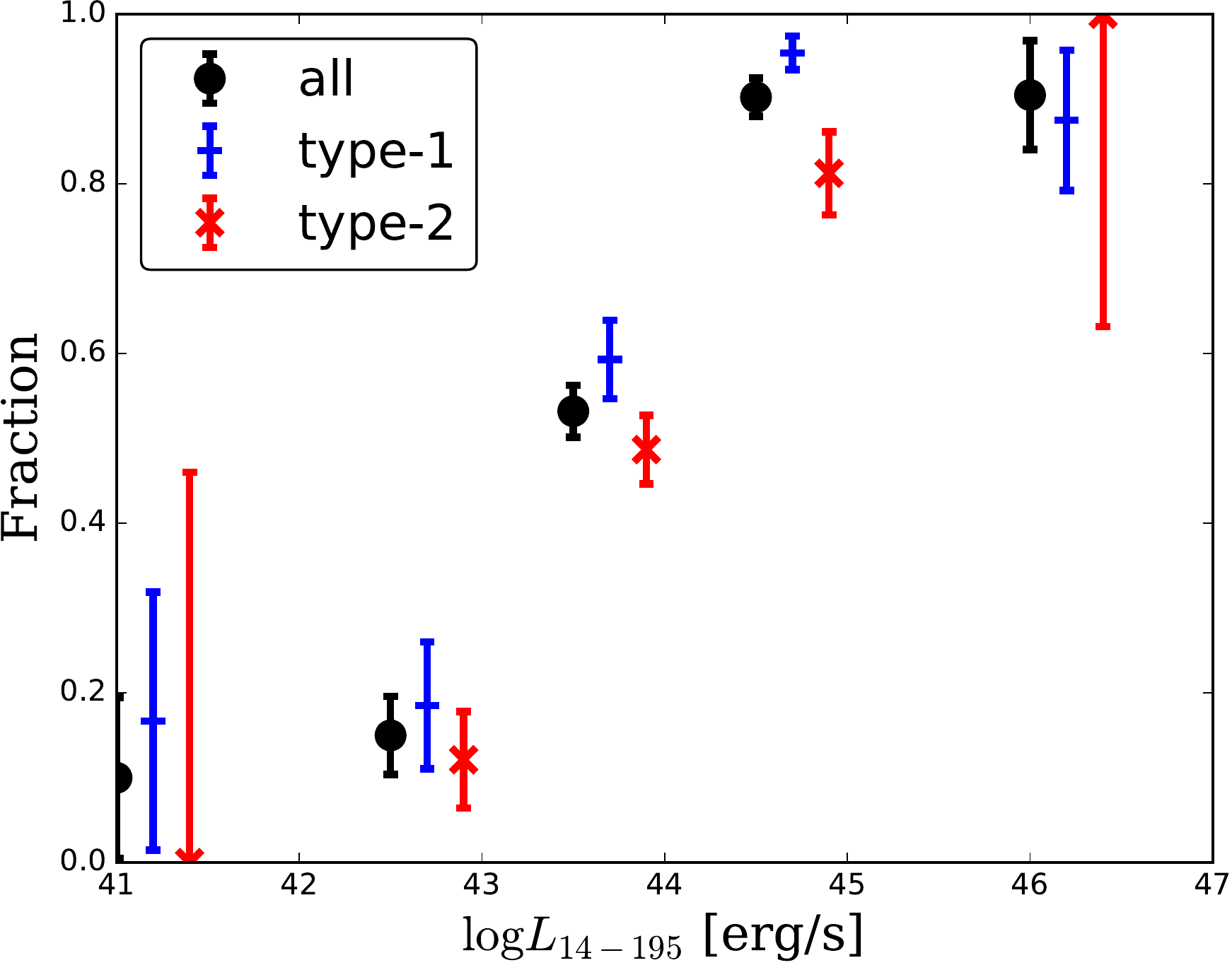}
\caption{
Fraction of AGN that meet the IR color selections as a function of BAT luminosity (in logarithmic units)
highlighted with black filled circle (all), blue cross (type-1), and red cross (type-2) color.
(Top) color cut with $W1-W2>0.8$ by \cite{ste12}. (Bottom) color cut by \cite{mat12a}.}\label{fig:fracvsLx}
\end{center}
\end{figure}

IR color-color selection is useful to identify obscured AGN candidates
 and also efficient compared to other time-consuming methods such 
 as spectroscopical methods.
Figure~\ref{fig:IR2color} shows the distribution of AGN on the \textit{WISE} 
color-color plane.
Increasing levels of AGN contribution to the MIR emission have
 been shown in Figure~\ref{fig:IR2color_type} to move sources
  upwards in the plane with the color cut $W1 - W2 = 0.8$ \citep{ste12} 
  and also within the AGN wedge \citep{mat12a}.
It is clear that our objects do not always locate within the criteria above. 
As discussed in Section 3.1, lower luminosity sources could have 
non-negligible level of contamination from the host galaxies in the 
NIR and MIR bands.
To check it quantitatively,  we divide the sample into subgroups of
 luminosities, then calculate the detection rate.
Figure~\ref{fig:fracvsLx} shows the detection rate of AGN using 
the thresholds of \cite{ste12} (top) and \cite{mat12a} (bottom), respectively.
The detection rate increases drastically at $L_{14-195}>10^{43}$~erg~s$^{-1}$
 and most ($>80$\%) sources can be selected using the IR color-color methods
at $L_{14-195}>10^{44}$~erg~s$^{-1}$. 
Thus, while the IR color-color methods are highly effective at high
 luminosities ($L_{14-195}>10^{44}$~erg~s$^{-1}$),
searching for faint AGN with $L_{14-195}<10^{44}$~erg~s$^{-1}$
 with near IR color-color methods should be complemented with
  other AGN identification methods such as hard ($E>2$~keV) 
  X-rays \citep[e.g.,][]{lam15}. 

Figure~\ref{fig:fracvsLx} shows that type-1 and type-2 AGN do 
not show any significant difference.
It indicates that the detection rate does not originate from the
 different AGN population, such as the effects of the suppressions
  of NIR SEDs in type-2 AGN due to heavier obscuration by the 
  torus clumps \citep{ram11}, but more likely by the dilution from
   the host galaxy stellar direct emission which causes blue $W1-W2$
    colors \citep[e.g.,][]{ste05, ris06, san08, ima10,ich14}.
The same trend is also reported in \cite{tob14}, which used the 
\textit{WISE}-matched SDSS AGN selected by the BPT diagram \citep{bal81}.
They showed that \textit{WISE} color-method efficiency 
increases with $L_{22~\mu{\rm m}}$.
\cite{kaw16b} also reported that hard X-ray selected 
low-luminosity AGN cannot be found using the IR 
color selections above. 
These results are consistent with what is shown in
 Figure~\ref{fig:fracvsLx} by considering that most 
 sources are at $L_{\rm 14-195}< 10^{44}$~erg~s$^{-1}$.
The same trend is also reported from the X-ray 
studies of Compton-thick AGN \citep{gan15, tan16}. 
They reported that secure Compton-thick AGN in the local
 universe do not preferentially locate within the AGN cut
  or wedge at $L_{\rm 14-195}< 10^{43}$~erg~s$^{-1}$.
However, even for the luminous AGN, if they are heavily obscured AGN such as
buried AGN, NIR and even MIR absorption may play a role to locate them outside
 the AGN cut or wedge \citep[e.g.,][]{hai14, ima16}.
In addition, some authors \citep[e.g.,][]{sat14, sec15} find that the fraction of AGN 
by \textit{WISE} color selection is highest at lower stellar masses and drops 
dramatically in higher mass galaxies, 
suggesting the stellar mass (or, Eddington-ratio) is another key parameter affecting
 the success rate  of the IR color selections as well as the X-ray luminosity discussed above.

\section{CONCLUSIONS}
We have compiled the IR (3--500~$\mu$m) counterparts of a nearby
 complete flux limited 604 AGN sources detected in the 70-month 
 integration of the \textit{Swift}/BAT all-sky survey in the 14--195 keV band.
Utilizing the IR catalogs obtained from \textit{WISE}, \textit{AKARI},
 \textit{IRAS}, and \textit{Herschel}, we identified 604, 560, 601, and 
 402 counterpart in the any IR, NIR, MIR, and FIR band, respectively.
For our discussion, the detected sources are divided into two AGN 
types based on $N_{\rm H}$ with a boundary of 
$N_{\rm H} = 10^{22}$~cm$^{-2}$.
Our results are summarized as follows:

\begin{enumerate}
\item We find a good luminosity correlation between the MIR
 and ultra hard X-ray band over 5 orders of magnitude
 ($41< \log (L_{14-195}/ {\rm erg}~{\rm s}^{-1})<46$). 
 Using the linear relation of  $\log (L_{\rm MIR}/10^{43}
 ~{\rm erg}~{\rm s}^{-1}) = a + b \log (L_{14-195}/10^{43}~
 {\rm erg}~{\rm s}^{-1})$, the slope $b=0.96-0.98$ is obtained 
 for the whole sample and $b=1.05-1.07$ for the high luminosity
  sample ($L_{14-195}>10^{43}$~erg~s$^{-1}$).
This value is consistent with those obtained by high spatial
 resolution MIR image observations of X-ray selected catalogs. 
 Whereas the slope is shallower than that obtained from the 
 sample of high-$z$ optically selected luminous AGN. 
 This indicates that X-ray emission could be saturated than 
 MIR ones in the high-luminosity end.

\item We find a rising trend between bolometric AGN power
 and FIR over 5 orders of magnitude in the individual plots. 
 The slope is consistent with that obtained by \cite{net09} as 
 well as \cite{mat15a}. The binned analysis also shows that 
 mean $L_{\rm bol}$ as a function of $L_{\rm FIR}$ shows
  the rising trend, which is consistent with the individual plot
   analysis. However, the mean $L_{\rm FIR}$ as a function 
   of $L_{\rm bol}$ shows a flattened trend. This seemingly
    contradicting result could be originated from the difference
     of the dominant timescale between SF and AGN activity
      that SF and BH accretion is closely connected over 
      long timescales, but this relation can be hidden at lower
       $L_{\rm bol}$ due to the short-term AGN variability
       \citep[e.g., ][]{mul12, che13, hic14}

\item We find a small number of FIR pure-AGN candidates 
which have strong AGN luminosity with very weak SF 
contribution from their host galaxies.
These objects represent a good sample to 
construct the pure-AGN IR
SED including the FIR end. They could be good
candidates to study AGN feedback since they might
be in the stage that SF activity is suppressed due to 
energy output from the AGN.

\item Using the correction from MIR to bolometric luminosity
 ratio to covering factor by \cite{sta16}, we find the covering
  factor decreases with bolometric luminosities, confirming
   the luminosity-dependent unified model.

\item  We find that the efficiency of the \textit{WISE} 
color-color cuts proposed by \cite{ste12} and \cite{mat12a} 
is highly AGN luminosity dependent. These methods cannot
 completely pick up local X-ray selected low-luminosity AGN 
 with $L_{14-195}<10^{44}$~erg~s$^{-1}$, while the color-color
  cut methods efficiently pick up most AGN with $L_{14-195}>10^{44}$~erg~s$^{-1}$.
 
\end{enumerate}



\acknowledgments

We thank the anonymous referee for a very careful reading of the manuscript and
numerous helpful suggestions that greatly strengthened the paper.
This research has made use of the NASA/IPAC Extragalactic Database (NED) which
 is operated by the Jet Propulsion Laboratory, California Institute of Technology, 
 under contract with the National Aeronautics and Space Administration.
 This research has made use of ``Aladin sky atlas'' developed at CDS, Strasbourg Observatory, France
 \citep{bon00, boc14}.
 K.I. thanks the Department of Astronomy at Kyoto university, where a part of the research was conducted.
 K.I. and Y.U. acknowledge support from JSPS Grant-in-Aid for Scientific Research (grant number 40756293: KI, 26400228: YU).
C.R. acknowledge financial support from the CONICYT-Chile grants ``EMBIGGEN'' Anillo ACT1101, 
FONDECYT 1141218, and Basal-CATA PFB--06/2007.
Part of this was financially supported by the Grant-in-Aid for JSPS fellow for young researchers (PD; KI. KM; DC1; TK.).

\clearpage
%

\onecolumngrid
\newpage
\setlength{\topmargin}{3.0cm}
\setlength{\tabcolsep}{0.020in} 
\floattable
\rotate
\renewcommand{\arraystretch}{1.1}
\begin{deluxetable}{clllllcccccccccccccccccccccccccc}
\thispagestyle{empty}
  \tablefontsize{\footnotesize}
  \tabletypesize{\scriptsize}
  \tablecaption{IR and 14--195~keV X-ray Properties of the \textit{Swift}/BAT 70-Month AGN Catalog  \label{tab:table2}}
  \tablewidth{0pt}
  \tablehead{
      \colhead{(1)} & \colhead{(2)} & \colhead{(3)} & \colhead{(4)} & \colhead{(5)} & \colhead{(6)} & \colhead{(7)} & \colhead{(8)} & \colhead{(9)} & \colhead{(10)} & \colhead{(11)} & \colhead{(12)} & \colhead{(13)} & \colhead{(14)} & \colhead{(15)} & \colhead{(16)} & \colhead{(17)} & \colhead{(18)} & \colhead{(19)} & \colhead{(20)} & \colhead{(21)} & \colhead{(22)} & \colhead{(23)} & \colhead{(24)} & \colhead{(25)} & \colhead{(26)} & \colhead{(27)} & \colhead{(28)}
\\
      \colhead{No.} & \colhead{Name} & \colhead{$z$} & \colhead{$f_{3.4}$} & \colhead{$f_{4.6}$} & \colhead{$f_{12}$} & \colhead{$f_{22}$} & \colhead{$f_{70}$} & \colhead{$f_{90}$} & \colhead{$f_{140}$} & \colhead{$f_{160}$} & \colhead{$f_{250}$} & \colhead{$f_{350}$} & \colhead{$f_{500}$} &  \colhead{IR Catalog} & \colhead{$L_{3.4}$} & \colhead{$L_{4.6}$} & \colhead{$L_{12}$} & \colhead{$L_{22}$} & \colhead{$L_{70}$} & \colhead{$L_{90}$} & \colhead{$L_{140}$} & \colhead{$L_{160}$} & \colhead{$L_{250}$} & \colhead{$L_{350}$} & \colhead{$L_{500}$} & \colhead{$C_{\rm T,12}$} & \colhead{$C_{\rm T,22}$}
    }
    \startdata
\thispagestyle{empty}
2 & Fairall 1203  & 0.058400 & $\cdots$ & $\cdots$ & $\cdots$ & $\cdots$ & 0.196 & $\cdots$ & $\cdots$ & $\cdots$ & $\cdots$ & $\cdots$ & $\cdots$ & (X,X,If,X,X) &  $\cdots$ &  $\cdots$ &  $\cdots$ &  $\cdots$ & 43.87 &  $\cdots$ &  $\cdots$ &  $\cdots$ &  $\cdots$ &  $\cdots$ &  $\cdots$ & $\cdots$ & $\cdots$  \\
3 & NGC 7811  & 0.025500 & 0.015 & 0.015 & 0.052 & 0.103 & 0.456 & $\cdots$ & $\cdots$ & $\cdots$ & $\cdots$ & $\cdots$ & $\cdots$ & (W,W,If,X,X) & 43.31 & 43.17 & 43.28 & 43.32 & 43.50 &  $\cdots$ &  $\cdots$ &  $\cdots$ &  $\cdots$ &  $\cdots$ &  $\cdots$ & 0.404 & 0.418  \\
6 & Mrk 335  & 0.025800 & 0.072 & 0.099 & 0.163 & 0.312 & 0.309 & $\cdots$ & $\cdots$ & 0.150 & 0.068 & $\cdots$ & $\cdots$ & (W,W,H,X,H) & 43.98 & 43.99 & 43.79 & 43.81 & 43.30 &  $\cdots$ &  $\cdots$ & 42.63 & 42.09 &  $\cdots$ &  $\cdots$ & 0.507 & 0.515  \\
7 & 2MASX J00091156-0036551  & 0.073300 & 0.005 & 0.008 & $\cdots$ & $\cdots$ & $\cdots$ & $\cdots$ & $\cdots$ & $\cdots$ & $\cdots$ & $\cdots$ & $\cdots$ & (X,X,X,X,X) & 43.80 & 43.85 &  $\cdots$ &  $\cdots$ &  $\cdots$ &  $\cdots$ &  $\cdots$ &  $\cdots$ &  $\cdots$ &  $\cdots$ &  $\cdots$ & $\cdots$ & $\cdots$  \\
10 & 2MASX J00210753-1910056  & 0.095600 & 0.011 & 0.013 & 0.023 & 0.054 & $\cdots$ & $\cdots$ & $\cdots$ & $\cdots$ & $\cdots$ & $\cdots$ & $\cdots$ & (W,W,X,X,X) & 44.35 & 44.30 & 44.12 & 44.23 &  $\cdots$ &  $\cdots$ &  $\cdots$ &  $\cdots$ &  $\cdots$ &  $\cdots$ &  $\cdots$ & 0.222 & 0.230  \\
14 & 2MASX J00264073-5309479  & 0.062900 & 0.009 & 0.009 & 0.013 & 0.024 & $\cdots$ & $\cdots$ & $\cdots$ & $\cdots$ & $\cdots$ & $\cdots$ & $\cdots$ & (W,W,X,X,X) & 43.89 & 43.77 & 43.49 & 43.50 &  $\cdots$ &  $\cdots$ &  $\cdots$ &  $\cdots$ &  $\cdots$ &  $\cdots$ &  $\cdots$ & 0.218 & 0.218  \\
16 & [HB89] 0026+129  & 0.14200 & 0.013 & 0.017 & 0.028 & 0.048 & $\cdots$ & $\cdots$ & $\cdots$ & $\cdots$ & $\cdots$ & $\cdots$ & $\cdots$ & (W,W,X,X,X) & 44.78 & 44.78 & 44.58 & 44.55 &  $\cdots$ &  $\cdots$ &  $\cdots$ &  $\cdots$ &  $\cdots$ &  $\cdots$ &  $\cdots$ & 0.228 & 0.226  \\
17 & ESO112-6  & 0.029000 & 0.010 & 0.007 & 0.025 & 0.049 & 1.019 & 1.183 & $\cdots$ & $\cdots$ & $\cdots$ & $\cdots$ & $\cdots$ & (W,W,Ip,A,X) & 43.22 & 42.95 & 43.09 & 43.11 & 43.96 & 43.88 &  $\cdots$ &  $\cdots$ &  $\cdots$ &  $\cdots$ &  $\cdots$ & 0.297 & 0.302  \\
19 & 2MASX J00341665-7905204  & 0.074000 & 0.022 & 0.033 & 0.082 & 0.152 & $\cdots$ & $\cdots$ & $\cdots$ & $\cdots$ & $\cdots$ & $\cdots$ & $\cdots$ & (W,W,X,X,X) & 44.41 & 44.46 & 44.44 & 44.44 &  $\cdots$ &  $\cdots$ &  $\cdots$ &  $\cdots$ &  $\cdots$ &  $\cdots$ &  $\cdots$ & 0.490 & 0.491  \\
20 & 2MASX J00343284-0424117  & 0.21300 & 0.001 & 0.003 & 0.010 & 0.027 & $\cdots$ & $\cdots$ & $\cdots$ & $\cdots$ & $\cdots$ & $\cdots$ & $\cdots$ & (W,W,X,X,X) & 44.19 & 44.45 & 44.51 & 44.69 &  $\cdots$ &  $\cdots$ &  $\cdots$ &  $\cdots$ &  $\cdots$ &  $\cdots$ &  $\cdots$ & 0.219 & 0.226  \\

    \enddata
    \thispagestyle{empty}
\tablenotetext{}{Notes.-- Infrared to X-ray properties of \textit{Swift}/BAT 70-month AGN located at $|b|>10^{\circ}$.
(1) source No. in \cite{bau13};
(2) object name;
(3) redshift;
(4)--(14) IR flux density ($f_{\nu}$) at 3.4, 4.6, 12, 22, 70, 90, 140, 160, 250, 350, and 500~$\mu$m in units of Jy;
(15) IR reference catalogs for 12~$\mu$m, 22~$\mu$m, 70~$\mu$m, 90~$\mu$m, and 160~$\mu$m:
A  $=$ AKARI PSC,
H  $=$ Herschel/PACS,
If $=$ IRAS Faint Source Catalog,
Ip $=$ IRAS Point Source Catalog,
W  $=$ WISE,
X  $=$ Non Detection;
(16)--(26) logarithmic IR luminosities ($\log \lambda L_{\lambda}$) at 3.4, 4.6, 12, 22, 70, 90, 140, 160, 250, 350, and 500~$\mu$m in units of erg~s$^{-1}$;
(27)--(28) covering factor based on the correction of \cite{sta16} using the MIR information at 12 and 22~$\mu$m, respectively.\\
(This table is available in its entirety in the machine-readable format. A portion is shown here for guidance regarding its form and content.)
}\label{tab:IRcatalog}
\end{deluxetable}
\setlength{\topmargin}{0in}

\newpage

\begin{table*}
\begin{center}
\scriptsize
\caption{Correlation Parameters between the IR and the 14--195~keV luminosity}\label{tab:LIRvsLx}
\begin{tabular}{rrcccccccccccccc}
\hline
\hline
\multicolumn{1}{c}{Band} &
\multicolumn{1}{c}{$N$} &
\multicolumn{1}{c}{$\rho_{L}$} &
\multicolumn{1}{c}{$\rho_{f}$} &
\multicolumn{1}{c}{$P_{L}$} &
\multicolumn{1}{c}{$P_{f}$} &
\multicolumn{1}{c}{$a$} & 
\multicolumn{1}{c}{$b$}  &
\\
(1) & (2) & (3) & (4) & (5) & (6) & (7) & (8)  \\
\hline
3.4 & 549 & 0.860 & 0.543 & $\checkmark$ & $\checkmark$ & $ 0.000 \pm0.028 $ & $ 0.890 \pm0.030 $ \\
4.6 & 548 & 0.869 & 0.541 & $\checkmark$ & $\checkmark$ & $ -0.128 \pm0.026 $ & $ 0.977 \pm0.027 $ \\
12 & 596 & 0.815 & 0.547 & $\checkmark$ & $\checkmark$ & $ -0.099 \pm0.024 $ & $ 0.963 \pm0.022 $ \\
22 & 592 & 0.789 & 0.516 & $\checkmark$ & $\checkmark$ & $ 0.024 \pm0.024 $ & $ 0.979 \pm0.022 $ \\
70 & 388 & 0.526 & 0.329 & $\checkmark$ & $\checkmark$ & $ 0.235 \pm0.035 $ & $ 0.943 \pm0.036 $ \\
90 & 241 & 0.612 & 0.343 & $\checkmark$ & $\checkmark$ & $ 0.407 \pm0.032 $ & $ 0.904 \pm0.035 $ \\
140 & 89 & 0.710 & 0.324 & $\checkmark$ & $\checkmark$ & $ 0.683 \pm0.045 $ & $ 0.867 \pm0.045 $ \\
160 & 229 & 0.244 & 0.271 & $\checkmark$ & $\checkmark$ & $ 0.098 \pm0.044 $ & $ 0.884 \pm0.046 $ \\
250 & 213 & 0.260 & 0.306 & $\checkmark$ & $\checkmark$ & $ -0.459 \pm0.050 $ & $ 0.908 \pm0.046 $ \\
350 & 170 & 0.417 & 0.277 & $\checkmark$ & $\checkmark$ & $ -0.788 \pm0.050 $ & $ 0.821 \pm0.060 $ \\
500 & 107 & 0.507 & 0.391 & $\checkmark$ & $\checkmark$ & $ -1.163 \pm0.046 $ & $ 0.715 \pm0.061 $ \\

 \hline
\end{tabular}\\
Notes.--- Correlation properties between 14--195 keV X-ray luminosity and infrared luminosities.
 (1) IR Band with a unit of $\mu$m; 
 (2) number of sample; 
 (3) the Spearman’s Rank coefficient for luminosity correlations ($\rho_L$); 
 (4) the Spearman’s Rank coefficient for flux--flux correlations ($\rho_f$);
 (5) the standard Student t-test null significance level for luminosity correlations ($P_L$). $\checkmark$ represents $P_L<0.01$;
 (6) the standard Student t-test null significance level for flux--flux correlations ($P_f$). $\checkmark$ represents $P_f<0.01$;
 (7) regression intercept (a) and its 1$\sigma$ uncertainty; 
 (8) slope value ($b$) and its 1$\sigma$ uncertainty. Equation is represented as $Y = a + bX$.
\end{center}
\end{table*}


\setlength{\topmargin}{3.0cm}
\setlength{\tabcolsep}{0.020in}
\floattable
\rotate
\renewcommand{\arraystretch}{1.1}
\begin{deluxetable}{cccccccc}
\thispagestyle{empty}
  \tablefontsize{\footnotesize}
  \tabletypesize{\scriptsize}
  \tablecaption{Equations of the luminosity correlation between the IR and X-Ray band}\label{tab:LIRvsLx_wlit}
  \tablewidth{0pt}
  \tablehead{
      \colhead{(1)} & \colhead{(2)} & \colhead{(3)} & \colhead{(4)} & \colhead{(5)} & \colhead{(6)} & \colhead{(7)} & \colhead{(8)}\\
      \colhead{IR band} & \colhead{X-ray band} & \colhead{Equation} & \colhead{$z$ range} & \colhead{$L_{\rm X}$ range} & \colhead{Selection} & \colhead{AGN type} & \colhead{Reference}
    }
    \startdata
\thispagestyle{empty}
12~$\mu$m & 14--195~keV &  $\log \frac{L_{12 \mu{\rm m}}}{10^{43}~{\rm erg/s}} = (-0.10 \pm 0.02) + (0.96 \pm 0.02) \log \frac{L_{\rm 14-195}}{10^{43}~{\rm erg/s}}$  &$z<0.3$ & $41<\log L_{\rm 14-195} < 46$ & X-ray & both & this study\\
22~$\mu$m & 14--195~keV &  $\log \frac{L_{22 \mu{\rm m}}}{10^{43}~{\rm erg/s}} = (0.02 \pm 0.02) + (0.98 \pm 0.02) \log \frac{L_{\rm 14-195}}{10^{43}~{\rm erg/s}}$  &$z<0.3$ & $41<\log L_{\rm 14-195} < 46$ & X-ray & both & this study\\
12~$\mu$m & 14--195~keV &  $\log \frac{L_{12 \mu{\rm m}}}{10^{43}~{\rm erg/s}} = (-0.21 \pm 0.03) + (1.05 \pm 0.03) \log \frac{L_{\rm 14-195}}{10^{43}~{\rm erg/s}}$  &$z<0.3$ & $43<\log L_{\rm 14-195} < 46$ & X-ray & both & this study\\
22~$\mu$m & 14--195~keV &  $\log \frac{L_{22 \mu{\rm m}}}{10^{43}~{\rm erg/s}} = (-0.09 \pm 0.03) + (1.07 \pm 0.03) \log \frac{L_{\rm 14-195}}{10^{43}~{\rm erg/s}}$  &$z<0.3$ & $43<\log L_{\rm 14-195} < 46$ & X-ray & both & this study\\
6~$\mu$m & 2--10~keV & $\log \frac{L_{12 \mu{\rm m}}}{10^{41}~{\rm erg/s}} \simeq 2.1 \times 10^{-2} \left(512 -\sqrt{ 2.2 \times 10^6 - 4.7\times 10^4 \log \frac{L_{\rm 2-10}}{10^{41}~{\rm erg/s}}}\right)$ &$1.5<z<4.7$& $45<\log L_{\rm 2-10} < 46.2$ & optical & type-1 & \cite{ste15}\\
6~$\mu$m & 2--10~keV & $\log \frac{L_{6 \mu{\rm m}}}{10^{43}~{\rm erg/s}} = 0.40 + 1.39 \log \frac{L_{\rm 2-10}}{10^{43}~{\rm erg/s}}$ & $0.2<z<4$  & $42<\log L_{\rm 2-10} < 46$ & X-ray & type-1 & \cite{fio09}\\
12~$\mu$m & 2--10~keV & $\log \frac{L_{12 \mu{\rm m}}}{10^{43}~{\rm erg/s}} = (0.19 \pm 0.05) + (1.11 \pm 0.07) \log \frac{L_{\rm 2-10}}{10^{43}~{\rm erg/s}}$ &$z<0.1$ & $41<\log L_{\rm 2-10} < 45$ & X-ray & both & \cite{gan09}\\
12~$\mu$m & 2--10~keV & $\log \frac{L_{12 \mu{\rm m}}}{10^{43}~{\rm erg/s}} = 0.30 + 0.99 \log \frac{L_{\rm 2-10}}{10^{43}~{\rm erg/s}}$ &$0.05<z<2.8$& $42<\log L_{\rm 2-10} < 46$ & X-ray & both & \cite{mat15b}\\
12~$\mu$m & 2--10~keV & $\log \frac{L_{12 \mu{\rm m}}}{10^{43}~{\rm erg/s}} = (0.33 \pm 0.04) + (0.97 \pm 0.03) \log \frac{L_{\rm 2-10}}{10^{43}~{\rm erg/s}}$ &$z<0.3$& $40<\log L_{\rm 2-10} < 46$ & X-ray & both & \cite{asm15}\\

    \enddata
    \thispagestyle{empty}
\tablenotetext{}{
Notes.--- Correlation properties between 14--195 keV X-ray luminosity and infrared luminosities.
 (1) IR Band with a unit of $\mu$m;
 (2) X-ray Band with a unit of $\mu$m;
 (3) Equation of the luminosity correlation between IR and X-ray;
 (4) $z$ range of the sample for each study;
 (5) X-ray luminosity range of the sample for each study;
 (6) The energy band used to select the parent AGN sample;
 (7) The main AGN type in the sample. ``type-1'' represents type-1 AGN, and ``both'' represents both type-1 and type-2 AGN;
 (8) Reference
}\label{tab:IRXcatalog}
\end{deluxetable}
\setlength{\topmargin}{0in}


\begin{table}
\begin{center}
\scriptsize
\caption{Standard Deviation of the IR to the expected slope}
\begin{tabular}{rrcc}
\hline
\hline
\multicolumn{1}{c}{Sample} &
\multicolumn{1}{c}{$N$} &
\multicolumn{1}{c}{$\sigma$}
\\
(1) & (2) & (3)\\
\hline
3.4$\ \mu$m all & 549 & 0.3377 $\pm$ 0.0204 \\
type-1 & 279 & 0.3255 $\pm$ 0.0276 \\
type-2 & 270 & 0.3323 $\pm$ 0.0287 \\
4.6$\ \mu$m all & 548 & 0.3544 $\pm$ 0.0214 \\
type-1 & 281 & 0.3339 $\pm$ 0.0282 \\
type-2 & 267 & 0.3617 $\pm$ 0.0314 \\
12$\ \mu$m all & 596 & 0.3937 $\pm$ 0.0228 \\
type-1 & 307 & 0.3702 $\pm$ 0.0299 \\
type-2 & 289 & 0.4176 $\pm$ 0.0348 \\
22$\ \mu$m all & 592 & 0.4266 $\pm$ 0.0248 \\
type-1 & 302 & 0.3941 $\pm$ 0.0321 \\
type-2 & 290 & 0.4572 $\pm$ 0.0380 \\
70$\ \mu$m all & 388 & 0.5496 $\pm$ 0.0395 \\
type-1 & 173 & 0.5173 $\pm$ 0.0558 \\
type-2 & 215 & 0.5659 $\pm$ 0.0547 \\
90$\ \mu$m all & 241 & 0.4665 $\pm$ 0.0426 \\
type-1 & 87 & 0.4566 $\pm$ 0.0696 \\
type-2 & 154 & 0.4727 $\pm$ 0.0540 \\
140$\ \mu$m all & 89 & 0.4291 $\pm$ 0.0647 \\
type-1 & 30 & 0.4593 $\pm$ 0.1206 \\
type-2 & 59 & 0.4161 $\pm$ 0.0773 \\
160$\ \mu$m all & 229 & 0.5860 $\pm$ 0.0549 \\
type-1 & 100 & 0.5816 $\pm$ 0.0827 \\
type-2 & 129 & 0.5906 $\pm$ 0.0738 \\
250$\ \mu$m all & 213 & 0.6155 $\pm$ 0.0598 \\
type-1 & 101 & 0.5973 $\pm$ 0.0845 \\
type-2 & 112 & 0.6270 $\pm$ 0.0842 \\
350$\ \mu$m all & 170 & 0.5316 $\pm$ 0.0578 \\
type-1 & 75 & 0.5245 $\pm$ 0.0862 \\
type-2 & 95 & 0.5362 $\pm$ 0.0782 \\
500$\ \mu$m all & 107 & 0.4487 $\pm$ 0.0616 \\
type-1 & 40 & 0.4589 $\pm$ 0.1039 \\
type-2 & 67 & 0.4458 $\pm$ 0.0776 \\

  \hline
\end{tabular}\\
Notes.--- 
 (1) IR Band with a unit of $\mu$m; 
 (2) number of sample; 
 (3) standard deviation ($\sigma$) from the expected value $L_{\rm IR}^{\rm (slope)}$ given the slope in $L_{\rm IR}$ and $L_{14-195}$.
\end{center}\label{tab:LIRovLx}
\end{table}


\appendix

\section{Flux correlation between MIR bands and 14--195~keV}

\begin{figure*}
\begin{center}
\includegraphics[width=8.2cm]{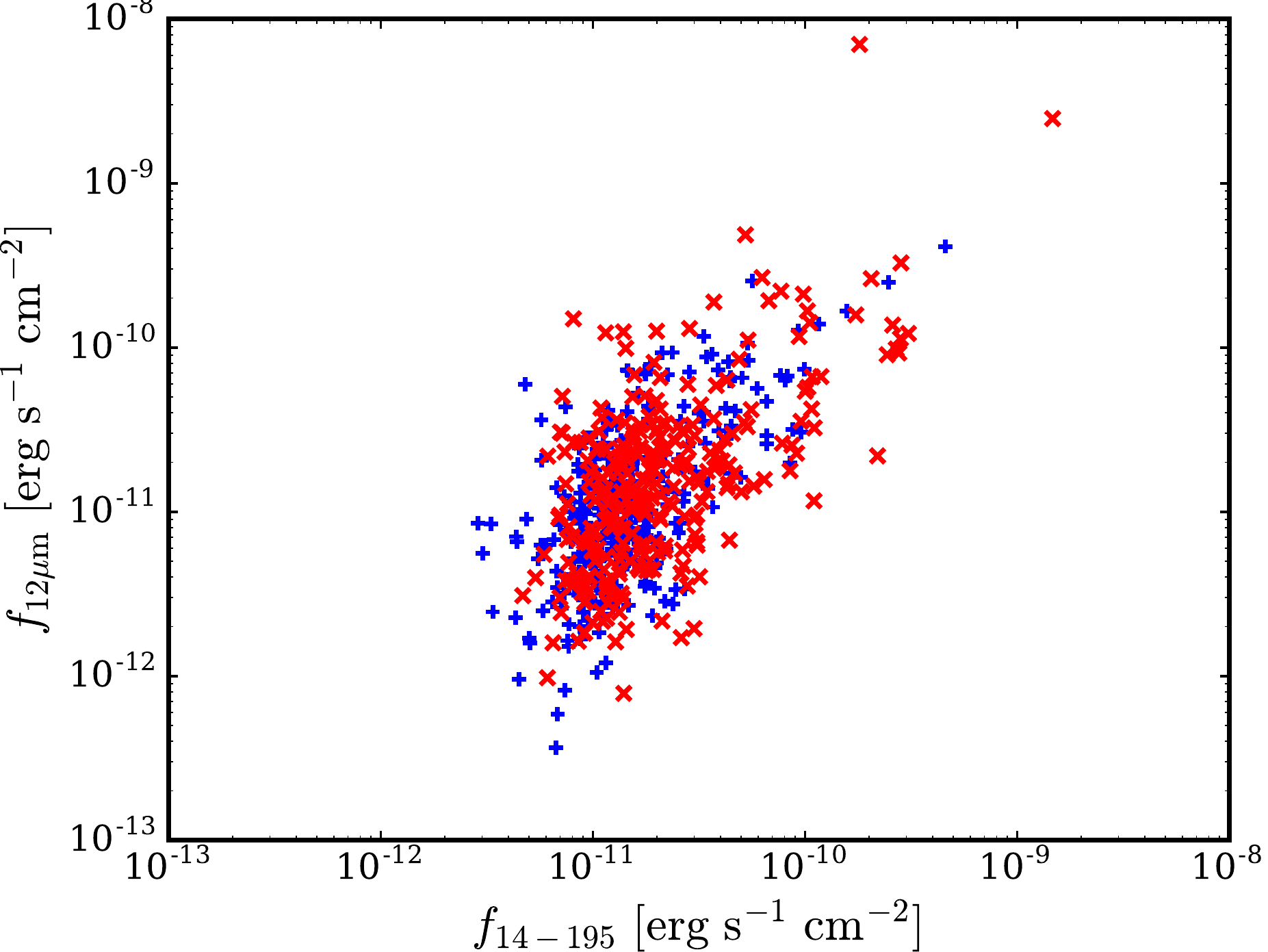}~
\includegraphics[width=8.2cm]{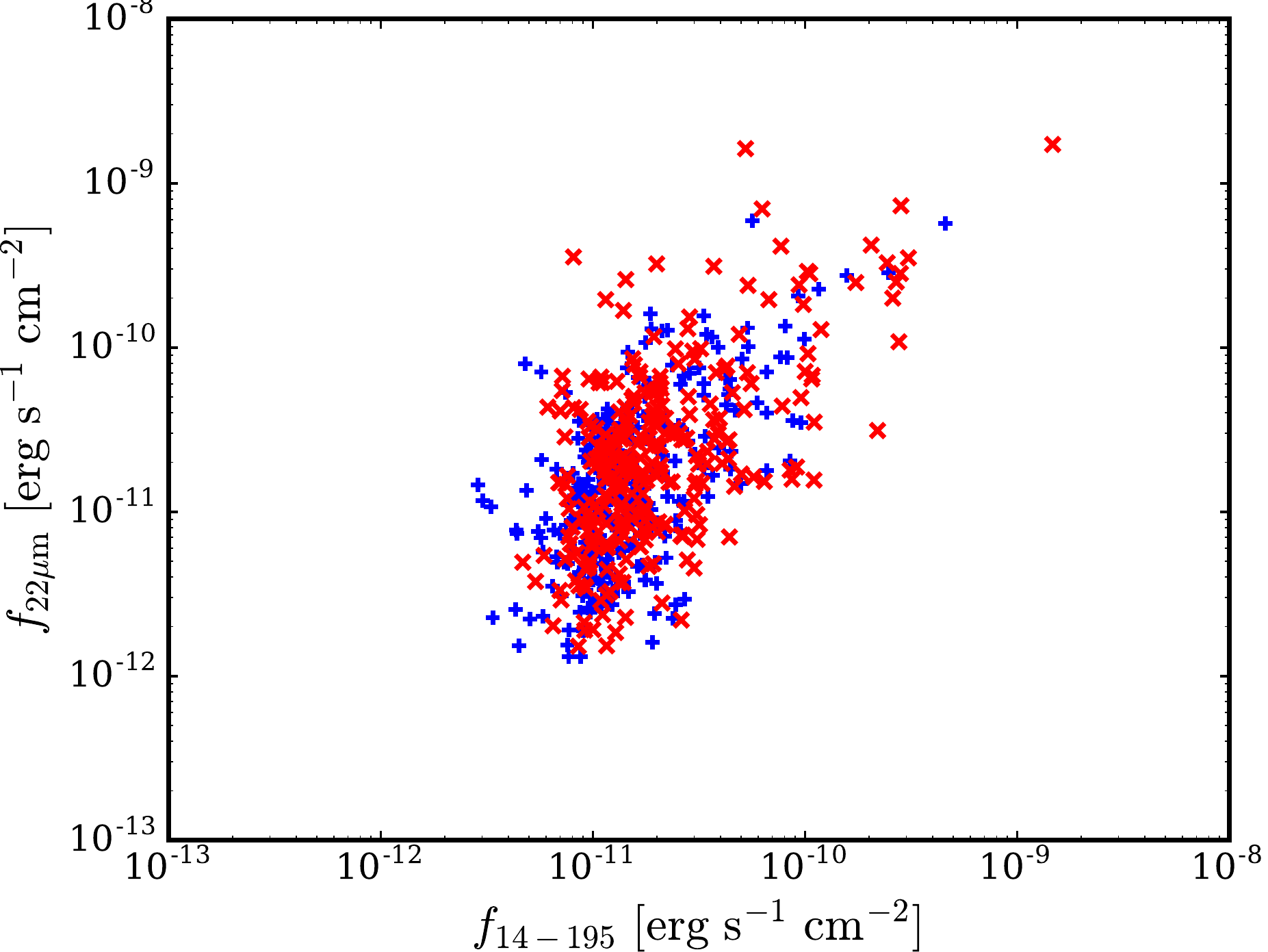}
\caption{
Correlation between the fluxes at each MIR band and $L_{14-195~{\rm keV}}$ for the 12 and 22~$\mu$m band.
Blue/red color represents type-1/-2, respectively.}\label{fig:fIRvsfx_supp}
\end{center}
\end{figure*}

Figure~\ref{fig:fIRvsfx_supp} shows the flux correlations between
 the MIR (12 and 22~$\mu$m) and 14--195~keV.
This shows that there is a clear correlation between two bands 
even in the flux-flux plots.
The figure also clearly shows that our sample is X-ray flux limited.
There is a clear sharp decline of the number of AGN in the sample at faint 14--195~keV
 fluxes,  especially at $f_{14-195}< 10^{-11}$~erg~s$^{-1}$~cm$^{-2}$.
 The MIR detection limits for these sources are typically $2.5 \times 10^{-13}$~erg~s$^{-1}$~cm$^{-2}$
 in the 12~$\mu$m and $8.1 \times 10^{-13}$~erg~s$^{-1}$~cm$^{-2}$ in the 22~$\mu$m, which
are below the detected fluxes from the sources. Therefore the sample is effectively complete
 in the MIR and the selection is dominated by the X-ray flux limits
 as discussed in Section~\ref{sect:LxvsLIR}.

\listofchanges

\end{document}